\documentclass[twocolumn]{aastex631}

\newcommand{\mytilde}{\raise.19ex\hbox{$\scriptstyle\sim$}}

\usepackage{amsmath}

\begin{document}

\title{Near-IR Weak-lensing (NIRWL) Measurements in the CANDELS Fields. II. Mass Mapping and Overdensity Characterization}

\correspondingauthor{Kyle Finner}
\email{kfinner@caltech.edu}

\author[0000-0002-4462-0709]{Kyle Finner}
\affiliation{IPAC, California Institute of Technology, 1200 E California Blvd., Pasadena, CA 91125, USA}
\author[0000-0003-1954-5046]{Bomee Lee}
\affiliation{Korea Astronomy and Space Science Institute (KASI), 776 Daedeokdae-ro, Yuseong-gu, Daejeon 34055, Republic of Korea}

\author[0000-0001-7583-0621]{Ranga-Ram Chary}
\affiliation{SPACE Institute, University of California, 603 Charles E. Young Drive East, Los Angeles, CA 90095-1567}

\author[0000-0003-2508-0046]{Giuseppe Congedo}
\affiliation{ Institute for Astronomy, University of Edinburgh, Royal Observatory, Blackford Hill, Edinburgh, EH9 3HJ, United Kingdom}


\author[0000-0002-2550-5545]{Kim HyeongHan} 
\affiliation{Department of Physics, Duke University, Durham, NC 27708, USA}

\author[0000-0002-5751-3697]{M. James Jee}
\affiliation{Department of Astronomy, Yonsei University, 50 Yonsei-ro, Seodaemun-gu, Seoul 03722, Republic of Korea}
\affiliation{Department of Physics and Astronomy, University of California, Davis, One Shields Avenue, Davis, CA 95616, USA}

\author[0000-0001-6999-4718]{Peter Taylor} 
\affiliation{Center for Cosmology and Astroparticle Physics, The Ohio State University, 191 West Woodruff Ave, Columbus, Ohio 43210, USA}
\affiliation{Department of Physics, The Ohio State University, 191 West Woodruff Ave, Columbus, Ohio 43210, USA}
\affiliation{Department of Astronomy, The Ohio State University, 140 West 18th Ave, Columbus, Ohio 43210, USA}



\begin{abstract}
The Hubble Space Telescope Cosmic Assembly Near-infrared Deep Extragalactic Legacy Survey (CANDELS) fields offer an exceptional combination of depth, spatial resolution, and area for identifying a shear-selected sample of dark matter overdensities. We present the first near-infrared (NIR) weak-lensing (WL) analysis of the 0.23 square degrees covered by the HST CANDELS fields: COSMOS, UDS, EGS, GOODS-N, and GOODS-S. Leveraging the high sensitivity of HST NIR imaging to distant galaxies, we achieve a WL source galaxy density of $\sim170$ galaxies arcmin$^{-2}$. Our analysis identifies 12 shear-selected overdensities spanning masses from $M_{200}=(0.2$--$2.2)\times10^{14}\ M_\odot$, with a median mass of $M_{200}=5.5\times10^{13}\ M_\odot$, demonstrating the strong capability of NIR WL for measuring low-mass systems. The systems lie in the redshift range $0.22<z<0.9$, with a mean redshift of $z=0.68$. We utilize multiwavelength data to confirm the nature of the overdensities. Seven of the overdensities have diffuse X-ray emission reported in the literature, with X-ray centroids that are spatially consistent with our WL peaks, confirming their nature as collapsed structures. We find that our WL detections broadly follow the expected X-ray luminosity--WL mass scaling relations. By stacking the tangential shear of all detections, we determine the average radial mass density profile and find that it is well fit by an NFW model with fitted concentration and mass of $4.9\pm2.1$ and $M_{200}=1.3\pm0.3\times10^{14}\ M_\odot$, respectively. These results serve as a precursor to NIR WL science with the Roman High Latitude Wide Area Survey.

\end{abstract}



\section{Introduction} \label{sec:intro}
In the $\Lambda$ cold dark matter ($\Lambda$CDM) framework, the large-scale structure of the Universe forms through hierarchical processes \citep{1974press-schecter, 1984blumenthal, 1985davis}. Initial small fluctuations in the primordial matter distribution are amplified by gravity and collapse to form the first structures. These smaller structures subsequently merge and accrete more matter, giving rise to the structure of the cosmic web observed today. Investigating the distribution and evolution of these matter overdensities is therefore crucial for validating the $\Lambda$CDM model and for gaining deeper insights into the fundamental constituents and evolution of the Universe.


The detection of overdensities has been predominantly achieved through X-ray imaging \citep[e.g.,][]{2001ebeling, 2016pierre, 2024bulbul}, millimeter observations of the Sunyaev Zel'dovich (SZ) effect \citep[e.g.,][]{2017carmen, 2021hilton, 2024bleem}, and galaxy surveys \citep[e.g.,][]{1958abell, 2001gladders, 2008wilson, 2016rykoff}, which are dependent on the baryon content or dynamical state of structures. A different approach to detecting overdense regions of the universe is utilizing weak gravitational lensing (WL), which provides a direct probe of the total matter distribution without these dependencies. WL uses the distortion, or shear, of background galaxy images to detect the foreground mass. Ground-based surveys have proven useful for detecting shear-selected overdensities via their WL signal \citep[e.g.,][]{2002wittman, 2006wittman, 2002miyazaki, 2007miyazaki, 2007schirmer, 2007gavazzi, 2014utsumi, 2021oguri}. A limiting factor in these surveys is the spatial resolution ($\sim$0\farcs5 at best, due to seeing) and sky brightness from the ground, which together limit the galaxy source density. Space-based surveys can push beyond this source density limit through improved spatial resolution. For example, \cite{2007leauthaud} and \cite{2007massey} performed a WL analysis of the COSMOS survey footprint using the Hubble Space Telescope (HST) Advanced Camera for Surveys (ACS) data to map and characterize the large-scale structure. Notably, the COSMOS WL survey footprint ($\sim$ 1.64 deg$^2$ for the WL analysis) is small compared to ground-based surveys. However, deep fields with high-resolution imaging provide a substantially higher number density of background galaxies, enabling the detection and characterization of lower-mass overdensities at higher redshifts. For example, WL has been applied to HST NIR imaging to measure galaxy clusters at $z>1.5$ \citep{2017jee, 2020finner, 2025kim}, and JWST is enabling studies of the most distant clusters \citep{2026scofield}.

The Cosmic Assembly Near-infrared Deep Extragalactic Legacy Survey \citep[CANDELS;][]{2011grogin, 2011koekemoer} fields, with their exceptional depth, high-resolution imaging from the HST, and extensive multi-wavelength coverage, provide a unique environment for WL studies. The deep, infrared imaging provides a high number density of background galaxies \citep{2018lee} and, combined with accurate photometric redshifts, enables a sensitive search for localized mass concentrations through WL. In \cite{2023cfinner}, we created WL quality mosaic images\footnote{https://archive.stsci.edu/hlsp/candelsnirwl} of the CANDELS fields in the F160W filter that minimized astrometry and distortion effects that can arise in aligning and stacking HST imaging. Furthermore, we derived robust PSF models and quantified systematic effects such as undersampling, that could impact our WL analysis.  

In this work, we perform a WL analysis of the five CANDELS fields to search for and characterize shear-selected overdensities. In section \ref{sec:data}, we describe the data set and our method of source detection. In section \ref{sec:methods}, we cover PSF modeling (\ref{sec:psf}), shape measurements (\ref{sec:shape_measurements}), source selection (\ref{sec:source_selection}), mass mapping (\ref{sec:mass_mapping}), identification of WL overdensities (\ref{sec:identification}), and mass estimation (\ref{sec:mass_estimation}). We present the results of our analysis in Section \ref{sec:results} and discuss their significance in Section \ref{sec:discussion}. We conclude in Section \ref{sec:conclusions}.

We fix the cosmology of our analysis to a flat $\Lambda$CDM with $h=0.7$, $\Omega_m=0.3$, and $\Omega_\Lambda=0.7$. Masses are reported as $M_\Delta$ enclosed within $R_\Delta$, where $R_\Delta$ is the radius at which the average density within a spherical overdensity is $\Delta$ times the critical density of the Universe. Magnitudes are reported in the AB system.

\section{Data Sets and Source Extraction} \label{sec:data}
Mosaic science images for the five CANDELS fields were generated as described in detail in the companion paper by \cite{2023cfinner}. Briefly, these mosaic images were created from the \textit{HST} F160W frames that form a contiguous footprint for each field. Using the Drizzle pipeline \citep{2012drizzle}, the individual frames were drizzled using a Gaussian kernel to mosaics with a pixel scale of $50~mas$. Companion weight images were saved from the drizzle step and root-mean-square (rms) images were created. These five CANDELS F160W mosaics are utilized for galaxy shape measurements in this study.

Photometry was performed with the \textsc{SExtractor} \citep{1996bertin} software in dual-image mode where the weight and rms images were provided to the \texttt{WEIGHT\_IMAGE} keyword as weights for the detection and measurements, respectively. Our goal was to detect as many faint and small, but real, sources as possible so that galaxy shape measurement becomes the limiting factor. In the \textsc{SExtractor} parameter file, we set the \texttt{DETECT\_MINAREA}=5 pixels, \texttt{DETECT\_THRESH}=1.5, \texttt{DEBLEND\_NTHRESH}=32, \texttt{DEBLEND\_MINCONT}=0.005, and use the 2-pixel Gaussian convolution filter. This detection technique results in approximately 40,000 
objects per field. 



\section{Methods} \label{sec:methods}
\subsection{PSF Modeling} \label{sec:psf}
The distortion caused by PSF variations and pixelization, is a systematic in measuring the shape distortion introduced by WL. In \cite{2023cfinner}, we developed PSF models using a principal component analysis of stars in the CANDELS fields and extensively investigated the systematics that could impact our study. We showed that the level of systematic shape bias is of the order 0.02, and that careful selection of background galaxies can aid in relieving the most impactful systematic, which was determined to be from PSF undersampling. For this study, PSF models are generated on a 31 pixel by 31 pixel grid and applied via forward modeling during galaxy shape fitting.

\subsection{Galaxy Shape Measurements} \label{sec:shape_measurements}
The WL signal is a coherent distortion of background galaxy images by the gravitational potential of foreground masses. In the WL regime, a statistical analysis of galaxy shapes can be used to recover the WL signal. This is achieved by averaging the ellipticities of galaxies, which under no shear and randomly oriented position angles should equal zero.

Our galaxy shape measurement pipeline follows the Shapes From Iterative Training \citep[SFIT;][]{2013jee} technique, which fits a PSF-convolved two-dimensional elliptical Gaussian model to galaxy image cutouts and calibrates the resulting shape measurements using image simulations. Full details of the shape-fitting pipeline are described in \citep{2017finner, 2020finner}. The elliptical Gaussian model has seven parameters: the background level ($A_0$), the amplitude ($A_1$), the centroid position ($x$, $y$), the complex ellipticity components ($e_1$, $e_2$), and the semi-major axis size ($a$). The ellipticity and its components are defined as
\begin{align}
    e &= \frac{a-b}{a+b}, \\
    e_1 &= e\cos(2\theta), \\
    e_2 &= e\sin(2\theta),
\end{align}
where $b$ is the semi-minor axis and $\theta$ is the position angle of the ellipse.

The best-fit model parameters are obtained by minimizing the $\chi^2$ statistic,
\begin{equation}
\chi^2 = \sum \left(\frac{I - P \circledast G}{\sigma_{\mathrm{rms}}}\right)^2,
\end{equation}
with \textsc{MPFIT} \citep{2009markwardt}, where $I$ is the image cutout, $P$ is the PSF model, $G$ is the Gaussian model, and $\sigma_{\mathrm{rms}}$ is the rms image cutout. We construct galaxy cutouts from the F160W mosaics with a size equal to seven times the \texttt{A\_IMAGE} measurement from \textsc{SExtractor}. This choice mitigates severe truncation bias \citep[see][]{2018mandelbaum} at the expense of including some nearby galaxies. During the fit, we down-weight the pixels belonging to these nearby galaxies by artificially increasing their $\sigma_{\mathrm{rms}}$ values, causing the fitting algorithm to ignore them. To improve the convergence of the model fitting, we fix the centroid and background parameters to their \textsc{SExtractor} values.

\subsection{Source Selection} \label{sec:source_selection}
Source selection for this study is different than a typical WL study of overdensities where the lens distance is known. It is also different than a cosmic shear study because our goal is to detect significant overdensities and maximize the WL signal from them. Therefore, we have developed an iterative source selection technique that can find overdensities and then maximize their signal by re-selecting source galaxies based on the estimated lens redshift. Here, we describe the steps of source selection that are common to each iteration and we provide further detail on the iteration process in Section \ref{sec:mass_mapping}.

Due to the extensive prior observations of the CANDELS fields, we have the luxury of photometric redshift measurements for a large number of galaxies in each field. In this study, we choose the 3D-HST photometric redshift catalog \citep{2014skelton} because it provides the most abundant spectroscopic redshift measurements in the fields along with robust photometric redshifts. We join our shape measurement catalogs to the 3D-HST catalogs. The resulting catalogs contain 30528, 31324, 32142, 31465, and 34657 galaxies with number densities of 152, 150, 183, 178, and 171 arcmin$^{-2}$ for the COSMOS, UDS, GOODS-S (GS), GOODS-N (GN), and EGS, respectively (summarized in Table \ref{tab:field_summary}).

We enforce quality cuts on the WL shape measurements to ensure that only well-fit sources are used. We remove poorly fit sources from the catalogs by requiring the \textsc{MPFIT} status to be 1 (a ``good'' fit). From past experience and based on our demonstration of the impact of undersampling \citep{2023cfinner}, we know that the ellipticity of small sources can have a shape bias (about 2\% for the undersampling effect). Therefore, we restrict the sources to have a measured semi-minor axis, $b$, greater than 0.5 pixels \citep[for more details on size constraints see][]{2013jee}. To remove sources that are too elliptical to be real galaxies, we constrain the catalog to only objects with measured $e<0.9$. The final quality constraint that we put on the catalog is to include only galaxies with ellipticity uncertainty $de<0.3$. 

We applied the LensMC \citep{2024lensmc} galaxy-shape measurement pipeline as an alternative ellipticity-fitting method. This pipeline has recently been applied to Euclid observations and shown to perform well relative to other ellipticity pipelines \citep{2025schrabback}. We found that the LensMC ellipticity measurements are in excellent agreement with those from SFIT. However, under our ellipticity uncertainty cut, LensMC retains fewer usable source galaxies, reducing the source density and consequently the sensitivity of the WL maps to the detection of overdensities. We therefore proceed with SFIT for the present analysis and leave a more detailed LensMC analysis to future work.

\begin{table}[t]
\centering
\caption{Summary of weak-lensing source catalogs and peak detections by field.}
\label{tab:field_summary}
\begin{tabular}{lcccc}
\hline\hline
Field & $N_{\rm gal}$ & Source Density & Peak Detections \\
      &               & (arcmin$^{-2}$) &                \\
\hline
COSMOS  & 30528 & 152 & 6 \\
UDS     & 31324 & 150 & 3 \\
GOODS-S & 32142 & 183 & 2 \\
GOODS-N & 31465 & 178 & 1 \\
EGS     & 34657 & 171 & 0 \\
\hline
\end{tabular}
\end{table}

\subsection{Mass Mapping} \label{sec:mass_mapping}
A key product of a WL analysis is a map of the projected mass distribution, which is sometimes referred to as the dimensionless surface mass density or convergence. The convergence, $\kappa$, is a component of the gravitational lensing effect that isotropically magnifies the light from background galaxies. The other component is the shear, $\gamma$, which anisotropically distorts galaxy images. To create a map of the projected matter distribution, \cite{1993kaiser} formulated a method of convolution to proceed from shear to convergence and vice versa. We utilize a similar technique called \textsc{FIATMAP} \citep{2006wittman, 2023wittman, 2024stancioli} that includes the following weighting function for galaxies based on their projected distance, $r$, from a given pixel position in the map:

\begin{equation}
W(r)
= \frac{1 - e^{-r^2/(2 r_i^2)}}{r^2}
  \, e^{-r^2/(2 r_o^2)} ,
\end{equation}
where $r_i$ and $r_o$ are the inner and outer radii of the filter. The inner radius cutoff prevents extreme statistical fluctuations from affecting the measurement and the outer radius cutoff suppresses noise. We set the inner and outer radii to $50''$ and $100''$, respectively. This choice balances the need to suppress shape-noise fluctuations while retaining sensitivity to cluster-scale mass structure.

For each convergence map that we produce, we convert it into an estimate of the $S/N$ through bootstrapping \citep[e.g.,][]{2012clowe, 2026okabe}. Specifically, we resample the source-galaxy catalog with replacement 1000 times and reconstruct a convergence map for each realization. The pixel-by-pixel rms among these 1000 convergence realizations is adopted as the noise map, which is then used to compute the $S/N$ map.

We also use these 1000 bootstrap realizations to quantify the uncertainty in the WL mass centroid by measuring the scatter in the location of the convergence peak. For each overdensity identified in Section~\ref{sec:identification}, we identify the peak in each bootstrap mass map that lies within $5$~arcmin of the corresponding peak in the original mass map. We then use the distribution of these bootstrap peak positions to estimate the centroid uncertainty. The centroid uncertainties are included in Table \ref{tab:overdensities}.

\begin{figure}[ht]
    \centering
    \includegraphics[width=0.5\textwidth]{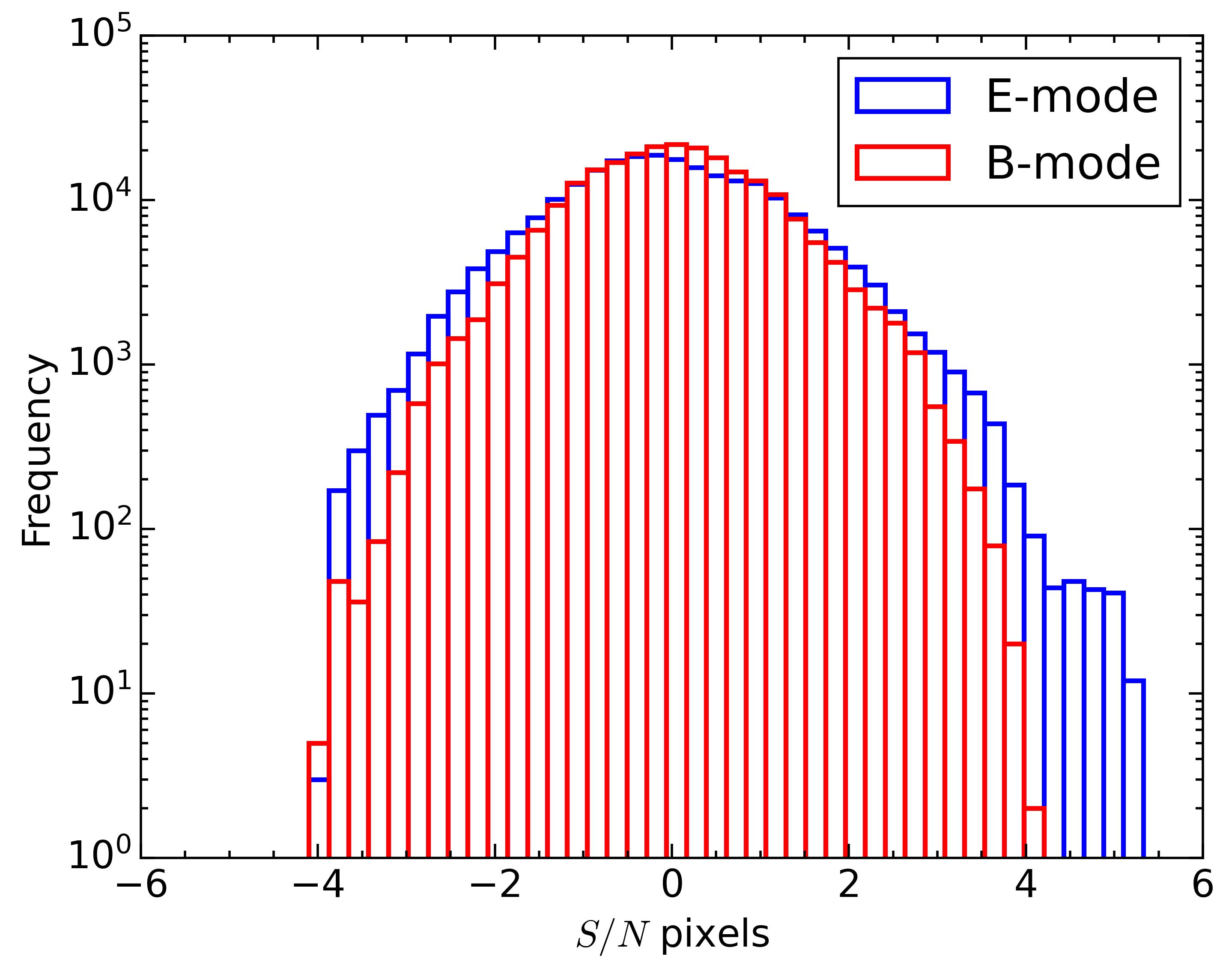}
    \caption{Histogram of WL $S/N$ pixels of the COSMOS field for a source catalog with photo-z~$>0.6$. The narrower $B$-mode distribution is expected. The positive $E$-mode tail indicates the detection of the lensing signal. Candidate overdensities are selected by further investigating the $E$-mode pixels that lie above $3\sigma$.}
    \label{fig:SN_hist_cosmos}
\end{figure}

\subsection{Identification of WL Overdensities} \label{sec:identification}
Identification of overdensities in WL maps is a difficult task because spurious peaks can occur, especially at low $S/N$. We aim to present as many bona fide detections as possible as well as some candidates that should have future scrutiny. Here, we describe the steps of our identification of overdensities and the criteria that we use to identify them.

We take the stance that a priori the WL overdensities are unknown. Therefore, we are blindly searching for WL signal. The WL signal (distortion) caused by an overdensity at a given redshift is imparted on the images of all galaxies behind the lens according to the lensing efficiency equation $\beta = d_{ls}/d_s$, where $d_{ls}$ is the angular diameter distance from the lens to the source and $d_s$ is the angular diameter distance from observer to source. For a first pass detection of candidate overdensities, we generate convergence maps at a series of photometric redshift lower bounds in the range $0.2 < z < 2.2$ with intervals of $\delta z=0.2$. For example, the first mass map is generated from a source catalog that contains galaxies that have passed the source selection criteria (Section \ref{sec:source_selection}) and have $z>0.2$. The result of this step is a $S/N$ map at each of the 10 redshift intervals.

\begin{figure*}[ht]
    \centering
    \includegraphics[width=\textwidth]{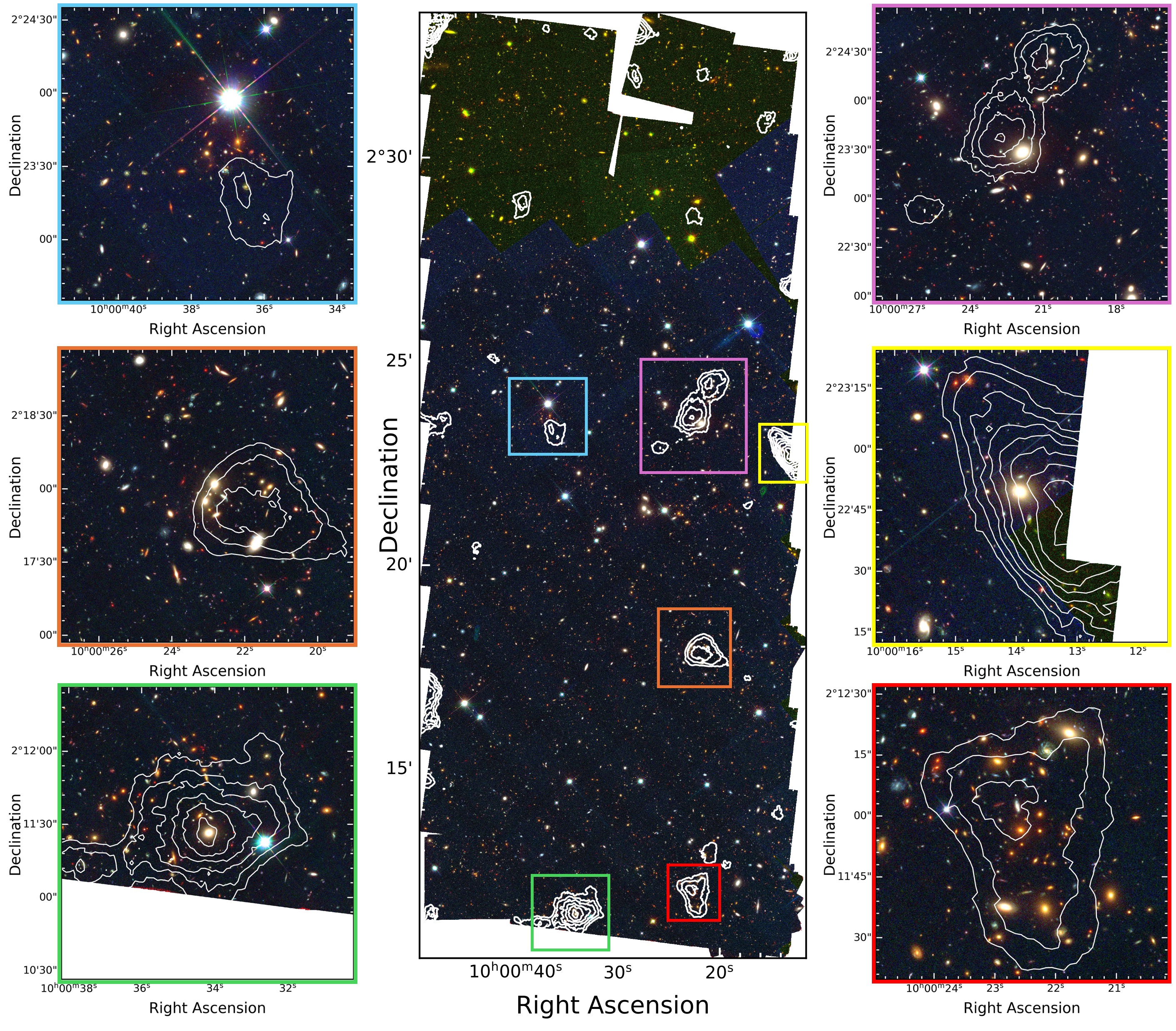}
    \caption{WL $S/N$ contours (white) plotted over the COSMOS color image. The WL $S/N$ map was created from a source catalog with a redshift cut~$>0.6$. Contour levels start at $2.5\sigma$ and increase in intervals of $0.5\sigma$. The six best examples of candidate overdensities at this redshift cut are highlighted with a colored box and a zoomed-in version of the plot is provided. Further optimization of the overdensities are done on a case-by-case basis to maximize the lensing signal. }
    \label{fig:cosmos_SN_map}
\end{figure*}

FIATMAP outputs both the WL signal $E$-mode maps and $B$-mode maps. The $B$-mode maps are a valuable test for systematics and a direct estimate of the expected noise distribution. Figure~\ref{fig:SN_hist_cosmos} presents the $E$- and $B$-mode $S/N$ histograms of the COSMOS field for a source catalog with redshift $z>0.6$. The $B$-mode $S/N$ distribution is centered near zero and exhibits a modest broadening ($\sigma=1.07$). After standardizing by the measured mean and variance, the empirical tail fractions are close to those expected for a unit Gaussian, particularly at $\gtrsim3\sigma$, with a small excess of pixels at the $2$--$2.5\sigma$ level. The $E$-mode distribution is broader than the $B$-mode, reflecting the presence of signal beyond pure noise. Inspecting the positive tail of the $S/N$ distribution, we observe a clear excess of $E$-mode pixels beginning around $S/N \simeq 3$, with the $E$-mode completely dominating at $S/N>4$. These pixels likely correspond to WL signals of sufficient significance to be considered candidate overdensities.

Previous ground-based shear-selected overdensity studies have shown that such samples are subject to non-negligible contamination from noise peaks and line-of-sight projections, even at relatively high $S/N$. For example, \cite{2007schirmer} selected shear peaks with $S/N>4.5$ and found that only 65\% were associated with galaxy overdensities, while \cite{2007miyazaki} identified optical or X-ray counterparts for 80\% of their $S/N>4$ detections. These results imply a residual contamination fraction of $\sim20$--35\%, which can be reduced by improved image depth and resolution \citep{2018miyazaki}.

Motivated by the observed excess in the $E$-mode distribution and the known contamination in shear-selected samples, we select pixels with $S/N>3$ in the $E$-mode as locations warranting further analysis rather than as a definitive catalog of halos.

Figure \ref{fig:cosmos_SN_map} presents a CANDELS COSMOS WL $S/N$ map for a background galaxy catalog with a redshift lower bound of $z>0.6$. We highlight the six best examples of $S/N$ peaks that contain $E$-modes with $S/N > 3$ from the histogram of Figure \ref{fig:SN_hist_cosmos}. The WL signal of these candidate overdensities is not yet optimized, but indicates that an overdensity may be present. To continue with the identification of overdensities, the right ascension and declination of the $S/N$ peak for each of these candidates are accumulated in a table that is provided into the second round of identification. This procedure is performed for each of the 10 redshift intervals.

Because WL peaks can arise from noise or projected structure, independent confirmation using luminous tracers is required. In the second round of identification, we aim to verify that luminous tracers of the overdensities exist. We perform two additional verification steps that are visualized in Figures \ref{fig:NIRWLJ100022+0212} through \ref{fig:NIRWLJ123645+6216}. First, we plot a redshift histogram of all galaxies within $5$~arcmins of the WL $S/N$ peak (bottom right of Figure \ref{fig:NIRWLJ100022+0212} for example). This size corresponds to a radius $\sim0.5$ Mpc at $z=0.2$ and $\sim1.2$ Mpc at $z=1$, which is sufficiently large to capture the localized overdensity of galaxies. The redshift histogram contains both photometric and spectroscopic redshifts. We search for distinct, sharp peaks in the redshift distribution. For previously known overdensities, the spectroscopic redshifts tend to identify a common redshift because of the targeted nature of the measurements. If a spike in the redshift histogram is found, we calculate the median $z$ from the galaxies with $z\pm0.1$ of the peak. We then refine the identification of candidate overdensity galaxies by selecting those within a range of $z\pm0.03$ of the median value. These candidate members are visualized on a color cutout of the candidate overdensity with the WL $S/N$ contours overlaid (top left of Figure \ref{fig:NIRWLJ100022+0212} for example), which allows us to evaluate the morphology and color of the redshift-identified galaxies that exist in the lensing overdensity.

Our second test is to plot the candidate galaxies in a color-magnitude diagram. Many of the fields have substantial filter coverage that enables a color measurement that brackets the expected 4000~\AA~break. The 4000~\AA~break is a feature of evolved galaxies that gives rise to a red sequence in the color-magnitude diagram \citep{1977visvanathan, 1992bower} and has also been found in low-mass groups \citep{1998zabludoff}. We evaluate all of the information from these verification steps and compile a list of WL overdensities.

At this stage, we have identified candidate overdensities from the WL maps and verified that they have luminous tracers. It is now time to optimize the WL signal for each overdensity. We take the median redshift that was determined during the identification of the overdensity as a starting point for generating a new $S/N$ map that has background galaxy selection optimized. Background sources are selected by requiring $z_{\mathrm{phot}} > z_{{\mathrm median}} + 0.1$. This buffer provides a conservative margin to suppress contamination by galaxies scattered near or below the lens redshift. For the 3D-HST/CANDELS catalog, the typical photo-$z$ scatter is of order $0.02(1+z)$ for bright galaxies, but increases toward fainter magnitudes. The $\Delta z = 0.1$ buffer is several times larger than the nominal scatter for much of the sample and is a compromise between minimizing shear dilution and maintaining sufficient source density for the WL measurement. A new WL $S/N$ mass map is generated from this optimized source catalog and the process is repeated for each overdensity. These optimized source catalogs per overdensity are used for the remainder of the analysis of the individual overdensities.


\subsection{Mass Estimates} \label{sec:mass_estimation}
After the identification of the WL overdensities, we determine a mass for the overdensity by fitting the tangential shear. The tangential reduced shear is formulated as 

\begin{equation}
    g_{+} = -g_1 \cos{2\phi} - g_2\sin{2\phi},
\end{equation}
where $g_1$ and $g_2$ are the complex components of the reduced shear $g=\gamma/(1-\kappa)$ and $\phi$ is the azimuthal angle from the horizontal axis that bisects the center of the cluster. In most cases, we center the measurements of the tangential shear on the BCG because it has a well-defined centroid and has been identified as the best tracer of the center of mass \citep{2012george}. However, when noted in the text of Section \ref{sec:results}, a different center may be used. Miscentering is a known systematic in WL analysis and can lead to biases in mass estimation of the order 10\% \citep{2018lee_mass_bias, 2022sommer}. 

The method that we employ to fit the mass of each overdensity starts from a spherical NFW halo and models the tangential shear following the formalism in \cite{2000wright}. The best-fit model is determined by comparing the expected (modeled) shear to the source galaxy shape measurements at each galaxy position. The NFW model can be parameterized by the mass and concentration. However, these two parameters are highly degenerate and typically cannot be constrained simultaneously at low S/N. Therefore, we decide to only fit the mass of the cluster and fix the concentration to the \cite{2019diemer} concentration--mass ($c$--$M$) relation. 

\subsection{X-ray Overdensities from Literature} \label{sec:xray_lit}
The CANDELS fields have abundant X-ray observations that have been used to identify diffuse X-ray emission. To match diffuse X-ray emission to our WL-detected overdensities, we enforce a distance criterion of 200 kpc at the redshift of the overdensity. For each of the X-ray detections that are matched to our WL detected overdensities, we provide X-ray positions in the respective figures.

For the GS and UDS fields, X-ray groups are identified in \cite{2010finoguenov, 2015finoguenov}. They report the X-ray luminosity $L_x$ in the $0.1-2.4$~keV band. In the COSMOS field, \cite{2007finoguenov},   \cite{2011george}, and \cite{2019gozaliasl} provide tables of X-ray detected groups. We use the \cite{2019gozaliasl} X-ray luminosities, which are reported in the $0.1-2.5$~keV energy band. The GN field has X-ray identification in \cite{2002bauer} and in \cite{2013pentericci}. However, there is no X-ray detection for our WL detected overdensity in the GN field. This is discussed in more detail in Section 4. Finally, in the EGS, we find no significant overdensities in our WL study. \cite{2013erfanianfar} compiled a catalog of X-ray detections for galaxy groups that covers the EGS field. Only 3 of their X-ray sources overlap with the CANDELS EGS footprint and their X-ray luminosities are low. 

\section{Results} \label{sec:results}
The following sections summarize the 12 overdensities that are detected through our WL analysis of the five CANDELS fields. Table \ref{tab:overdensities} gives an overview of the detections including positions, redshifts, and mass estimates. Section \ref{sec:stacked_shear} presents a stacked tangential shear analysis and Section \ref{sec:xray_relation} places the systems onto the X-ray luminosity--mass scaling relation. We then provide a description of each system in Section \ref{sec:individual_systems}.

\begin{table*}[]
\caption{Summary of 12 WL detected overdensities.}
\label{tab:overdensities}
\begin{minipage}{0.2\linewidth}
\begin{tabular}{lccccccccc}
\hline
\hline
Name    & Field           & R.A.       & Dec.    & $z$ & $M_{200}$ & $M_{500}$ & $S/N$ & Cent. Unc. & $L_\mathrm{x}$  \\
& & & & & $10^{14}$ $M_\odot$ & $10^{14}$ $M_\odot$ & & arcsec & $10^{42}$ ergs s$^{-1}$ \\
\hline

NIRWL J100022+0212 & COSMOS & 10:00:22 & 02:11:56  & 0.685 &  $0.5\pm0.3$ & $0.4\pm0.3$ & 4.2 & 16.5 & $4.1\pm1.2^a$ \\
NIRWL J100034+0211 & COSMOS & 10:00:34 & 02:11:26  & 0.359 & $0.5\pm0.2$ & $0.4\pm0.2$ & 5.9 &  7.7 &  \\
NIRWL J100022+0217 & COSMOS & 10:00:22 & 02:17:58  & 0.366 & $0.6\pm0.2$ & $0.5\pm0.2$ & 4.2 & 11.3 & $1.8\pm0.4^a$ \\
NIRWL J100021+0223 & COSMOS & 10:00:21 & 02:23:29  & 0.222 & $0.4\pm0.2$ & $0.4\pm0.2$ & 4.3 & 16.6 &  $2.6\pm0.2^a$ \\
NIRWL J100035+0223 & COSMOS & 10:00:35 & 02:23:36  & 0.895 & $0.7\pm0.4$ & $0.6\pm0.3$ & 3.3 & 20.2 &  $10.0\pm2.3^a$  \\
NIRWL J100021+0231 & COSMOS & 10:00:20 & 02:31:52  & 0.885 & $1.5\pm0.5$ & $1.3\pm0.4$ & 3.9 & 20.0 & $23.5\pm3.2^a$ \\

NIRWL J021737-0513 & UDS & 02:17:37 & -05:13:29 & 0.648  & $1.8\pm0.4$ & $1.6\pm0.3$  & 5.5 & 13.1 & $42.0\pm3.0^b$ \\
NIRWL J021808-0515 & UDS & 02:18:08 & -05:15:40 & 0.680  & $0.3\pm0.3$  & $0.3\pm0.2$  & 3.3 & 32.3 &  \\
NIRWL J021709-0508 & UDS & 02:17:09 & -05:18:06 & 0.624  & $0.2\pm0.2$  & $0.2\pm0.1$  & 4.2 & 17.9 &  \\

NIRWL J033209-2742 & GOODS-S & 03:32:09 & -27:42:47 & 0.727  & $0.3\pm0.3$ & $0.3\pm0.2$ & 5.2 & 14.6 & $3.2\pm0.6^c$ \\
NIRWL J033239-2752 & GOODS-S & 03:32:39 & -27:52:28  & 0.759 & $2.2\pm0.5$ & $1.9\pm0.4$ & 5.3 & 10.2 & \\

NIRWL J123645+6216 & GOODS-N & 12:36:45 & 62:16:23  & 0.861 & $1.1\pm0.3$ & $0.9\pm0.3$ & 6.2 & 13.4 & \\

\hline
\end{tabular}

\footnotesize
\begin{tabular}{@{}l@{}}
X-ray luminosity sources are: $^{a}$\cite{2019gozaliasl}
$^{b}$\cite{2010finoguenov}
$^{c}$\cite{2015finoguenov}
\end{tabular}
\end{minipage}
\end{table*}

\subsection{Stacked Tangential Shear} \label{sec:stacked_shear}
WL provides a direct probe of the projected mass distribution of the shear-selected systems in this study. However, these low-mass systems are limited by their low WL signal and shape noise. To overcome these limitations, we construct a stacked radial density profile by averaging the tangential shear signal to obtain high-fidelity measurements of the ensemble distribution. Stacking reduces statistical fluctuations in individual profiles and improves constraints on the ensemble mass distribution.

We construct a stacked radial density profile following the methodology of \cite{2014umetsu}. This approach permits the combination of shear signals of multiple systems without imposing model-dependent scaling assumptions. For each system, we take the tangential shear profile $g_+$, as described in Section \ref{sec:mass_estimation}, and convert the shear measurements into excess surface-density contrast

\begin{equation}
    \Delta\Sigma(r) = \Sigma_{crit}g_+(r),
\end{equation}
where the lensing critical density is

\begin{equation}
    \Sigma_{crit} = \frac{c^2}{4 \pi G} \frac{1}{\beta d_l},
\end{equation}
with $G$ the gravitational constant, $c$ the speed of light, $\beta$ the lensing efficiency, and $d_l$ the angular diameter distance to the lens. We apply the same physical radial bins to all overdensities and calculate the stacked excess surface density as an inverse-variance-weighted average,

\begin{equation}
    \left<\Delta\Sigma(r)\right> = \frac{\sum\limits_{i}w_j(r)\Delta\Sigma_j(r)}{\sum\limits_{i}w_j(r)}
\end{equation}
where weights $w_j(r)=\sigma^{-2}(r)$ include shape noise and measurement variance added in quadrature.

\begin{figure}[ht]
\centering
    \includegraphics[width=0.48\textwidth]{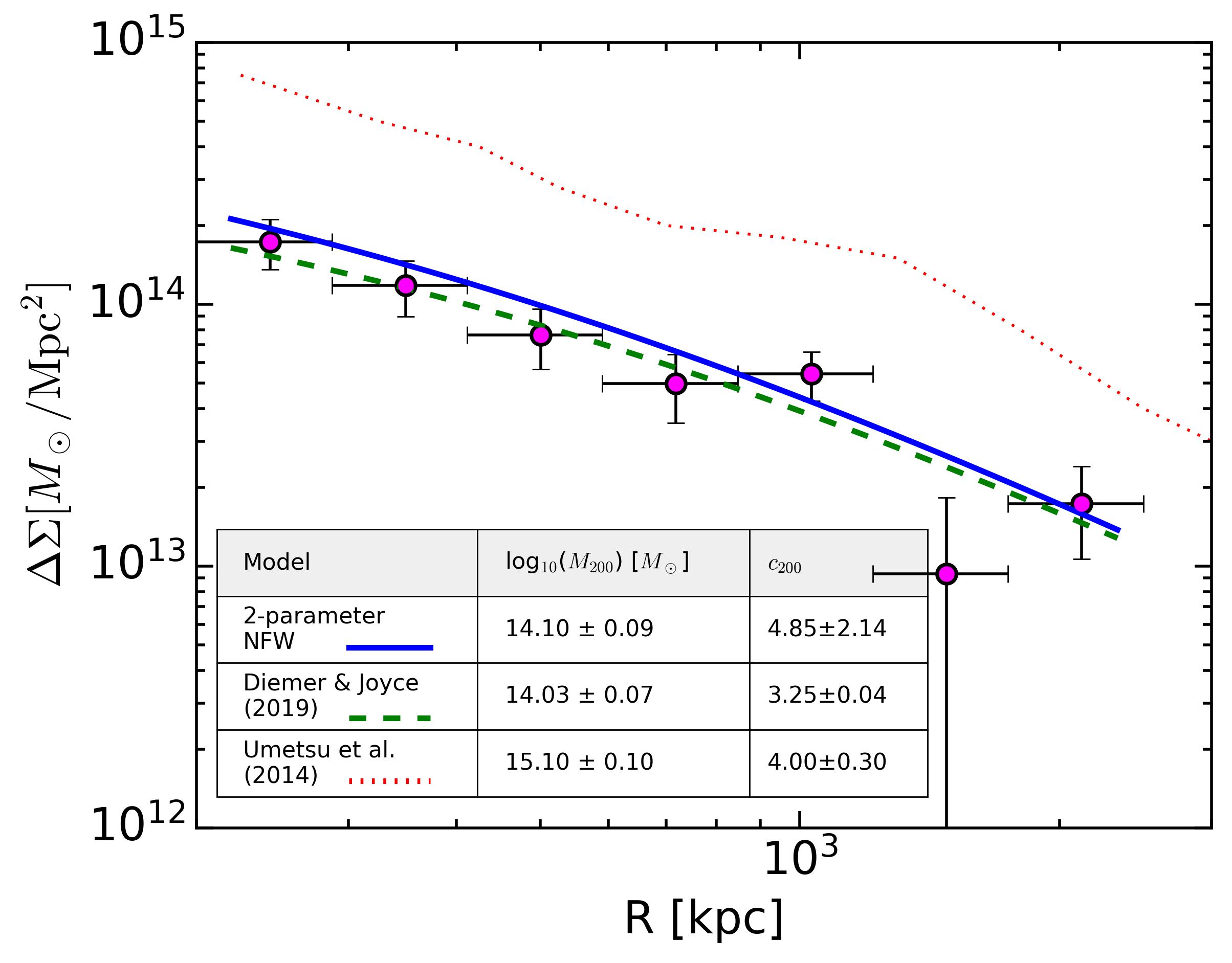}
    \caption{Excess surface mass density from the stacked tangential shear of the 12 overdensities. The measurements were stacked as described in Section \ref{sec:stacked_shear}. The blue line shows an NFW fit with both mass and concentration as free parameters. When fitting the $c$--$M$ relation of \cite{2019diemer} (green curve) a consistent mass is found but with lower concentration. For comparison, we include the stacked radial density profile from \cite{2014umetsu}, which covers a much higher mass range of clusters at lower redshifts.  The higher mass sample shows a similar slope to our low-mass sample.}
    \label{fig:stacked_shear}
\end{figure}

The stacked density profile of our 12 shear-selected overdensities is presented in Figure \ref{fig:stacked_shear}. The overdensities have an average redshift $\left<z\right>\sim0.68$. The magenta circles show the robust detection of the shear and expected monotonic decline until close to 1 Mpc, where the noise starts to play a significant role. We fit two relations to the density profile. We first fit a two-parameter NFW model with mass and concentration free (blue curve in Figure \ref{fig:stacked_shear}). The best-fit NFW model to the average density profile gives a mass of $M_{200} = 1.3\pm0.3\times10^{14}\ M_\odot$ with a concentration of $c=4.9\pm2.1$. Because of the mass and concentration degeneracy, we also perform a single-parameter NFW fit using the fixed $c$--$M$ relation of \cite{2019diemer}. The single-parameter fit (dashed green curve) leads to a consistent mass but a lower concentration, which is visually apparent from the green curve being flatter than the blue. Finally, the red dotted curve presents the measurements from \cite{2014umetsu} for a high-mass sample of galaxy clusters, which are roughly an order of magnitude more massive than our sample and at a lower average redshift.  

\subsection{X-ray Luminosity--WL Mass Scaling Relation} \label{sec:xray_relation}

We examine the relationship between WL mass estimates and X-ray luminosity for the shear-selected overdensities identified in this study. The NIRWL sample probes the low-mass end of the cluster regime ($\sim10^{14}\ M_\odot$) and extends into the group-scale mass range.

To connect the WL detections to the intracluster medium (ICM), we cross-matched the 12 NIRWL overdensities with published deep X-ray observations (see Section \ref{sec:xray_lit}) in the CANDELS fields. Of the 12 systems, seven have reported diffuse X-ray detections. Their X-ray luminosities are listed in Table~\ref{tab:overdensities}.

\begin{figure}[ht]
    \centering
    \includegraphics[width=\linewidth]{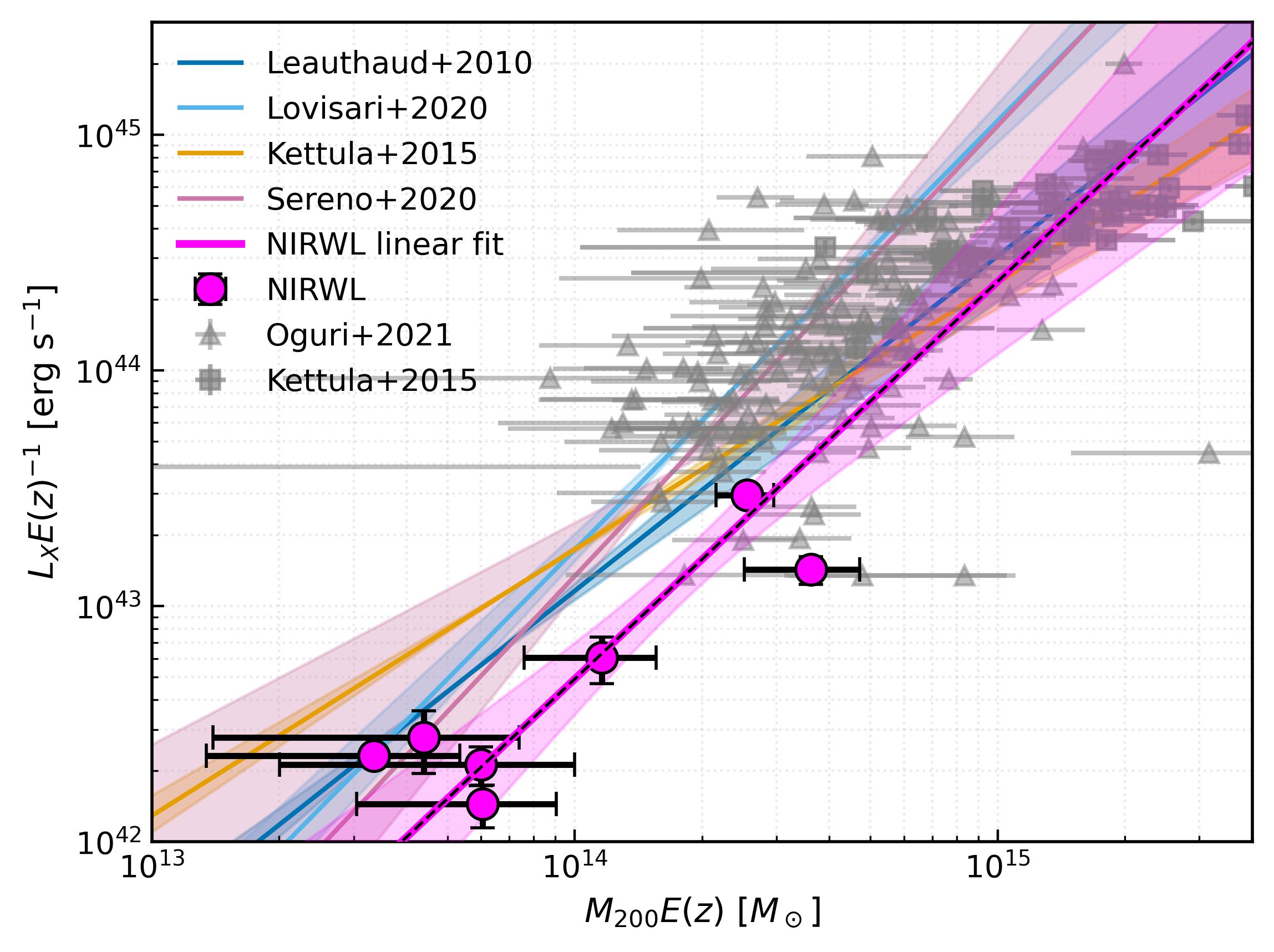}
    \caption{Mass versus X-ray luminosity scaling relation. NIRWL overdensities are marked with magenta circles. Shear-selected overdensities from \cite{2021oguri} are denoted with gray triangles and the high-mass clusters from \cite{2015ketulla} are marked with gray squares. Four $L_x$--$M$ scaling relations are included from literature with the bands showing the parameter uncertainty but not the intrinsic scatter.}
    \label{fig:mass-xray}
\end{figure}

Figure~\ref{fig:mass-xray} presents the X-ray luminosity as a function of WL mass for these systems. For comparison, we include the high-mass cluster sample from Table~C1 of \cite{2015ketulla} and the HSC-SSP shear-selected sample from \cite{2021oguri}. We also overlay four published $L_X$--$M$ scaling relations: three calibrated using WL masses \citep{2010leauthaud,2015ketulla,2020sereno} and one calibrated using hydrostatic X-ray masses \citep{2020lovisari}. The X-ray luminosities from \cite{2015ketulla} and \cite{2020lovisari} are measured in the same energy band as our sample (0.1--2.4~keV), whereas the \cite{2020sereno} luminosities are derived in the 0.5--2.0~keV band and are therefore expected to be systematically lower.

We fit the X-ray mass scaling relation for our NIRWL overdensities in log-log space using orthogonal distance regression, accounting for uncertainties in both mass and X-ray luminosity. Using a pivot mass of $M_{\rm piv} = 3 \times 10^{14}\,M_\odot$, we find a best-fit slope of $\alpha=1.69 \pm 0.40$ and a normalization of $\log_{10}(L_0/{\rm erg\ s^{-1}})=43.32 \pm 0.12$, such that
\begin{equation}
L_X E(z)^{-1}
=
L_0
\left(
\frac{M_{200} E(z)}{M_{\mathrm piv}}
\right)^{\alpha},
\end{equation}
 with a reduced chi-square of $\chi^2_\nu = 1.35$. We further discuss this result in Section \ref{sec:xray_discussion}.

\subsection{Individual Systems} \label{sec:individual_systems}

\subsubsection{COSMOS} \label{sec:cosmos}
COSMOS is our most productive field, with six WL overdensity detections. We also identify two additional WL signals just beyond the edge of the CANDELS/COSMOS footprint and confirm that they correspond to X-ray groups; however, we do not include them in our analysis because the lensing peaks lie outside the field boundary. 

Our analysis is not the first to map the lensing signal in the COSMOS field. \cite{2007massey} (hereafter M07) produced WL maps of the larger HST ACS coverage of COSMOS and focused on the large scale structure. More recently, \cite{2026scognamiglio} (hereafter S26) used the JWST COSMOS-Web imaging to provide an updated mass map. Our CANDELS analysis is limited to a smaller footprint, however, we achieve a background galaxy density similar to the JWST analysis and produce mass maps at a higher resolution. We describe how our six detections compare to the past two mass maps at the end of the section. 

\begin{figure*}[ht]
    \centering
    \includegraphics[width=\textwidth]{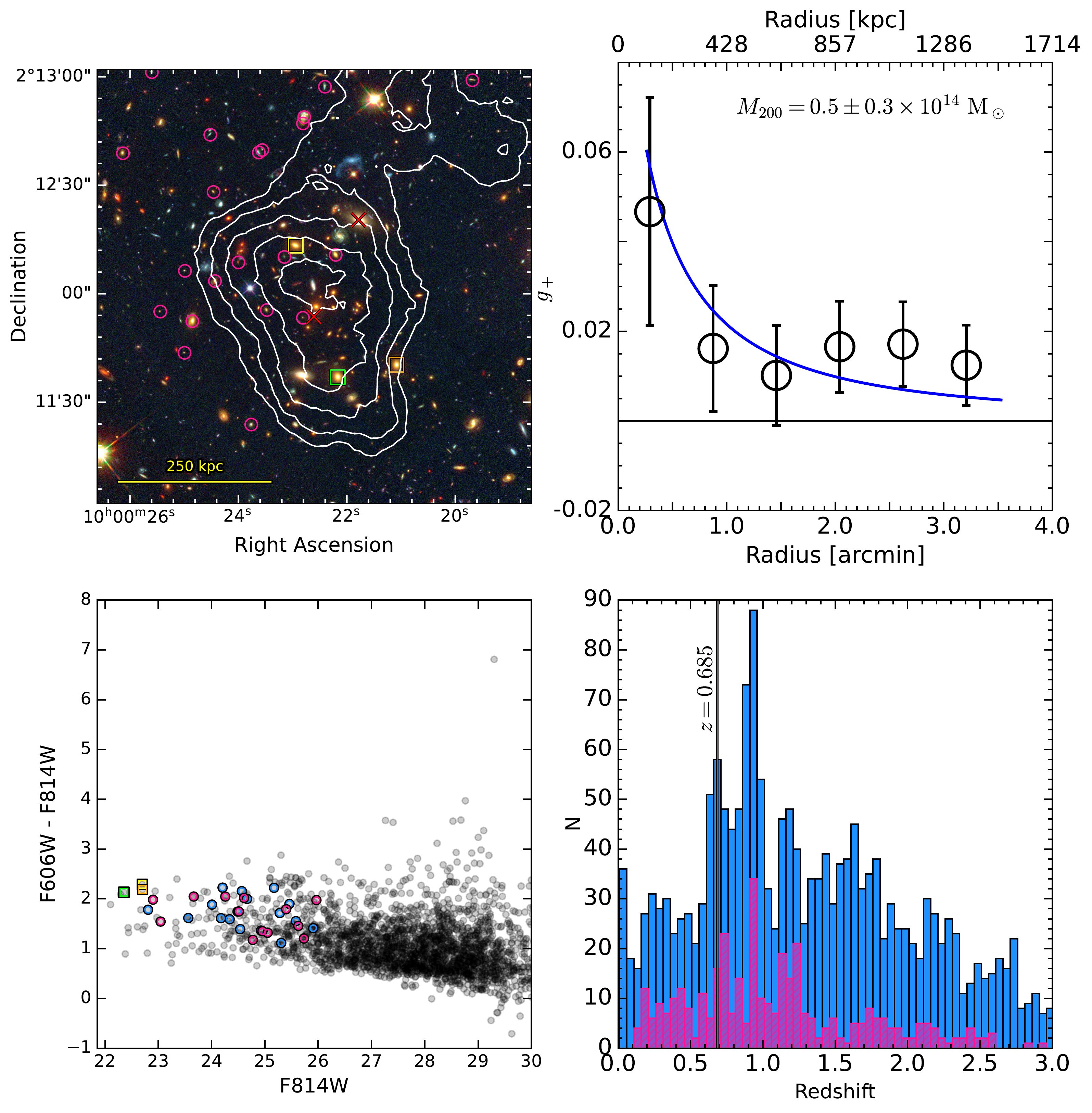}
    \caption{NIRWL J100022+0212 - COSMOS. Top left: Mass map over CANDELS color image with brightest (green square), second brightest (yellow square), and third brightest (orange square) galaxies highlighted. Additional spectroscopically selected galaxies that are candidates for the overdensity are marked with magenta circles. Red X's mark the locations of X-ray defined groups. The elongated mass map and bright galaxies positioned at the ends of the mass are characteristic of merging clusters. Top right: Tangential shear plot with fitted mass using the \cite{2019diemer} $c$--$M$ relation. Errorbars are standard errors on the mean. Bottom left: Color-magnitude diagram with candidate overdensity galaxies (spectroscopic = pink, photometric = blue) selected from the redshift histogram. Black circles are all galaxies along the line-of-sight within a radius of 10 arcmins of the WL $S/N$ peak. Bottom right: Redshift histogram with spectroscopic (pink) and photometric (blue) redshifts.  The average redshift of the galaxies that are selected in the overdensity is marked with a vertical yellow line.}
    \label{fig:NIRWLJ100022+0212}
\end{figure*}

\textbf{NIRWL J100022+0212} ($z=0.685$, $M_{200}=0.5\pm0.3\times10^{14}\ M_\odot$, Figure~\ref{fig:NIRWLJ100022+0212}): This system lies in the southern region of the COSMOS field (red box in Figure \ref{fig:cosmos_SN_map}) and is detected with a peak significance of $S/N=4.2$. It is not immediately obvious what redshift the overdensity lies at. From the redshift histogram, a peak is found in both photo-$z$ and spec-$z$ at $z\sim0.95$. The brightest galaxy at this redshift is within $1 \arcsec$ of the mass peak. A weaker peak in the redshift histogram is found at $z\sim0.69$. The two brightest elliptical galaxies at this redshift are located at opposing ends of the mass distribution. The overdensity is associated with an X-ray group ($z=0.686$) with a luminosity of $L_X=4.07\pm1.21\times10^{42}$ erg s$^{-1}$ \citep{2011george,2019gozaliasl}. We therefore choose to focus the analysis on the lower redshift overdensity and calculate the median redshift to be $z=0.685$. 

The overdensity shows an elongated mass distribution with brightest cluster galaxies at opposing ends and a centrally located X-ray brightness peak. This is typical for Bullet-like merging clusters \citep[e.g.,][]{2004markevitch, 2021finner, 2023finner_ciza, 2023wittman, 2025cha}. For the mass estimation of this cluster, we define the cluster center at the WL mass peak. An NFW fit yields a mass of $M_{200}=0.5\pm0.3\times10^{14}\ M_\odot$.

\textbf{NIRWL J100034$+$0211} ($z=0.359$, $M_{200}=0.5\pm0.2\times 10^{14}\ M_\odot$, Figure~\ref{fig:NIRWLJ100034+0211}): This overdensity is detected near the southern edge of the COSMOS field (green box in Figure \ref{fig:cosmos_SN_map}) with a peak WL significance of $S/N=5.9$ that is centered on a bright elliptical galaxy with a spectroscopic redshift of $z=0.359$. Spectroscopic redshifts within 5\arcmin\ of the central galaxy reveal only a weak enhancement at $z=0.359$, suggesting that the structure is sparsely populated. Despite this, the tangential shear profile shows a significant detection, and an NFW fit yields a total mass of $M_{200}=0.5\pm0.2\times10^{14}\ M_\odot$. Consistent with the limited spectroscopic membership, the color-magnitude diagram does not display a well-defined red sequence. 

The redshift histogram further reveals a more prominent background structure at $z\sim0.95$ in photometric redshifts. The galaxies at this redshift visibly clustered in the northern half of the image cutout. However, a WL $S/N$ map constructed using source galaxies with $z>1$ shows no significant lensing signal. The proximity of the detection to the CANDELS footprint edge may bias the mass reconstruction due to reduced source density on the southern side. No X-ray counterpart is identified for this overdensity.

\textbf{NIRWL J100022+0217} ($z=0.366$, $M_{200}=0.6\pm0.2\times10^{14}\ M_\odot$), Figure~\ref{fig:NIRWLJ100022+0217}): This is the final detection in the southern half of the CANDELS COSMOS field (orange box in Figure \ref{fig:cosmos_SN_map}). The WL signal is detected with a peak significance of $S/N=4.2$ and lies within $5\arcsec$ of two bright elliptical galaxies with spectroscopic redshifts of $z=0.366$. The offset between the WL peak and the bright galaxies is not statistically significant. The tangential shear measured about the BCG shows a significant detection, and an NFW fit yields a mass of $M_{200}=0.6\pm0.2\times10^{14}\ M_\odot$. A well-defined red sequence is present for galaxies with $0.34<z<0.40$. The overdensity coincides with an X-ray group whose emission peak lies between the two brightest galaxies \citep{2011george,2019gozaliasl}.

\textbf{NIRWL J100021+0223} ($z=0.222$, $M_{200}=0.4\pm0.2\times10^{14}\ M_\odot$, Figure~\ref{fig:NIRWLJ100021+0223}): This overdensity is detected near the center of the CANDELS COSMOS field (purple box in Figure~\ref{fig:cosmos_SN_map}) with a peak significance of $S/N=4.3$. The WL peak is slightly offset from a bright elliptical galaxy with a spectroscopic redshift of $z=0.222$. The reconstructed mass distribution is elongated in the NW-SE direction, with a secondary peak toward the NW. While no single standout galaxy is present in the NW region, a population of galaxies is visible. The spectroscopic redshift histogram shows a strong concentration at $z=0.22$, and the selected members define a tight red sequence in the color-magnitude diagram. The tangential shear profile and NFW fit indicate a group-scale mass of $M_{200}=0.4\pm0.2\times10^{14}\ M_\odot$. An associated X-ray group is detected, with a center near the BCG and an X-ray luminosity of $L_X=2.59\pm0.20\times10^{42}$ erg s$^{-1}$.

\textbf{NIRWL J100035+0223} ($z=0.895$, $M_{200}=0.7\pm0.4\times10^{14}\ M_\odot$, Figure~\ref{fig:NIRWLJ100035+0223}): This lower-significance detection (blue box in Figure~\ref{fig:cosmos_SN_map}) corresponds to a well-defined spike in the redshift histogram at $z=0.895$. The system is also identified as an X-ray group \citep{2011george,2019gozaliasl}. The WL peak appears offset from the brightest galaxies, likely influenced by masking of background sources within $30\arcsec$ of a nearby bright star. The tangential shear signal is weak, and the resulting mass estimate is $M_{200}=0.7\pm0.4\times10^{14}\ M_\odot$.

\textbf{NIRWL J100021+0231} ($z=0.885$, $M_{200}=1.5\pm0.5\times10^{14}\ M_\odot$, Figure~\ref{fig:NIRWLJ100021+0231}): This overdensity is located near the northern boundary of the CANDELS COSMOS field and is detected at a redshift of $z=0.885$. Spectroscopic coverage in this region is limited, but the photometric redshift distribution shows a strong peak at this redshift. A prominent BCG is identified and is offset by $\sim15\arcsec$ from the WL peak, which is not a surprising offset for a cluster of this mass at $z=0.89$. The tangential shear profile yields a significant detection, with an NFW mass of $M_{200}=1.5\pm0.5\times10^{14}\ M_\odot$. An associated X-ray group is detected $\sim3\arcsec$ NW of the BCG with an X-ray luminosity of $L_X=2.35\pm0.32\times10^{43}$ erg s$^{-1}$.

\textbf{Comparison with M07 and S26 COSMOS mass maps.} The six significant overdensities detected in the CANDELS COSMOS field extend from the southern boundary, along the western half of the imaging, to the northern edge of the footprint. This overall distribution is broadly consistent with the mass maps presented by M07 and S26, both of which show an extended mass distribution spanning the western half of the CANDELS field. However, at the resolution that we have chosen to map the COSMOS field, we are presenting individual systems instead of a blended distribution.

In the southern portion of CANDELS/COSMOS, NIRWL J100022+0212 appears in both M07 and S26 as a subpeak embedded within the extended distribution, with the S26 map showing somewhat better agreement with our peak position. Further to the southeast, NIRWL J100034$+$0211 is detected in both M07 and S26 at lower convergence as a bump that blends into the more prominent signal associated with NIRWL J100022+0212. Both the M07 and S26 maps show WL signal that elongates to the north of NIRWL J100022+0212. In M07, this structure passes over NIRWL J100022+0217 without producing a distinct peak, whereas S26 shows a local enhancement at the position of NIRWL J100022+0217.

In both maps, the northward elongation terminates approximately 3\arcmin\ south of NIRWL J100021+0223. The peak seen in M07 and S26 at this location may represent a blend of NIRWL J100021+0223 with an overdensity located slightly outside the CANDELS footprint (see the yellow box in Figure~\ref{fig:cosmos_SN_map}), and possibly with NIRWL J100035+0223. Farther north, the M07 and S26 mass maps pass over NIRWL J100021+0231 and extend beyond the northern edge of the CANDELS footprint.

\subsubsection{UDS} \label{sec:UDS}

\textbf{NIRWL J021737-0513} ($z=0.648$, $M_{200}=1.8\pm0.4\times10^{14}\ M_\odot$, Figure~\ref{fig:NIRWLJ021737-0513}):
This is the most significant detection in the UDS field. The WL signal is elongated in the east–west direction, with the mass peak located near the BCG ($z=0.65$). The redshift histogram exhibits a prominent peak centered at $z=0.648$, and the selected galaxies define a clear red sequence in the color-magnitude diagram. An NFW fit to the tangential shear yields a mass of $M_{200}=1.8\pm0.4\times10^{14}\ M_\odot$. 

\cite{2010finoguenov} measured an X-ray luminosity of $42.0\pm3.0\times10^{42}$ erg s$^{-1}$ for this system, also known as XLSSC59. \cite{2020umetsu} included the cluster in a WL analysis of X-ray–selected clusters using Subaru Hyper-suprime-cam observations and measured a mass of $M_{200}=1.37^{+1.19}_{-0.68}\times10^{14}\ M_\odot$, consistent with our estimate. The BCG acts as a strong lens, producing two images of background galaxies at $z=1.1$ and $z=1.5$. The cluster is also reported to reside within a large-scale filament \citep{2018galametz}.

\textbf{NIRWL J021808-0515} ($z=0.680$, $M_{200}=0.3\pm0.3\times10^{14}\ M_\odot$, Figure~\ref{fig:NIRWLJ021808-0515}):
This second overdensity in UDS lies southeast of NIRWL J021737-0513. Its close three-dimensional proximity to that system makes it an interesting target for filament studies and \cite{2018galametz} identified it as part of the same large-scale filament. The photometric redshift distribution highlights an overdensity at $z=0.68$, and the selected galaxies trace a red sequence in the color-magnitude diagram. 

The WL detection has the lowest significance in our sample ($S/N=3.3$), and the NFW fit yields $M_{200}=0.3\pm0.3\times10^{14}\ M_\odot$. The low WL signal may reflect the low mass of the system and potential systematics from a nearby bright star and the proximity to the edge of the CANDELS UDS footprint.

\textbf{NIRWL J021709-0508} ($z=0.624$, $M_{200}=0.2\pm0.2\times10^{14}\ M_\odot$, Figure~\ref{fig:NIRWLJ021709-0508}):
This third UDS overdensity is detected near the northern edge of the imaging. The WL signal appears elongated, although its morphology may be affected by the proximity to the image boundary. The redshift histogram shows a clear peak at $z=0.624$, consistent with other structures in the UDS field. The detection coincides with galaxy group C4 identified by \cite{2018galametz}, which is part of the same large-scale structure as the two previous systems. An NFW fit to the tangential shear yields $M_{200}=0.2\pm0.2\times10^{14}\ M_\odot$.

\subsubsection{GOODS South} \label{sec:gs}

\textbf{NIRWL J033209-2742} ($z=0.727$, $M_{200}=0.3\pm0.3\times10^{14}\ M_\odot$, Figure~\ref{fig:NIRWLJ033209-2742}):
This overdensity is identified in the northern region of GOODS-S with a peak significance of $S/N=5.2$. The redshift histogram indicates a structure at $z=0.727$, and the BCG at this redshift is well aligned with the WL peak. Lower significance contours extend toward the northwest, tracing a second bright galaxy. An NFW fit centered on the BCG yields $M_{200}=0.3\pm0.3\times10^{14}\ M_\odot$. 

Although the detection significance is high and numerous galaxies are associated with the structure, the inferred mass is modest. The system is detected in X-rays and identified as a galaxy group by \cite{2015finoguenov}, with an X-ray luminosity of $3.2\pm0.6\times10^{42}$ erg s$^{-1}$, consistent with expectations from scaling relations.

\textbf{NIRWL J033239-2752} ($z=0.759$, $M_{200}=2.2\pm0.5\times10^{14}\ M_\odot$, Figure~\ref{fig:NIRWLJ033239-2752}):
This system lies in the southern region of GOODS-S and is detected with $S/N=5.3$. The WL signal is extended primarily along the east–west direction. Associating galaxies with this overdensity is challenging. The redshift histogram shows a peak near $z=0.7$, but galaxies at that redshift do not spatially coincide with the WL peaks. At $z=0.759$, the galaxy distribution more closely matches the mass map, although no strong redshift spike or well-defined red sequence is present.

An NFW fit to the tangential shear yields a mass of $M_{200}=2.2\pm0.5\times10^{14}\ M_\odot$. No X-ray counterpart is identified. Given the ambiguous optical associations, this system remains a candidate overdensity and warrants further investigation to assess possible systematics affecting the WL signal.

\subsubsection{GOODS North} \label{sec:gn}

\textbf{NIRWL J123645+6216} ($z=0.861$, $M_{200}=1.1\pm0.3\times10^{14}\ M_\odot$, Figure~\ref{fig:NIRWLJ123645+6216}):
This is the only robust detection in GOODS-N. The WL overdensity lies in the northern portion of the CANDELS GOODS-N footprint and is detected with a peak significance of $S/N=6.2$. The spectroscopic redshift distribution shows a clear overdensity at $z=0.861$. The brightest galaxy at this redshift is slightly offset to the west of the WL peak, while another bright galaxy to the east has $z=0.87$. A bright foreground galaxy at $z=0.5$ is also present in the area.

The spectroscopically selected member galaxies are spatially skewed toward the southwest of the WL peak and include the second- and third-brightest galaxies in the system. The tangential shear profile shows a significant detection, and the NFW fit yields $M_{200}=1.1\pm0.3\times10^{14}\ M_\odot$.

This cluster corresponds to the known overdensity ClG~1236+6215, first identified by \cite{1999barger} and spectroscopically confirmed by \cite{2001dawson}, who measured $z\sim0.85$. \cite{2014guennou} confirmed the redshift and suggested a possible foreground structure at $z\sim0.53$. Deep X-ray observations have yielded only upper limits; \cite{2002bauer} reported an upper limit of $2.1\times10^{42}$ erg s$^{-1}$ in the soft band, and \cite{2013pentericci} similarly reported a non-detection, suggesting that the system is underluminous in X-rays.

\subsubsection{EGS} \label{sec:egs}

We detect no significant overdensities in the EGS field. The EGS footprint is relatively narrow, spanning only $\sim6$ arcmin across, making it less favorable for WL searches of extended overdensities. Edge effects are particularly problematic in this geometry and may suppress sensitivity to mass peaks. In addition, EGS appears to host fewer overdensities than the other CANDELS fields, a conclusion supported by the X-ray observations.

\section{Discussion} \label{sec:discussion}

\subsection{Weak-Lensing Constraints in the Group Regime}

The overdensities identified in this study extend into the group-scale mass regime, with individual systems reaching $M_{200}\sim2\times10^{13}\ M_\odot$. WL measurements at these masses are inherently sensitive to shape noise and projection effects, making robust characterization challenging. For each overdensity, we fit the mass by assuming a fixed $c$--$M$ relation in Section \ref{sec:individual_systems} and, in Section \ref{sec:mass_estimation}, we performed a stacked analysis with both the mass and concentration fit. The detection of a statistically significant stacked signal demonstrates that the ensemble mass distribution of these systems is well constrained despite the modest signal of individual halos.

The stacked tangential shear profile yields a best-fit NFW mass of $M_{200}=1.3\pm0.3\times10^{14}\ M_\odot$ and a concentration of $c=4.9\pm2.1$. This concentration is consistent with theoretical concentration-mass relations at $z\sim0.6$ \citep{2019diemer}, indicating that the average properties of the sample align with expectations from $\Lambda$CDM halo growth.

Comparison with the stacked cluster sample of \cite{2014umetsu} shows that, although our sample is roughly an order of magnitude lower in mass, the radial slope of the excess surface density profile is similar. The comparable slopes and concentrations are broadly consistent with the expected evolution of the $c$--$M$ relation, in which the increase in concentration toward lower halo masses is partly offset by the decrease toward higher redshift. This is consistent with the view that group and cluster systems can be described by the same general halo profile, rather than requiring a qualitatively different internal structure.

\subsection{Connection to X-ray Scaling Relations} \label{sec:xray_discussion}
In Section~\ref{sec:xray_relation}, we used the seven NIRWL overdensities with reported diffuse X-ray counterparts to fit an $L_X$--$M$ relation. These systems extend well into the group regime and therefore provide a useful comparison to scaling relations that are often calibrated on more massive cluster samples.

The NIRWL systems follow the overall trend of published $L_X$--$M$ scaling relations and extend these comparisons into the lower-mass, group regime (see Figure~\ref{fig:mass-xray}). A formal comparison to individual literature relations, however, shows non-negligible offsets when only the quoted measurement uncertainties are considered. We interpret the agreement as qualitative rather than as a precise statistical match because the literature relations shown in Figure~\ref{fig:mass-xray} do not include intrinsic scatter, and because systematics such as sample selection, mass calibration, and X-ray bandpass differ among studies. Furthermore, some of the NIRWL detections may have projected line-of-sight overdensities, which would impact the relation. 

Among the relations presented, that of \cite{2010leauthaud}, derived for stacked groups and clusters in the COSMOS field, yields the smallest discrepancy with our data, although the mismatch remains non-negligible when evaluated using measurement errors alone ($\chi^2_\nu = 3.26$ and $p = 1.8 \times 10^{-3}$). The relations from \cite{2020lovisari}, \cite{2015ketulla}, and \cite{2020sereno} show larger discrepancies. We therefore do not over-interpret the ranking among the literature relations.

The NIRWL sample probes masses down to $M_{200}\sim2\times10^{13}\ M_\odot$, extending well into the group regime where increased scatter is expected. Even so, the broad alignment of the X-ray--detected subsample with previously published relations suggests that these systems follow the same overall mass--luminosity behavior seen in more massive clusters. This, in turn, indicates that the thermodynamic state of the intracluster medium is broadly consistent with gravitational scaling expectations despite the lower halo masses \citep{2010leauthaud,2020lovisari}.

The five overdensities without reported diffuse X-ray emission warrant particular attention. These systems may represent gas-poor groups or dynamically young structures, or their X-ray luminosities may simply fall below the current detection limits in the CANDELS fields. The presence of shear-selected systems without X-ray counterparts highlights the complementary selection functions of WL and ICM-based surveys and underscores the importance of multiwavelength observations for constructing more complete halo samples.

\subsection{Implications for Future IR Weak-Lensing Surveys}

The CANDELS fields cover only 0.23 deg$^2$, yet we identify twelve shear-selected overdensities spanning $0.22<z<0.9$. The achieved source density of $\sim170$ arcmin$^{-2}$ is substantially higher than that of wide-area ground-based surveys and demonstrates the sensitivity of space-based NIR imaging to low-mass halos. 

Upcoming surveys with Euclid and Roman will combine wide area with high spatial resolution and deep NIR imaging. Roman, in particular, should achieve source densities approaching those realized in CANDELS over thousands of square degrees. Our NIRWL study therefore serves as a proof of concept for IR WL detection of group-scale halos at intermediate and high redshift.

In addition to expanding the sample size by orders of magnitude, wide-area IR WL surveys will enable robust measurements of the mass function in the group regime, improved constraints on the $c$--$M$ relation, and detailed investigations of halo assembly across cosmic time. The detection of low-mass systems in CANDELS suggests that Roman-era surveys will extend shear-selected studies into regimes currently inaccessible to ground-based data.

\subsection{Reliability and Caveats of NIR Shear-Selected Overdensities}

Shear-selected overdensities are known to be susceptible to contamination from noise fluctuations and line-of-sight projections, and ground-based surveys have shown that shear peaks can include a non-negligible fraction of spurious detections \citep[e.g.,][]{2007schirmer,2007miyazaki}. In this study, however, the high source density achieved by the NIR data ($\sim$170 arcmin$^{-2}$) substantially reduces shape noise relative to typical ground-based surveys and improves the robustness of the reconstructed mass maps on arcmin scales. This is supported by the $B$-mode analysis, which closely follows a Gaussian expectation, indicating that the excess of $E$-mode pixels at $S/N\gtrsim3$ is unlikely to be driven by systematics alone. In addition, seven of the twelve overdensities have spatially consistent X-ray counterparts, reinforcing the conclusion that most of the detected peaks correspond to real collapsed structures.

However, there are some limitations to our study and the use of the CANDELS fields. The small footprint of the CANDELS fields restricts sensitivity to large-scale structure and can enhance edge effects in the mass reconstruction, particularly in elongated geometries such as EGS. Projection effects also cannot be fully eliminated in shear-selected samples and may contribute to the scatter in individual mass estimates. Finally, incomplete spectroscopic coverage in some fields limits our ability to definitively assess the nature of all overdensities, particularly those without X-ray emission or clear optical associations. Deeper spectroscopy and multiwavelength follow-up will therefore be required to better quantify the contamination fraction of NIR shear-selected samples at group mass scales. Nevertheless, the consistency between the ensemble WL profile, X-ray scaling relations, and theoretical expectations supports the robustness of our detections.

\section{Conclusions} \label{sec:conclusions}
We have conducted the first NIR WL analysis of all five HST/CANDELS fields. The exceptional depth and spatial resolution of the HST/WFC3 F160W imaging provided a high number density of galaxies ($\sim$ 170 arcmin$^{-2}$), which permitted a detailed search and characterization of shear-selected overdensities in these well-studied legacy fields. Utilizing PSF models and knowledge of systematics from our first work on the CANDELS fields \citep{2023cfinner}, we measured galaxy shapes using the SFIT shape measurement technique. We combined iterative source selection at photometric redshift intervals and mass-map reconstruction to identify significant WL shear signals. The selected candidate WL signals were compared with redshift overdensities and red sequences from color-magnitude diagrams to verify their reality. From this technique, we identified 12 significant overdensities that span $0.22 < z < 0.9$ and a mass range of $M_{200}\approx (0.2-2.2)\times10^{14}\ M_\odot$. 

Comparing the spatial location of the overdensities with the locations of diffuse X-ray emission from literature, we found that seven of them have X-ray counterparts. This demonstrates that these shear-selected systems correspond to genuine collapsed structures with baryonic content. We placed the overdensities onto well-known X-ray luminosity--mass scaling relations and found that these group and low-mass cluster overdensities are in broad agreement with the established scaling relations, which indicates that NIR WL measurements provide reliable halo masses for low-mass systems.

To achieve a high-significance detection of the average excess mass density profile, we stacked the tangential shear signals of the full sample. The stacked profile shows a significant detection of the surface mass density to $\sim$ 2 Mpc. Fitting an NFW model to the mass density profile, we found the average halo mass to be $M_{200}=1.3\pm0.3\times10^{14}\ M_\odot$ with a concentration of $c=4.9\pm2.1$. 

Our results demonstrate that NIR WL is an effective tool for detecting and characterizing overdensities across a wide redshift and mass range. The CANDELS fields are small regions of the sky compared to ongoing and upcoming surveys with telescopes such as Euclid, Roman, and Rubin. Exploiting the strong capabilities of high-resolution IR imaging for detecting distant galaxies, NIR WL studies will become powerful for studying overdensities at increasingly higher redshifts and lower masses with great precision.

\section{Software and third party data repository citations} \label{sec:cite}

\begin{acknowledgments}
K.F. would like to thank KASI for hosting him during the writing of this manuscript. Bomee Lee is supported by the National Research Foundation of Korea (NRF) grant funded by the Korea government (MSIT), 2022R1C1C1008695.

\end{acknowledgments}

%

\vspace{5mm}
\facilities{HST(WFC3)}


\software{astropy \citep{2022astropy},  
          Source Extractor \citep{1996bertin}
          }



\appendix
\setcounter{figure}{0}
\renewcommand{\thefigure}{A\arabic{figure}}
\section{Individual System Figures} \label{sec:appendix_systems}

We present the full set of WL maps, tangential shear profiles, color-magnitude diagrams, and redshift histograms for all overdensities in the sample.

\begin{figure*}[ht]
    \centering
    \includegraphics[width=\textwidth]{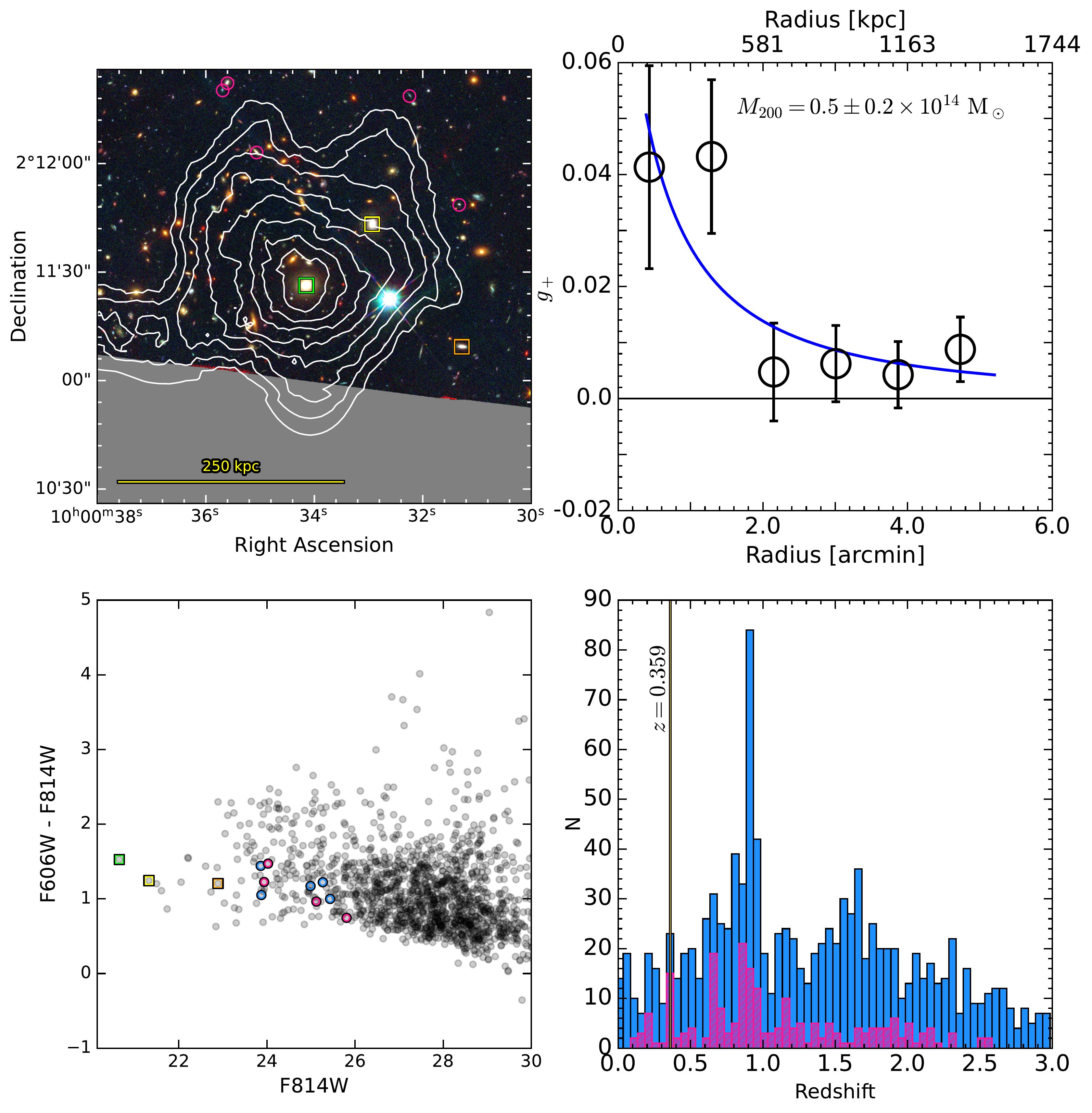}
    \caption{NIRWL J100034+0211 - COSMOS. Same as Figure \ref{fig:NIRWLJ100022+0212}. }
    \label{fig:NIRWLJ100034+0211}
\end{figure*}

\begin{figure*}[ht]
    \centering
    \includegraphics[width=\textwidth]{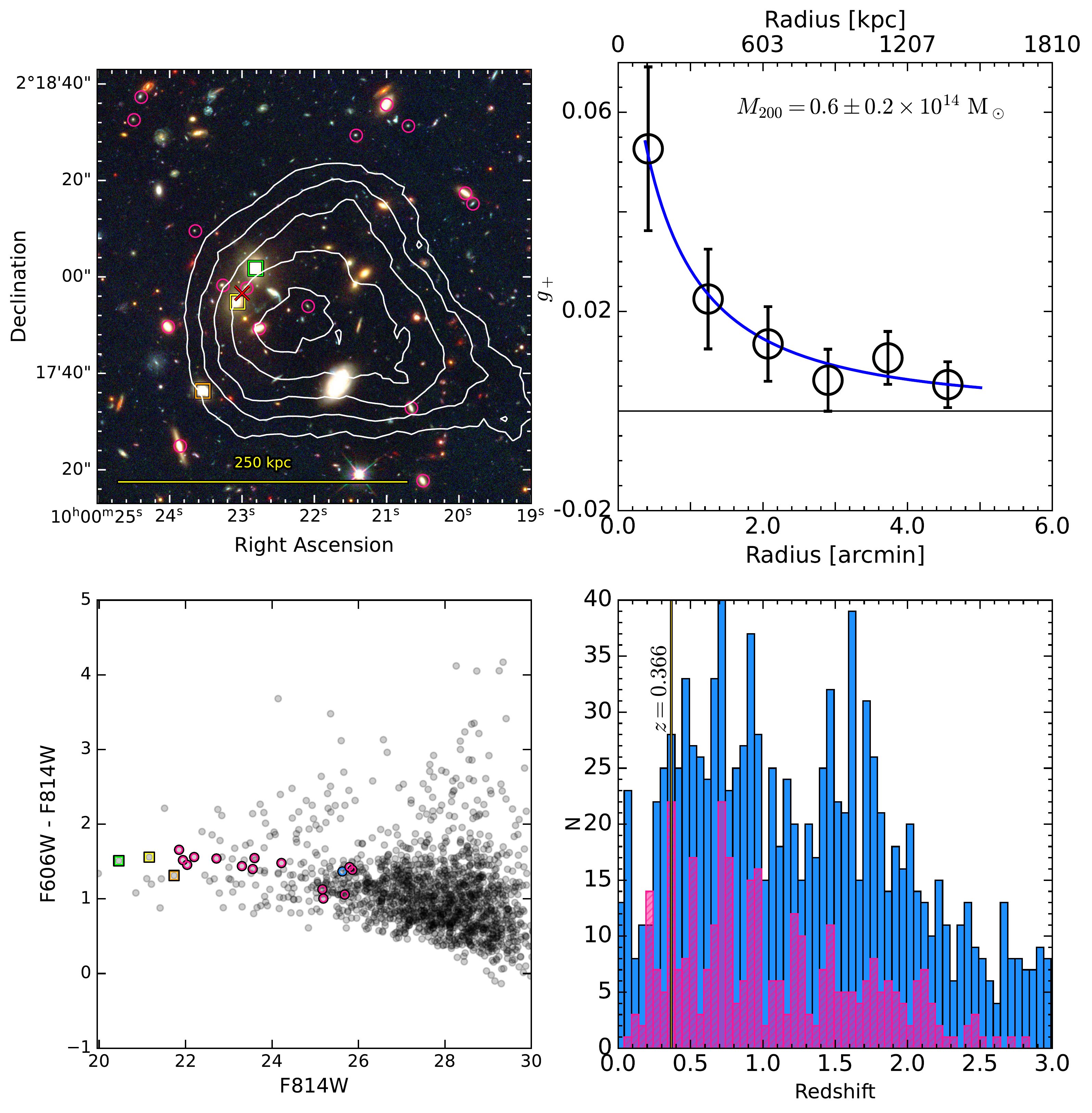}
    \caption{NIRWL J100022+0217 - COSMOS. Same as Figure \ref{fig:NIRWLJ100022+0212}. }
    \label{fig:NIRWLJ100022+0217}
\end{figure*}

\begin{figure*}[ht]
    \centering
    \includegraphics[width=\textwidth]{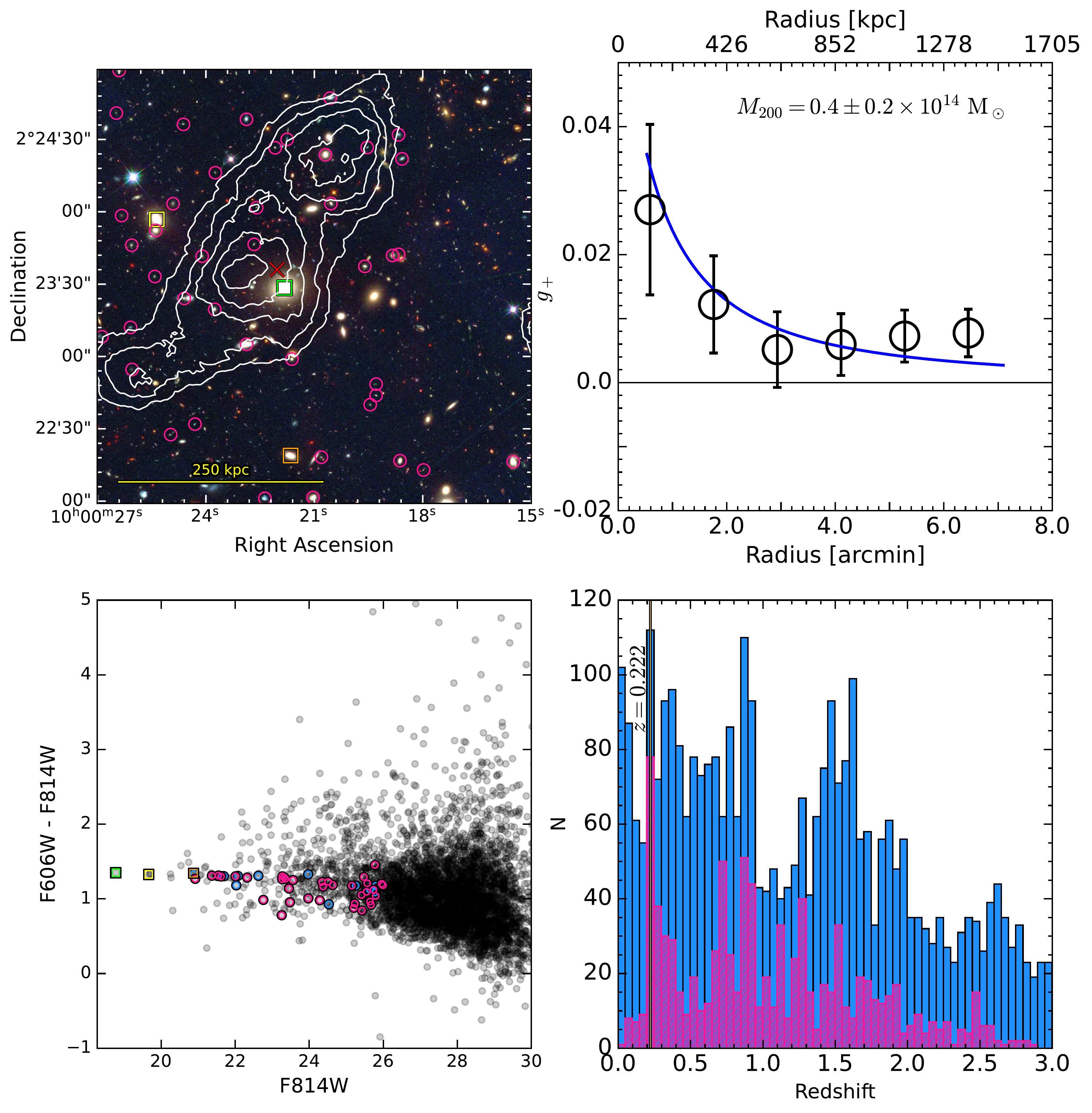}
    \caption{NIRWL J100021+0223 - COSMOS. Same as Figure \ref{fig:NIRWLJ100022+0212}. A low-$z$ group-scale overdensity with signs of substructure in the mass distribution.}
    \label{fig:NIRWLJ100021+0223}
\end{figure*}

\begin{figure*}[ht]
    \centering
    \includegraphics[width=\textwidth]{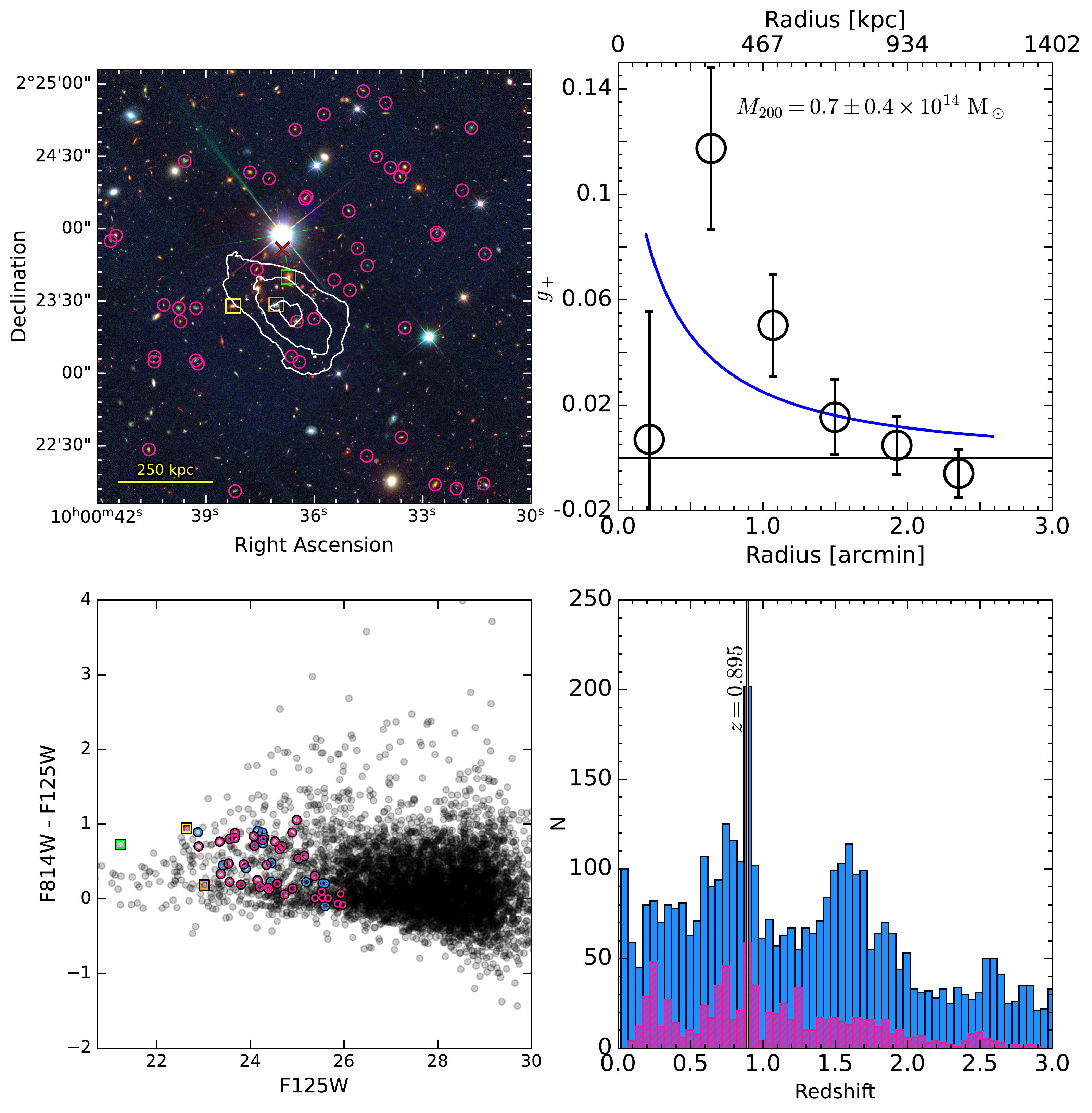}
    \caption{NIRWL J100035+0223 - COSMOS. Same as Figure \ref{fig:NIRWLJ100022+0212}.}
    \label{fig:NIRWLJ100035+0223}
\end{figure*}

\begin{figure*}[ht]
    \centering
    \includegraphics[width=\textwidth]{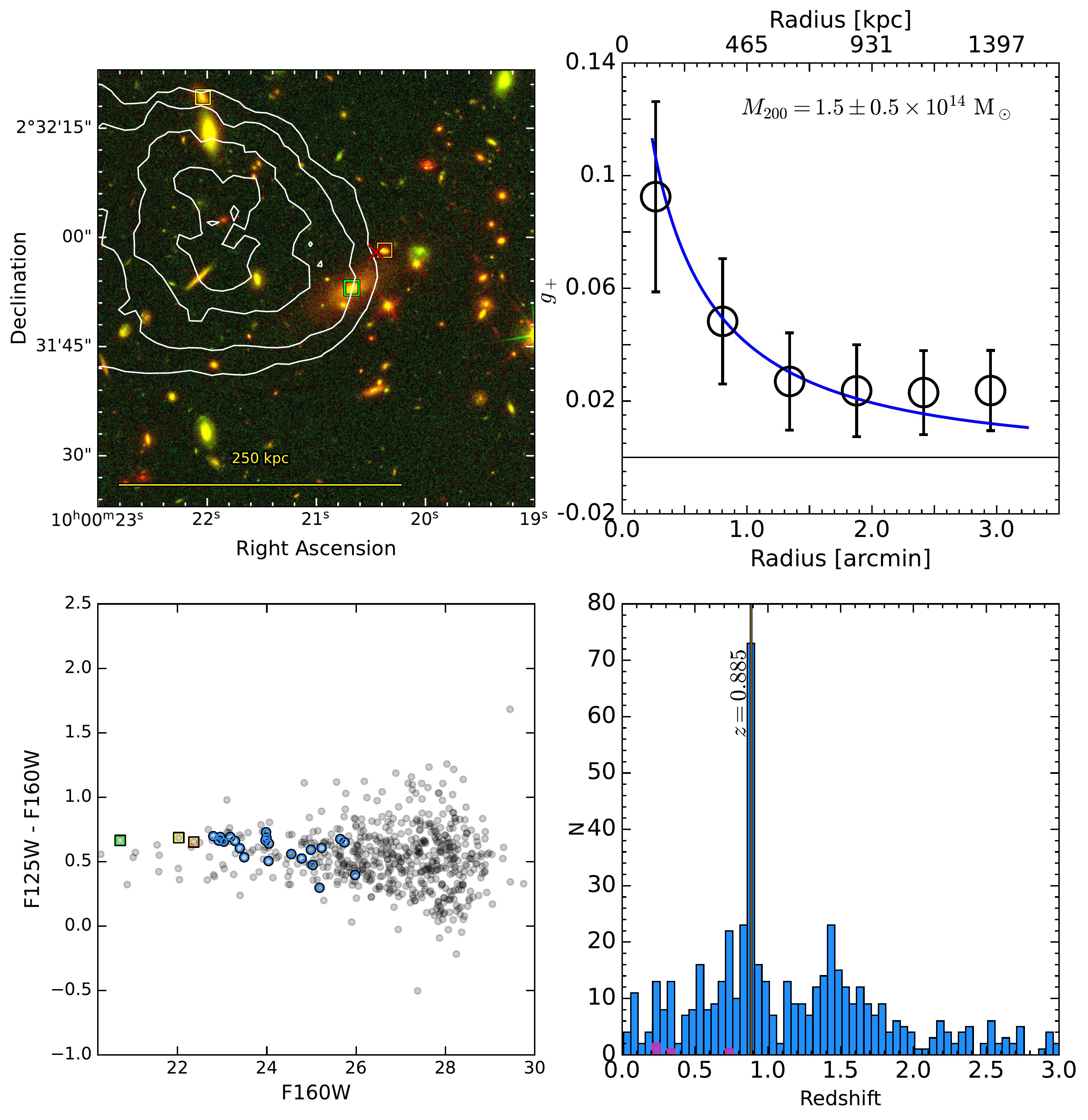}
    \caption{NIRWL J100021+0231 - COSMOS. Same as Figure \ref{fig:NIRWLJ100022+0212}. }
    \label{fig:NIRWLJ100021+0231}
\end{figure*}


\begin{figure*}[ht]
    \centering
    \includegraphics[width=\textwidth]{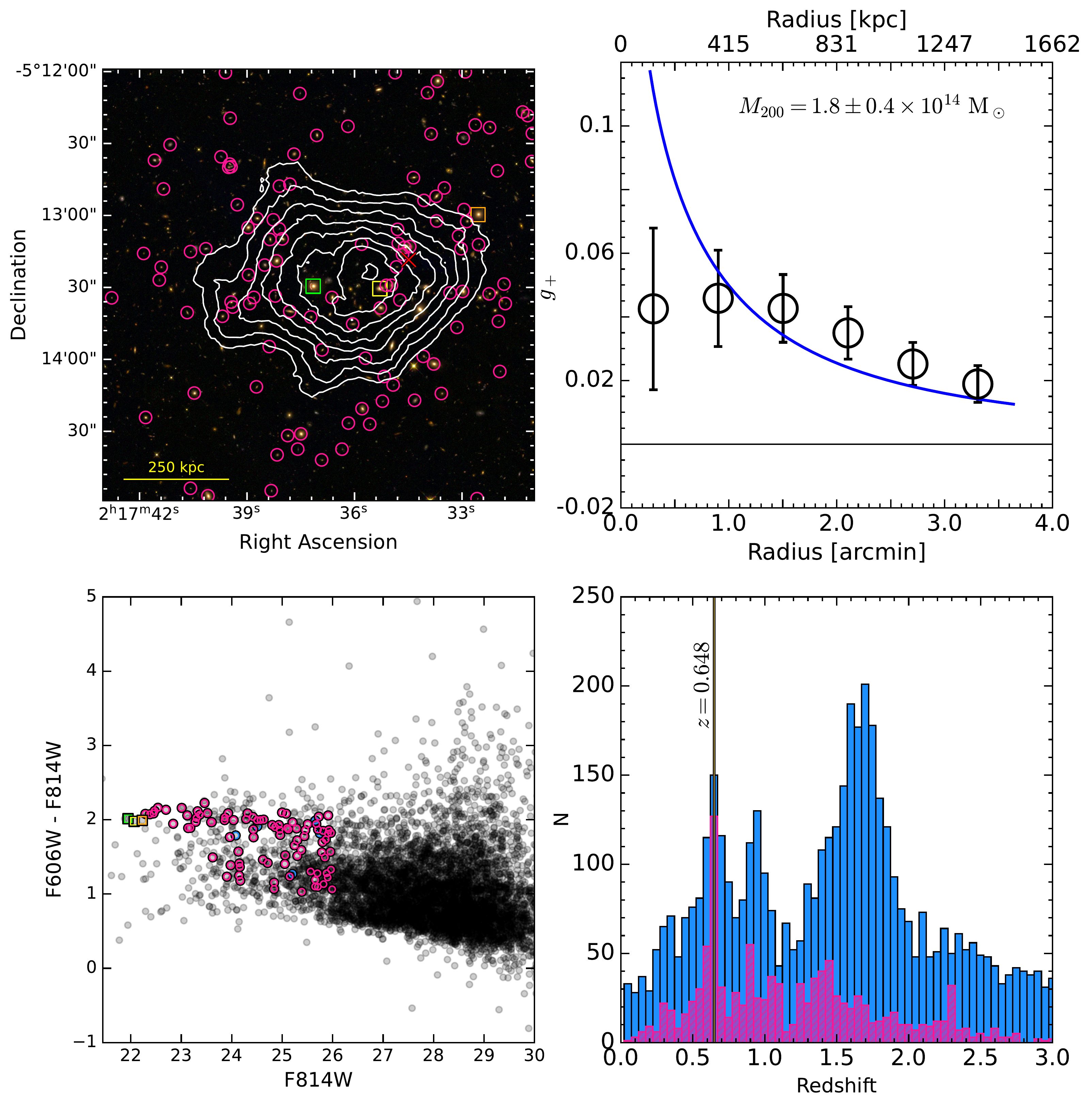}
    
    \caption{NIRWL J021737-0513 - UDS. Same as Figure \ref{fig:NIRWLJ100022+0212}. This is the most significant detection in the UDS field. The overdensity is a cluster at redshift 0.648. The dark matter distribution is elongated in an east-west direction.}
    \label{fig:NIRWLJ021737-0513}
\end{figure*}

\begin{figure*}[ht]
    \centering
    \includegraphics[width=\textwidth]{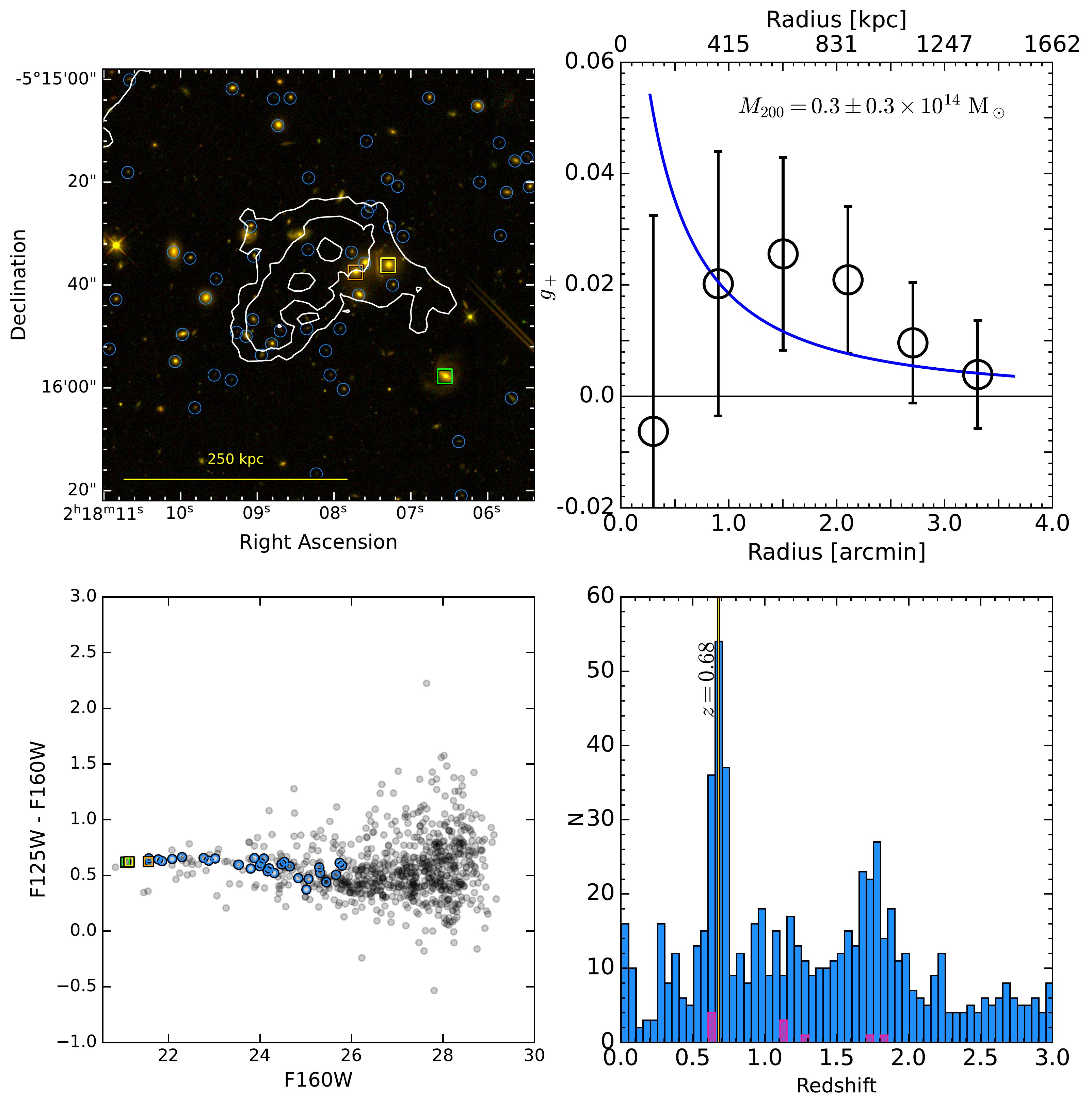}
    \caption{NIRWL J021808-0515 - UDS. Same as Figure \ref{fig:NIRWLJ100022+0212}. Spectroscopic coverage of this overdensity is limited to the brightest galaxies. We present the galaxies that are photometric candidates for membership. A candidate strong-lensing arc is located beside the bright galaxy marked with a yellow box.}
    \label{fig:NIRWLJ021808-0515}
\end{figure*}

\begin{figure*}[ht]
    \centering
    \includegraphics[width=\linewidth]{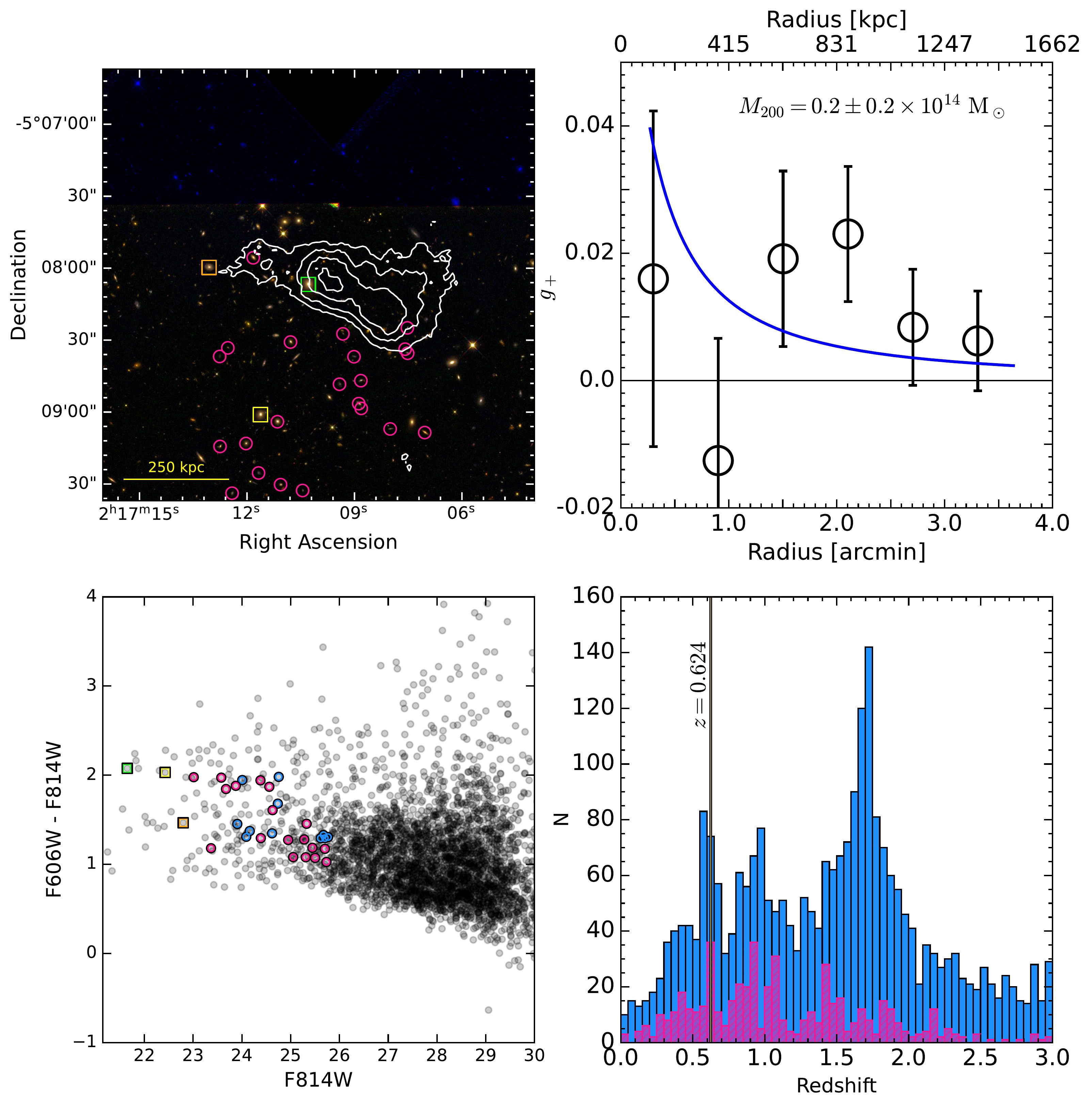}
    \caption{NIRWLJ021709-0508. UDS. Same as Figure \ref{fig:NIRWLJ100022+0212}. A low-significance detection in the UDS. }
    \label{fig:NIRWLJ021709-0508}
\end{figure*}


\begin{figure*}[ht]
    \centering
    \includegraphics[width=\textwidth]{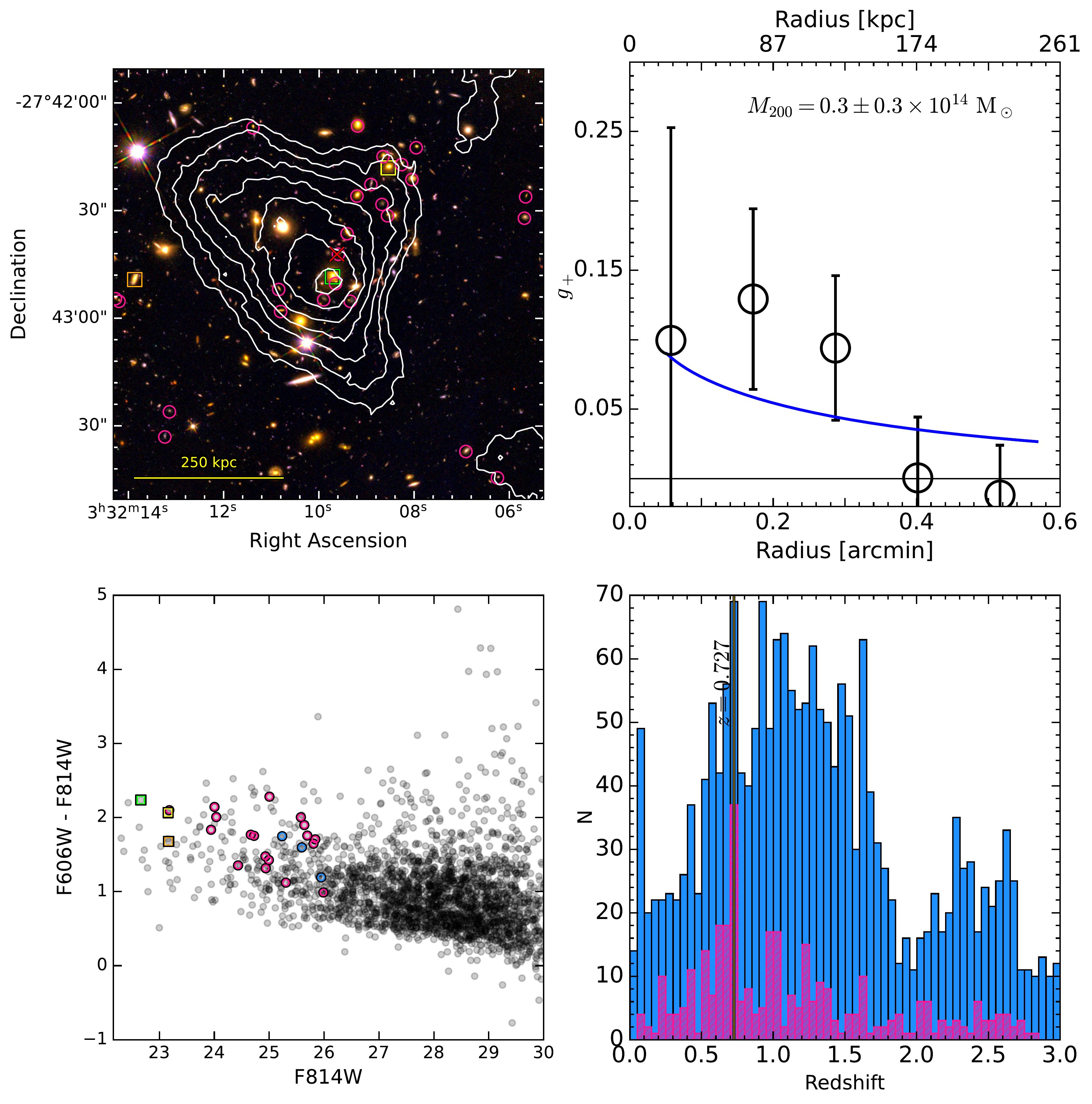}
    \caption{NIRWL J033209-2742 - GOODS-S. Same as Figure \ref{fig:NIRWLJ100022+0212}. The mass map peaks at the brightest galaxy in the region and elongates to the NW toward the second brightest galaxy.}
    \label{fig:NIRWLJ033209-2742}
\end{figure*}

\begin{figure*}[ht]
    \centering
    \includegraphics[width=\textwidth]{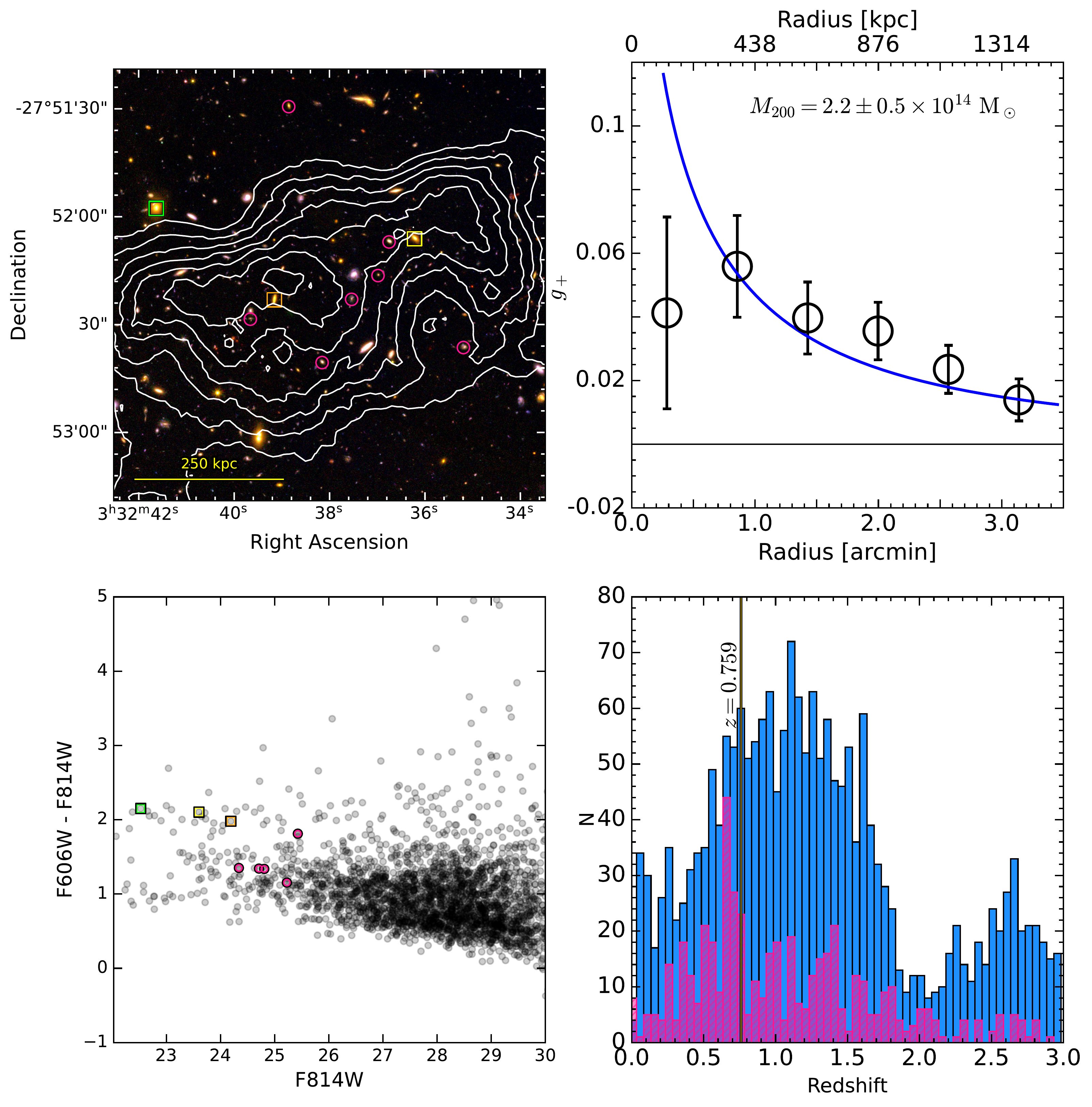}
    \caption{NIRWL J033239-2752 - GOODS-S. Same as Figure \ref{fig:NIRWLJ100022+0212}. The mass map has a single peak that lies beside the third BCG of the region. At lower significance the mass map shows signs of an extended structure.  }
    \label{fig:NIRWLJ033239-2752}
\end{figure*}

\begin{figure*}[ht]
    \centering
    \includegraphics[width=\textwidth]{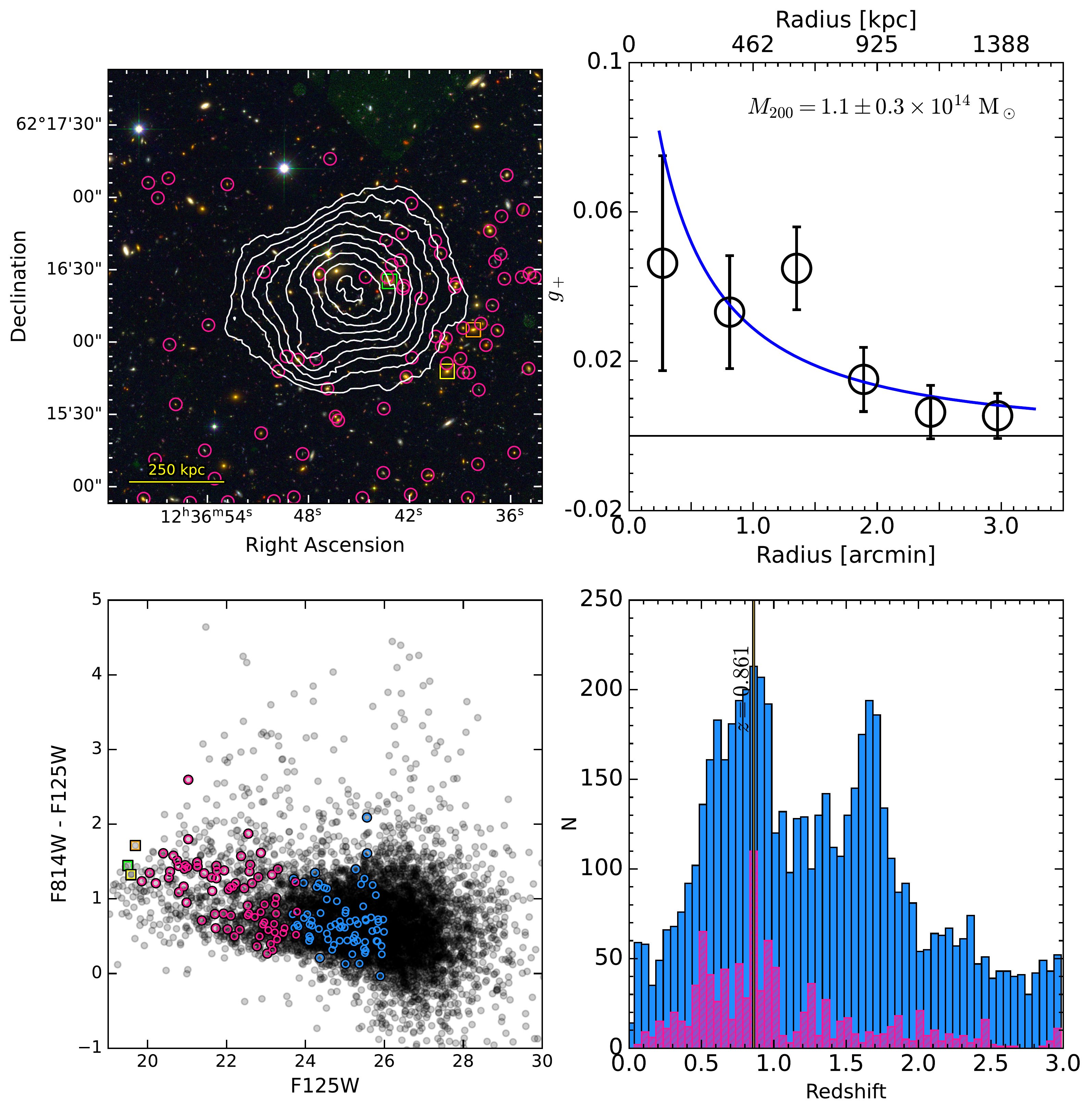}
    \caption{NIRWL J123645+6216 - GOODS-N. Same as Figure \ref{fig:NIRWLJ100022+0212}. The most significant detection in the GOODS-N field with a mean redshift of $z=0.861$. The peak of the mass map is near the BCG. A dense clump of galaxies is situated to the west of the BCG but do not coincide with significant WL signal.}
    \label{fig:NIRWLJ123645+6216}
\end{figure*}

\bibliography{sample631}{}

@ARTICLE{2024bulbul,
       author = {{Bulbul}, E. and {Liu}, A. and {Kluge}, M. and {Zhang}, X. and {Sanders}, J.~S. and {Bahar}, Y.~E. and {Ghirardini}, V. and {Artis}, E. and {Seppi}, R. and {Garrel}, C. and {Ramos-Ceja}, M.~E. and {Comparat}, J. and {Balzer}, F. and {B{\"o}ckmann}, K. and {Br{\"u}ggen}, M. and {Clerc}, N. and {Dennerl}, K. and {Dolag}, K. and {Freyberg}, M. and {Grandis}, S. and {Gruen}, D. and {Kleinebreil}, F. and {Krippendorf}, S. and {Lamer}, G. and {Merloni}, A. and {Migkas}, K. and {Nandra}, K. and {Pacaud}, F. and {Predehl}, P. and {Reiprich}, T.~H. and {Schrabback}, T. and {Veronica}, A. and {Weller}, J. and {Zelmer}, S.},
        title = "{The SRG/eROSITA All-Sky Survey. The first catalog of galaxy clusters and groups in the Western Galactic Hemisphere}",
      journal = {\aap},
         year = 2024,
        month = may,
       volume = {685},
          eid = {A106},
        pages = {A106},
          doi = {10.1051/0004-6361/202348264},
archivePrefix = {arXiv},
       eprint = {2402.08452},
 primaryClass = {astro-ph.CO},
       adsurl = {https://ui.adsabs.harvard.edu/abs/2024A&A...685A.106B}
}

@INPROCEEDINGS{2001gladders,
       author = {{Gladders}, M.~D. and {Yee}, H.~K.~C.},
        title = "{The Red-Sequence Cluster Survey.}",
    booktitle = {The New Era of Wide Field Astronomy},
         year = 2001,
       editor = {{Clowes}, Roger and {Adamson}, Andrew and {Bromage}, Gordon},
       series = {Astronomical Society of the Pacific Conference Series},
       volume = {232},
        month = jan,
        pages = {126},
          doi = {10.48550/arXiv.astro-ph/0011073},
archivePrefix = {arXiv},
       eprint = {astro-ph/0011073},
 primaryClass = {astro-ph},
       adsurl = {https://ui.adsabs.harvard.edu/abs/2001ASPC..232..126G}
}

@ARTICLE{1958abell,
       author = {{Abell}, George O.},
        title = "{The Distribution of Rich Clusters of Galaxies.}",
      journal = {\apjs},
         year = 1958,
        month = may,
       volume = {3},
        pages = {211},
          doi = {10.1086/190036},
       adsurl = {https://ui.adsabs.harvard.edu/abs/1958ApJS....3..211A}
}

@INPROCEEDINGS{2008wilson,
       author = {{Wilson}, G. and {Muzzin}, A. and {Lacy}, M. and {Yee}, H. and {Surace}, J. and {Lonsdale}, C. and {Hoekstra}, H. and {Majumdar}, S. and {Gilbank}, D. and {Gladders}, M.},
        title = "{Clusters of Galaxies at 1 < z< 2 : The Spitzer Adaptation of the Red-Sequence Cluster Survey}",
    booktitle = {Infrared Diagnostics of Galaxy Evolution},
         year = 2008,
       editor = {{Chary}, R. -R. and {Teplitz}, H.~I. and {Sheth}, K.},
       series = {Astronomical Society of the Pacific Conference Series},
       volume = {381},
        month = mar,
        pages = {210},
       adsurl = {https://ui.adsabs.harvard.edu/abs/2008ASPC..381..210W}
}

@ARTICLE{1974press-schecter,
       author = {{Press}, William H. and {Schechter}, Paul},
        title = "{Formation of Galaxies and Clusters of Galaxies by Self-Similar Gravitational Condensation}",
      journal = {\apj},
         year = 1974,
        month = feb,
       volume = {187},
        pages = {425-438},
          doi = {10.1086/152650},
       adsurl = {https://ui.adsabs.harvard.edu/abs/1974ApJ...187..425P}
}

@ARTICLE{1984blumenthal,
       author = {{Blumenthal}, G.~R. and {Faber}, S.~M. and {Primack}, J.~R. and {Rees}, M.~J.},
        title = "{Formation of galaxies and large-scale structure with cold dark matter.}",
      journal = {\nat},
         year = 1984,
        month = oct,
       volume = {311},
        pages = {517-525},
          doi = {10.1038/311517a0},
       adsurl = {https://ui.adsabs.harvard.edu/abs/1984Natur.311..517B}
}

@ARTICLE{1985davis,
       author = {{Davis}, M. and {Efstathiou}, G. and {Frenk}, C.~S. and {White}, S.~D.~M.},
        title = "{The evolution of large-scale structure in a universe dominated by cold dark matter}",
      journal = {\apj},
         year = 1985,
        month = may,
       volume = {292},
        pages = {371-394},
          doi = {10.1086/163168},
       adsurl = {https://ui.adsabs.harvard.edu/abs/1985ApJ...292..371D}
}

@ARTICLE{1977visvanathan,
       author = {{Visvanathan}, N. and {Sandage}, A.},
        title = "{The color - absolute magnitude relation for E and S0 galaxies. I. Calibration and tests for universality using Virgo and eight other nearby clusters.}",
      journal = {\apj},
         year = 1977,
        month = aug,
       volume = {216},
        pages = {214-226},
          doi = {10.1086/155464},
       adsurl = {https://ui.adsabs.harvard.edu/abs/1977ApJ...216..214V}
}

@ARTICLE{1992bower,
       author = {{Bower}, R.~G. and {Lucey}, J.~R. and {Ellis}, R.~S.},
        title = "{Precision photometry of early-type galaxies in the Coma and Virgo clusters : a test of the universality of the colour-magnitude relation - II. Analysis.}",
      journal = {\mnras},
         year = 1992,
        month = feb,
       volume = {254},
        pages = {601},
          doi = {10.1093/mnras/254.4.601},
       adsurl = {https://ui.adsabs.harvard.edu/abs/1992MNRAS.254..601B}
}

@ARTICLE{1998zabludoff,
       author = {{Zabludoff}, Ann I. and {Mulchaey}, John S.},
        title = "{The Properties of Poor Groups of Galaxies. I. Spectroscopic Survey and Results}",
      journal = {\apj},
         year = 1998,
        month = mar,
       volume = {496},
       number = {1},
        pages = {39-72},
          doi = {10.1086/305355},
archivePrefix = {arXiv},
       eprint = {astro-ph/9708132},
 primaryClass = {astro-ph},
       adsurl = {https://ui.adsabs.harvard.edu/abs/1998ApJ...496...39Z}
}

@ARTICLE{2016rykoff,
       author = {{Rykoff}, E.~S. and {Rozo}, E. and {Hollowood}, D. and {Bermeo-Hernandez}, A. and {Jeltema}, T. and {Mayers}, J. and {Romer}, A.~K. and {Rooney}, P. and {Saro}, A. and {Vergara Cervantes}, C. and {Wechsler}, R.~H. and {Wilcox}, H. and {Abbott}, T.~M.~C. and {Abdalla}, F.~B. and {Allam}, S. and {Annis}, J. and {Benoit-L{\'e}vy}, A. and {Bernstein}, G.~M. and {Bertin}, E. and {Brooks}, D. and {Burke}, D.~L. and {Capozzi}, D. and {Carnero Rosell}, A. and {Carrasco Kind}, M. and {Castander}, F.~J. and {Childress}, M. and {Collins}, C.~A. and {Cunha}, C.~E. and {D'Andrea}, C.~B. and {da Costa}, L.~N. and {Davis}, T.~M. and {Desai}, S. and {Diehl}, H.~T. and {Dietrich}, J.~P. and {Doel}, P. and {Evrard}, A.~E. and {Finley}, D.~A. and {Flaugher}, B. and {Fosalba}, P. and {Frieman}, J. and {Glazebrook}, K. and {Goldstein}, D.~A. and {Gruen}, D. and {Gruendl}, R.~A. and {Gutierrez}, G. and {Hilton}, M. and {Honscheid}, K. and {Hoyle}, B. and {James}, D.~J. and {Kay}, S.~T. and {Kuehn}, K. and {Kuropatkin}, N. and {Lahav}, O. and {Lewis}, G.~F. and {Lidman}, C. and {Lima}, M. and {Maia}, M.~A.~G. and {Mann}, R.~G. and {Marshall}, J.~L. and {Martini}, P. and {Melchior}, P. and {Miller}, C.~J. and {Miquel}, R. and {Mohr}, J.~J. and {Nichol}, R.~C. and {Nord}, B. and {Ogando}, R. and {Plazas}, A.~A. and {Reil}, K. and {Sahl{\'e}n}, M. and {Sanchez}, E. and {Santiago}, B. and {Scarpine}, V. and {Schubnell}, M. and {Sevilla-Noarbe}, I. and {Smith}, R.~C. and {Soares-Santos}, M. and {Sobreira}, F. and {Stott}, J.~P. and {Suchyta}, E. and {Swanson}, M.~E.~C. and {Tarle}, G. and {Thomas}, D. and {Tucker}, D. and {Uddin}, S. and {Viana}, P.~T.~P. and {Vikram}, V. and {Walker}, A.~R. and {Zhang}, Y. and {DES Collaboration}},
        title = "{The RedMaPPer Galaxy Cluster Catalog From DES Science Verification Data}",
      journal = {\apjs},
         year = 2016,
        month = may,
       volume = {224},
       number = {1},
          eid = {1},
        pages = {1},
          doi = {10.3847/0067-0049/224/1/1},
archivePrefix = {arXiv},
       eprint = {1601.00621},
 primaryClass = {astro-ph.CO},
       adsurl = {https://ui.adsabs.harvard.edu/abs/2016ApJS..224....1R}
}

@ARTICLE{2017carmen,
       author = {{Rodr{\'\i}guez-Gonz{\'a}lvez}, C. and {Chary}, R.~R. and {Muchovej}, S. and {Melin}, J.-B. and {Feroz}, F. and {Olamaie}, M. and {Shimwell}, T.},
        title = "{CARMA observations of massive Planck -discovered cluster candidates at z {\ensuremath{\gtrsim}} 0.5 associated with WISE overdensities: breaking the size-flux degeneracy}",
      journal = {\mnras},
         year = 2017,
        month = jan,
       volume = {464},
       number = {2},
        pages = {2378-2395},
          doi = {10.1093/mnras/stw2392},
archivePrefix = {arXiv},
       eprint = {1505.01132},
 primaryClass = {astro-ph.CO},
       adsurl = {https://ui.adsabs.harvard.edu/abs/2017MNRAS.464.2378R}
}

@ARTICLE{2016pierre,
       author = {{Pierre}, M. and {Pacaud}, F. and {Adami}, C. and {Alis}, S. and {Altieri}, B. and {Baran}, N. and {Benoist}, C. and {Birkinshaw}, M. and {Bongiorno}, A. and {Bremer}, M.~N. and {Brusa}, M. and {Butler}, A. and {Ciliegi}, P. and {Chiappetti}, L. and {Clerc}, N. and {Corasaniti}, P.~S. and {Coupon}, J. and {De Breuck}, C. and {Democles}, J. and {Desai}, S. and {Delhaize}, J. and {Devriendt}, J. and {Dubois}, Y. and {Eckert}, D. and {Elyiv}, A. and {Ettori}, S. and {Evrard}, A. and {Faccioli}, L. and {Farahi}, A. and {Ferrari}, C. and {Finet}, F. and {Fotopoulou}, S. and {Fourmanoit}, N. and {Gandhi}, P. and {Gastaldello}, F. and {Gastaud}, R. and {Georgantopoulos}, I. and {Giles}, P. and {Guennou}, L. and {Guglielmo}, V. and {Horellou}, C. and {Husband}, K. and {Huynh}, M. and {Iovino}, A. and {Kilbinger}, M. and {Koulouridis}, E. and {Lavoie}, S. and {Le Brun}, A.~M.~C. and {Le Fevre}, J.~P. and {Lidman}, C. and {Lieu}, M. and {Lin}, C.~A. and {Mantz}, A. and {Maughan}, B.~J. and {Maurogordato}, S. and {McCarthy}, I.~G. and {McGee}, S. and {Melin}, J.~B. and {Melnyk}, O. and {Menanteau}, F. and {Novak}, M. and {Paltani}, S. and {Plionis}, M. and {Poggianti}, B.~M. and {Pomarede}, D. and {Pompei}, E. and {Ponman}, T.~J. and {Ramos-Ceja}, M.~E. and {Ranalli}, P. and {Rapetti}, D. and {Raychaudury}, S. and {Reiprich}, T.~H. and {Rottgering}, H. and {Rozo}, E. and {Rykoff}, E. and {Sadibekova}, T. and {Santos}, J. and {Sauvageot}, J.~L. and {Schimd}, C. and {Sereno}, M. and {Smith}, G.~P. and {Smol{\v{c}}i{\'c}}, V. and {Snowden}, S. and {Spergel}, D. and {Stanford}, S. and {Surdej}, J. and {Valageas}, P. and {Valotti}, A. and {Valtchanov}, I. and {Vignali}, C. and {Willis}, J. and {Ziparo}, F.},
        title = "{The XXL Survey. I. Scientific motivations - XMM-Newton observing plan - Follow-up observations and simulation programme}",
      journal = {\aap},
         year = 2016,
        month = jun,
       volume = {592},
          eid = {A1},
        pages = {A1},
          doi = {10.1051/0004-6361/201526766},
archivePrefix = {arXiv},
       eprint = {1512.04317},
 primaryClass = {astro-ph.CO},
       adsurl = {https://ui.adsabs.harvard.edu/abs/2016A&A...592A...1P}
}

@ARTICLE{2020sereno,
       author = {{Sereno}, Mauro and {Umetsu}, Keiichi and {Ettori}, Stefano and {Eckert}, Dominique and {Gastaldello}, Fabio and {Giles}, Paul and {Lieu}, Maggie and {Maughan}, Ben and {Okabe}, Nobuhiro and {Birkinshaw}, Mark and {Chiu}, I. -Non and {Fujita}, Yutaka and {Miyazaki}, Satoshi and {Rapetti}, David and {Koulouridis}, Elias and {Pierre}, Marguerite},
        title = "{XXL Survey groups and clusters in the Hyper Suprime-Cam Survey. Scaling relations between X-ray properties and weak lensing mass}",
      journal = {\mnras},
         year = 2020,
        month = mar,
       volume = {492},
       number = {3},
        pages = {4528-4545},
          doi = {10.1093/mnras/stz3425},
archivePrefix = {arXiv},
       eprint = {1912.02827},
 primaryClass = {astro-ph.CO},
       adsurl = {https://ui.adsabs.harvard.edu/abs/2020MNRAS.492.4528S}
}

@ARTICLE{2020lovisari,
       author = {{Lovisari}, Lorenzo and {Schellenberger}, Gerrit and {Sereno}, Mauro and {Ettori}, Stefano and {Pratt}, Gabriel W. and {Forman}, William R. and {Jones}, Christine and {Andrade-Santos}, Felipe and {Randall}, Scott and {Kraft}, Ralph},
        title = "{X-Ray Scaling Relations for a Representative Sample of Planck-selected Clusters Observed with XMM-Newton}",
      journal = {\apj},
         year = 2020,
        month = apr,
       volume = {892},
       number = {2},
          eid = {102},
        pages = {102},
          doi = {10.3847/1538-4357/ab7997},
archivePrefix = {arXiv},
       eprint = {2002.11740},
 primaryClass = {astro-ph.CO},
       adsurl = {https://ui.adsabs.harvard.edu/abs/2020ApJ...892..102L}
}

@ARTICLE{2010leauthaud,
       author = {{Leauthaud}, Alexie and {Finoguenov}, Alexis and {Kneib}, Jean-Paul and {Taylor}, James E. and {Massey}, Richard and {Rhodes}, Jason and {Ilbert}, Olivier and {Bundy}, Kevin and {Tinker}, Jeremy and {George}, Matthew R. and {Capak}, Peter and {Koekemoer}, Anton M. and {Johnston}, David E. and {Zhang}, Yu-Ying and {Cappelluti}, Nico and {Ellis}, Richard S. and {Elvis}, Martin and {Giodini}, Stefania and {Heymans}, Catherine and {Le F{\`e}vre}, Oliver and {Lilly}, Simon and {McCracken}, Henry J. and {Mellier}, Yannick and {R{\'e}fr{\'e}gier}, Alexandre and {Salvato}, Mara and {Scoville}, Nick and {Smoot}, George and {Tanaka}, Masayuki and {Van Waerbeke}, Ludovic and {Wolk}, Melody},
        title = "{A Weak Lensing Study of X-ray Groups in the Cosmos Survey: Form and Evolution of the Mass-Luminosity Relation}",
      journal = {\apj},
         year = 2010,
        month = jan,
       volume = {709},
       number = {1},
        pages = {97-114},
          doi = {10.1088/0004-637X/709/1/97},
archivePrefix = {arXiv},
       eprint = {0910.5219},
 primaryClass = {astro-ph.CO},
       adsurl = {https://ui.adsabs.harvard.edu/abs/2010ApJ...709...97L}
}

@ARTICLE{2015ketulla,
       author = {{Kettula}, K. and {Giodini}, S. and {van Uitert}, E. and {Hoekstra}, H. and {Finoguenov}, A. and {Lerchster}, M. and {Erben}, T. and {Heymans}, C. and {Hildebrandt}, H. and {Kitching}, T.~D. and {Mahdavi}, A. and {Mellier}, Y. and {Miller}, L. and {Mirkazemi}, M. and {Van Waerbeke}, L. and {Coupon}, J. and {Egami}, E. and {Fu}, L. and {Hudson}, M.~J. and {Kneib}, J.~P. and {Kuijken}, K. and {McCracken}, H.~J. and {Pereira}, M.~J. and {Rowe}, B. and {Schrabback}, T. and {Tanaka}, M. and {Velander}, M.},
        title = "{CFHTLenS: weak lensing calibrated scaling relations for low-mass clusters of galaxies}",
      journal = {\mnras},
         year = 2015,
        month = aug,
       volume = {451},
       number = {2},
        pages = {1460-1481},
          doi = {10.1093/mnras/stv923},
archivePrefix = {arXiv},
       eprint = {1410.8769},
 primaryClass = {astro-ph.CO},
       adsurl = {https://ui.adsabs.harvard.edu/abs/2015MNRAS.451.1460K}
}

@ARTICLE{2018miyazaki,
       author = {{Miyazaki}, Satoshi and {Komiyama}, Yutaka and {Kawanomoto}, Satoshi and {Doi}, Yoshiyuki and {Furusawa}, Hisanori and {Hamana}, Takashi and {Hayashi}, Yusuke and {Ikeda}, Hiroyuki and {Kamata}, Yukiko and {Karoji}, Hiroshi and {Koike}, Michitaro and {Kurakami}, Tomio and {Miyama}, Shoken and {Morokuma}, Tomoki and {Nakata}, Fumiaki and {Namikawa}, Kazuhito and {Nakaya}, Hidehiko and {Nariai}, Kyoji and {Obuchi}, Yoshiyuki and {Oishi}, Yukie and {Okada}, Norio and {Okura}, Yuki and {Tait}, Philip and {Takata}, Tadafumi and {Tanaka}, Yoko and {Tanaka}, Masayuki and {Terai}, Tsuyoshi and {Tomono}, Daigo and {Uraguchi}, Fumihiro and {Usuda}, Tomonori and {Utsumi}, Yousuke and {Yamada}, Yoshihiko and {Yamanoi}, Hitomi and {Aihara}, Hiroaki and {Fujimori}, Hiroki and {Mineo}, Sogo and {Miyatake}, Hironao and {Oguri}, Masamune and {Uchida}, Tomohisa and {Tanaka}, Manobu M. and {Yasuda}, Naoki and {Takada}, Masahiro and {Murayama}, Hitoshi and {Nishizawa}, Atsushi J. and {Sugiyama}, Naoshi and {Chiba}, Masashi and {Futamase}, Toshifumi and {Wang}, Shiang-Yu and {Chen}, Hsin-Yo and {Ho}, Paul T.~P. and {Liaw}, Eric J.~Y. and {Chiu}, Chi-Fang and {Ho}, Cheng-Lin and {Lai}, Tsang-Chih and {Lee}, Yao-Cheng and {Jeng}, Dun-Zen and {Iwamura}, Satoru and {Armstrong}, Robert and {Bickerton}, Steve and {Bosch}, James and {Gunn}, James E. and {Lupton}, Robert H. and {Loomis}, Craig and {Price}, Paul and {Smith}, Steward and {Strauss}, Michael A. and {Turner}, Edwin L. and {Suzuki}, Hisanori and {Miyazaki}, Yasuhito and {Muramatsu}, Masaharu and {Yamamoto}, Koei and {Endo}, Makoto and {Ezaki}, Yutaka and {Ito}, Noboru and {Kawaguchi}, Noboru and {Sofuku}, Satoshi and {Taniike}, Tomoaki and {Akutsu}, Kotaro and {Dojo}, Naoto and {Kasumi}, Kazuyuki and {Matsuda}, Toru and {Imoto}, Kohei and {Miwa}, Yoshinori and {Suzuki}, Masayuki and {Takeshi}, Kunio and {Yokota}, Hideo},
        title = "{Hyper Suprime-Cam: System design and verification of image quality}",
      journal = {\pasj},
         year = 2018,
        month = jan,
       volume = {70},
          eid = {S1},
        pages = {S1},
          doi = {10.1093/pasj/psx063},
       adsurl = {https://ui.adsabs.harvard.edu/abs/2018PASJ...70S...1M}
}

@ARTICLE{2020finner,
       author = {{Finner}, Kyle and {James Jee}, M. and {Webb}, Tracy and {Wilson}, Gillian and {Perlmutter}, Saul and {Muzzin}, Adam and {Hlavacek-Larrondo}, Julie},
        title = "{Constraining the Mass of the Emerging Galaxy Cluster SpARCS1049+56 at z = 1.71 with Infrared Weak Lensing}",
      journal = {\apj},
         year = 2020,
        month = apr,
       volume = {893},
       number = {1},
          eid = {10},
        pages = {10},
          doi = {10.3847/1538-4357/ab7bdb},
archivePrefix = {arXiv},
       eprint = {2002.01956},
 primaryClass = {astro-ph.CO},
       adsurl = {https://ui.adsabs.harvard.edu/abs/2020ApJ...893...10F}
}

@ARTICLE{2025cha,
       author = {{Cha}, Sangjun and {Cho}, Boseong Young and {Joo}, Hyungjin and {Lee}, Wonki and {HyeongHan}, Kim and {Scofield}, Zachary P. and {Finner}, Kyle and {Jee}, M. James},
        title = "{A High-Caliber View of the Bullet Cluster through JWST Strong and Weak Lensing Analyses}",
      journal = {\apjl},
         year = 2025,
        month = jul,
       volume = {987},
       number = {1},
          eid = {L15},
        pages = {L15},
          doi = {10.3847/2041-8213/add2f0},
archivePrefix = {arXiv},
       eprint = {2503.21870},
 primaryClass = {astro-ph.GA},
       adsurl = {https://ui.adsabs.harvard.edu/abs/2025ApJ...987L..15C}
}

@ARTICLE{2023finner_ciza,
       author = {{Finner}, Kyle and {Randall}, Scott W. and {Jee}, M. James and {Blanton}, Elizabeth L. and {Cho}, Hyejeon and {Clarke}, Tracy E. and {Giacintucci}, Simona and {Nulsen}, Paul and {van Weeren}, Reinout},
        title = "{Hubble Space Telescope and Hyper-Suprime-Cam Weak-lensing Study of the Equal-mass Dissociative Merger CIZA J0107.7+5408}",
      journal = {\apj},
         year = 2023,
        month = jan,
       volume = {942},
       number = {1},
          eid = {23},
        pages = {23},
          doi = {10.3847/1538-4357/ac9fd3},
archivePrefix = {arXiv},
       eprint = {2210.12165},
 primaryClass = {astro-ph.GA},
       adsurl = {https://ui.adsabs.harvard.edu/abs/2023ApJ...942...23F}
}

@ARTICLE{2007massey,
       author = {{Massey}, Richard and {Rhodes}, Jason and {Ellis}, Richard and {Scoville}, Nick and {Leauthaud}, Alexie and {Finoguenov}, Alexis and {Capak}, Peter and {Bacon}, David and {Aussel}, Herv{\'e} and {Kneib}, Jean-Paul and {Koekemoer}, Anton and {McCracken}, Henry and {Mobasher}, Bahram and {Pires}, Sandrine and {Refregier}, Alexandre and {Sasaki}, Shunji and {Starck}, Jean-Luc and {Taniguchi}, Yoshi and {Taylor}, Andy and {Taylor}, James},
        title = "{Dark matter maps reveal cosmic scaffolding}",
      journal = {\nat},
         year = 2007,
        month = jan,
       volume = {445},
       number = {7125},
        pages = {286-290},
          doi = {10.1038/nature05497},
archivePrefix = {arXiv},
       eprint = {astro-ph/0701594},
 primaryClass = {astro-ph},
       adsurl = {https://ui.adsabs.harvard.edu/abs/2007Natur.445..286M}
}

@ARTICLE{2026scognamiglio,
       author = {{Scognamiglio}, Diana and {Leroy}, Gavin and {Harvey}, David and {Massey}, Richard and {Rhodes}, Jason and {Akins}, Hollis B. and {Brinch}, Malte and {Berman}, Edward and {Casey}, Caitlin M. and {Drakos}, Nicole E. and {Faisst}, Andreas L. and {Franco}, Maximilien and {Fung}, Leo W.~H. and {Gozaliasl}, Ghassem and {He}, Qiuhan and {Hatamnia}, Hossein and {Huff}, Eric and {Hogg}, Natalie B. and {Ilbert}, Olivier and {Kartaltepe}, Jeyhan S. and {Koekemoer}, Anton M. and {Jin}, Shouwen and {Lambrides}, Erini and {Leauthaud}, Alexie and {Lentz}, Zane D. and {Liu}, Daizhong and {Mahler}, Guillaume and {Maraston}, Claudia and {Martin}, Crystal L. and {McCleary}, Jacqueline and {Nightingale}, James and {Mobasher}, Bahram and {Paquereau}, Louise and {Pires}, Sandrine and {Robertson}, Brant E. and {Sanders}, David B. and {Scarlata}, Claudia and {Shuntov}, Marko and {Toni}, Greta and {von Wietersheim-Kramsta}, Maximilian and {Weaver}, John R.},
        title = "{An ultra-high-resolution map of (dark) matter}",
      journal = {Nature Astronomy},
         year = 2026,
        month = jan,
          doi = {10.1038/s41550-025-02763-9},
archivePrefix = {arXiv},
       eprint = {2601.17239},
 primaryClass = {astro-ph.CO},
       adsurl = {https://ui.adsabs.harvard.edu/abs/2026NatAs.tmp...25S}
}

@ARTICLE{2021finner,
       author = {{Finner}, Kyle and {HyeongHan}, Kim and {Jee}, M. James and {Wittman}, David and {Forman}, William R. and {van Weeren}, Reinout J. and {Golovich}, Nathan R. and {Dawson}, William A. and {Jones}, Alexander and {de Gasperin}, Francesco and {Jones}, Christine},
        title = "{Exemplary Merging Clusters: Weak-lensing and X-Ray Analysis of the Double Radio Relic, Merging Galaxy Clusters MACS J1752.0+4440 and ZWCL 1856.8+6616}",
      journal = {\apj},
         year = 2021,
        month = sep,
       volume = {918},
       number = {2},
          eid = {72},
        pages = {72},
          doi = {10.3847/1538-4357/ac0d00},
archivePrefix = {arXiv},
       eprint = {2010.02226},
 primaryClass = {astro-ph.GA},
       adsurl = {https://ui.adsabs.harvard.edu/abs/2021ApJ...918...72F}
}

@ARTICLE{2020umetsu,
       author = {{Umetsu}, Keiichi and {Sereno}, Mauro and {Lieu}, Maggie and {Miyatake}, Hironao and {Medezinski}, Elinor and {Nishizawa}, Atsushi J. and {Giles}, Paul and {Gastaldello}, Fabio and {McCarthy}, Ian G. and {Kilbinger}, Martin and {Birkinshaw}, Mark and {Ettori}, Stefano and {Okabe}, Nobuhiro and {Chiu}, I. -Non and {Coupon}, Jean and {Eckert}, Dominique and {Fujita}, Yutaka and {Higuchi}, Yuichi and {Koulouridis}, Elias and {Maughan}, Ben and {Miyazaki}, Satoshi and {Oguri}, Masamune and {Pacaud}, Florian and {Pierre}, Marguerite and {Rapetti}, David and {Smith}, Graham P.},
        title = "{Weak-lensing Analysis of X-Ray-selected XXL Galaxy Groups and Clusters with Subaru HSC Data}",
      journal = {\apj},
         year = 2020,
        month = feb,
       volume = {890},
       number = {2},
          eid = {148},
        pages = {148},
          doi = {10.3847/1538-4357/ab6bca},
archivePrefix = {arXiv},
       eprint = {1909.10524},
 primaryClass = {astro-ph.CO},
       adsurl = {https://ui.adsabs.harvard.edu/abs/2020ApJ...890..148U}
}

@ARTICLE{2010finoguenov,
       author = {{Finoguenov}, A. and {Watson}, M.~G. and {Tanaka}, M. and {Simpson}, C. and {Cirasuolo}, M. and {Dunlop}, J.~S. and {Peacock}, J.~A. and {Farrah}, D. and {Akiyama}, M. and {Ueda}, Y. and {Smol{\v{c}}i{\'c}}, V. and {Stewart}, G. and {Rawlings}, S. and {van Breukelen}, C. and {Almaini}, O. and {Clewley}, L. and {Bonfield}, D.~G. and {Jarvis}, M.~J. and {Barr}, J.~M. and {Foucaud}, S. and {McLure}, R.~J. and {Sekiguchi}, K. and {Egami}, E.},
        title = "{X-ray groups and clusters of galaxies in the Subaru-XMM Deep Field}",
      journal = {\mnras},
         year = 2010,
        month = apr,
       volume = {403},
       number = {4},
        pages = {2063-2076},
          doi = {10.1111/j.1365-2966.2010.16256.x},
archivePrefix = {arXiv},
       eprint = {0912.0039},
 primaryClass = {astro-ph.CO},
       adsurl = {https://ui.adsabs.harvard.edu/abs/2010MNRAS.403.2063F}
}

@ARTICLE{2007finoguenov,
       author = {{Finoguenov}, A. and {Guzzo}, L. and {Hasinger}, G. and {Scoville}, N.~Z. and {Aussel}, H. and {B{\"o}hringer}, H. and {Brusa}, M. and {Capak}, P. and {Cappelluti}, N. and {Comastri}, A. and {Giodini}, S. and {Griffiths}, R.~E. and {Impey}, C. and {Koekemoer}, A.~M. and {Kneib}, J. -P. and {Leauthaud}, A. and {Le F{\`e}vre}, O. and {Lilly}, S. and {Mainieri}, V. and {Massey}, R. and {McCracken}, H.~J. and {Mobasher}, B. and {Murayama}, T. and {Peacock}, J.~A. and {Sakelliou}, I. and {Schinnerer}, E. and {Silverman}, J.~D. and {Smol{\v{c}}i{\'c}}, V. and {Taniguchi}, Y. and {Tasca}, L. and {Taylor}, J.~E. and {Trump}, J.~R. and {Zamorani}, G.},
        title = "{The XMM-Newton Wide-Field Survey in the COSMOS Field: Statistical Properties of Clusters of Galaxies}",
      journal = {\apjs},
         year = 2007,
        month = sep,
       volume = {172},
       number = {1},
        pages = {182-195},
          doi = {10.1086/516577},
archivePrefix = {arXiv},
       eprint = {astro-ph/0612360},
 primaryClass = {astro-ph},
       adsurl = {https://ui.adsabs.harvard.edu/abs/2007ApJS..172..182F}
}

@ARTICLE{2013erfanianfar,
       author = {{Erfanianfar}, G. and {Finoguenov}, A. and {Tanaka}, M. and {Lerchster}, M. and {Nandra}, K. and {Laird}, E. and {Connelly}, J.~L. and {Bielby}, R. and {Mirkazemi}, M. and {Faber}, S.~M. and {Kocevski}, D. and {Cooper}, M. and {Newman}, J.~A. and {Jeltema}, T. and {Coil}, A.~L. and {Brimioulle}, F. and {Davis}, M. and {McCracken}, H.~J. and {Willmer}, C. and {Gerke}, B. and {Cappelluti}, N. and {Gwyn}, S.},
        title = "{X-Ray Groups of Galaxies in the AEGIS Deep and Wide Fields}",
      journal = {\apj},
         year = 2013,
        month = mar,
       volume = {765},
       number = {2},
          eid = {117},
        pages = {117},
          doi = {10.1088/0004-637X/765/2/117},
archivePrefix = {arXiv},
       eprint = {1302.5114},
 primaryClass = {astro-ph.CO},
       adsurl = {https://ui.adsabs.harvard.edu/abs/2013ApJ...765..117E}
}

@ARTICLE{2013pentericci,
       author = {{Pentericci}, L. and {Castellano}, M. and {Menci}, N. and {Salimbeni}, S. and {Dahlen}, T. and {Galametz}, A. and {Santini}, P. and {Grazian}, A. and {Fontana}, A.},
        title = "{The evolution of the AGN content in groups up to z \raisebox{-0.5ex}\textasciitilde 1}",
      journal = {\aap},
         year = 2013,
        month = apr,
       volume = {552},
          eid = {A111},
        pages = {A111},
          doi = {10.1051/0004-6361/201219759},
archivePrefix = {arXiv},
       eprint = {1302.2861},
 primaryClass = {astro-ph.CO},
       adsurl = {https://ui.adsabs.harvard.edu/abs/2013A&A...552A.111P}
}

@ARTICLE{2015finoguenov,
       author = {{Finoguenov}, A. and {Tanaka}, M. and {Cooper}, M. and {Allevato}, V. and {Cappelluti}, N. and {Choi}, A. and {Heymans}, C. and {Bauer}, F.~E. and {Ziparo}, F. and {Ranalli}, P. and {Silverman}, J. and {Brandt}, W.~N. and {Xue}, Y.~Q. and {Mulchaey}, J. and {Howes}, L. and {Schmid}, C. and {Wilman}, D. and {Comastri}, A. and {Hasinger}, G. and {Mainieri}, V. and {Luo}, B. and {Tozzi}, P. and {Rosati}, P. and {Capak}, P. and {Popesso}, P.},
        title = "{Ultra-deep catalog of X-ray groups in the Extended Chandra Deep Field South}",
      journal = {\aap},
         year = 2015,
        month = apr,
       volume = {576},
          eid = {A130},
        pages = {A130},
          doi = {10.1051/0004-6361/201323053},
archivePrefix = {arXiv},
       eprint = {1501.03506},
 primaryClass = {astro-ph.HE},
       adsurl = {https://ui.adsabs.harvard.edu/abs/2015A&A...576A.130F}
}

@ARTICLE{2017finner,
       author = {{Finner}, Kyle and {Jee}, M. James and {Golovich}, Nathan and
         {Wittman}, David and {Dawson}, William and {Gruen}, Daniel and
         {Koekemoer}, Anton M. and {Lemaux}, Brian C. and {Seitz}, Stella},
        title = "{MC$^{2}$: Subaru and Hubble Space Telescope Weak-lensing Analysis of the Double Radio Relic Galaxy Cluster PLCK G287.0+32.9}",
      journal = {\apj},
         year = "2017",
        month = "Dec",
       volume = {851},
       number = {1},
          eid = {46},
        pages = {46},
          doi = {10.3847/1538-4357/aa998c},
archivePrefix = {arXiv},
       eprint = {1710.02527},
 primaryClass = {astro-ph.CO},
       adsurl = {https://ui.adsabs.harvard.edu/abs/2017ApJ...851...46F}
}

@ARTICLE{2001ebeling,
       author = {{Ebeling}, H. and {Edge}, A.~C. and {Henry}, J.~P.},
        title = "{MACS: A Quest for the Most Massive Galaxy Clusters in the Universe}",
      journal = {\apj},
         year = 2001,
        month = jun,
       volume = {553},
       number = {2},
        pages = {668-676},
          doi = {10.1086/320958},
archivePrefix = {arXiv},
       eprint = {astro-ph/0009101},
 primaryClass = {astro-ph},
       adsurl = {https://ui.adsabs.harvard.edu/abs/2001ApJ...553..668E}
}

@ARTICLE{2017jee,
       author = {{Jee}, M. James and {Ko}, Jongwan and {Perlmutter}, Saul and
         {Gonzalez}, Anthony and {Brodwin}, Mark and {Linder}, Eric and
         {Eisenhardt}, Peter},
        title = "{First Weak-lensing Results from {\textquotedblleft}See Change{\textquotedblright}: Quantifying Dark Matter in the Two z {\ensuremath{\gtrsim}} 1.5 High-redshift Galaxy Clusters SPT-CL J2040-4451 and IDCS J1426+3508}",
      journal = {\apj},
         year = "2017",
        month = "Oct",
       volume = {847},
       number = {2},
          eid = {117},
        pages = {117},
          doi = {10.3847/1538-4357/aa88bc},
archivePrefix = {arXiv},
       eprint = {1709.04561},
 primaryClass = {astro-ph.CO},
       adsurl = {https://ui.adsabs.harvard.edu/abs/2017ApJ...847..117J}
}

@ARTICLE{2013jee,
       author = {{Jee}, M. James and {Tyson}, J. Anthony and {Schneider}, Michael D. and
         {Wittman}, David and {Schmidt}, Samuel and {Hilbert}, Stefan},
        title = "{Cosmic Shear Results from the Deep Lens Survey. I. Joint Constraints on {\ensuremath{\Omega}}$_{ M }$ and {\ensuremath{\sigma}}$_{8}$ with a Two-dimensional Analysis}",
      journal = {\apj},
         year = "2013",
        month = "Mar",
       volume = {765},
       number = {1},
          eid = {74},
        pages = {74},
          doi = {10.1088/0004-637X/765/1/74},
archivePrefix = {arXiv},
       eprint = {1210.2732},
 primaryClass = {astro-ph.CO},
       adsurl = {https://ui.adsabs.harvard.edu/abs/2013ApJ...765...74J}
}

@ARTICLE{1996bertin,
       author = {{Bertin}, E. and {Arnouts}, S.},
        title = "{SExtractor: Software for source extraction.}",
      journal = {\aaps},
         year = "1996",
        month = "Jun",
       volume = {117},
        pages = {393-404},
          doi = {10.1051/aas:1996164},
       adsurl = {https://ui.adsabs.harvard.edu/abs/1996A&AS..117..393B}
}

@ARTICLE{2019diemer,
       author = {{Diemer}, Benedikt and {Joyce}, Michael},
        title = "{An Accurate Physical Model for Halo Concentrations}",
      journal = {\apj},
         year = "2019",
        month = "Feb",
       volume = {871},
       number = {2},
          eid = {168},
        pages = {168},
          doi = {10.3847/1538-4357/aafad6},
       adsurl = {https://ui.adsabs.harvard.edu/abs/2019ApJ...871..168D}
}

@INPROCEEDINGS{2002wittman,
       author = {{Wittman}, David M. and {Tyson}, J. Anthony and {Dell'Antonio}, Ian P. and {Becker}, Andrew and {Margoniner}, Vera and {Cohen}, Judith G. and {Norman}, D. and {Loomba}, D. and {Squires}, G. and {Wilson}, Gillian and {Stubbs}, Christopher W. and {Hennawi}, J. and {Spergel}, David N. and {Boeshaar}, P. and {Clocchiatti}, A. and {Hamuy}, M. and {Bernstein}, G. and {Gonzalez}, A. and {Guhathakurta}, Puragra and {Hu}, W. and {Seljak}, U. and {Zaritsky}, Dennis},
        title = "{Deep lens survey}",
    booktitle = {Survey and Other Telescope Technologies and Discoveries},
         year = 2002,
       editor = {{Tyson}, J. Anthony and {Wolff}, Sidney},
       series = {Society of Photo-Optical Instrumentation Engineers (SPIE) Conference Series},
       volume = {4836},
        month = dec,
        pages = {73-82},
          doi = {10.1117/12.457348},
archivePrefix = {arXiv},
       eprint = {astro-ph/0210118},
 primaryClass = {astro-ph},
       adsurl = {https://ui.adsabs.harvard.edu/abs/2002SPIE.4836...73W}
}

@BOOK{2012drizzle,
       author = {{Gonzaga}, S.{et al.}},
        title = "{The DrizzlePac Handbook}",
         year = "2012",
       adsurl = {https://ui.adsabs.harvard.edu/abs/2012drzp.book.....G}
}

@ARTICLE{2018mandelbaum,
       author = {{Mandelbaum}, Rachel},
        title = "{Weak Lensing for Precision Cosmology}",
      journal = {\araa},
         year = "2018",
        month = "Sep",
       volume = {56},
        pages = {393-433},
          doi = {10.1146/annurev-astro-081817-051928},
archivePrefix = {arXiv},
       eprint = {1710.03235},
 primaryClass = {astro-ph.CO},
       adsurl = {https://ui.adsabs.harvard.edu/abs/2018ARA&A..56..393M}
}

@ARTICLE{1993kaiser,
       author = {{Kaiser}, Nick and {Squires}, Gordon},
        title = "{Mapping the Dark Matter with Weak Gravitational Lensing}",
      journal = {\apj},
         year = 1993,
        month = feb,
       volume = {404},
        pages = {441},
          doi = {10.1086/172297},
       adsurl = {https://ui.adsabs.harvard.edu/abs/1993ApJ...404..441K}
}

@INBOOK{2009markwardt,
       author = {{Markwardt}, C.~B.},
        title = "{Non-linear Least-squares Fitting in IDL with MPFIT}",
    booktitle = {Astronomical Data Analysis Software and Systems XVIII},
         year = "2009",
       editor = {{Bohlender}, D.~A. and {Durand}, D. and {Dowler}, P.},
       volume = {411},
       series = {Astronomical Society of the Pacific Conference Series},
        pages = {251},
       adsurl = {https://ui.adsabs.harvard.edu/abs/2009ASPC..411..251M}
}

@ARTICLE{2000wright,
       author = {{Wright}, Candace Oaxaca and {Brainerd}, Tereasa G.},
        title = "{Gravitational Lensing by NFW Halos}",
      journal = {\apj},
         year = 2000,
        month = may,
       volume = {534},
       number = {1},
        pages = {34-40},
          doi = {10.1086/308744},
       adsurl = {https://ui.adsabs.harvard.edu/abs/2000ApJ...534...34W}
}

@ARTICLE{2004markevitch,
       author = {{Markevitch}, M. and {Gonzalez}, A.~H. and {Clowe}, D. and
         {Vikhlinin}, A. and {Forman}, W. and {Jones}, C. and {Murray}, S. and
         {Tucker}, W.},
        title = "{Direct Constraints on the Dark Matter Self-Interaction Cross Section from the Merging Galaxy Cluster 1E 0657-56}",
      journal = {\apj},
         year = 2004,
        month = may,
       volume = {606},
       number = {2},
        pages = {819-824},
          doi = {10.1086/383178},
archivePrefix = {arXiv},
       eprint = {astro-ph/0309303},
 primaryClass = {astro-ph},
       adsurl = {https://ui.adsabs.harvard.edu/abs/2004ApJ...606..819M}
}

@ARTICLE{2006wittman,
       author = {{Wittman}, D. and {Dell'Antonio}, I.~P. and {Hughes}, J.~P. and {Margoniner}, V.~E. and {Tyson}, J.~A. and {Cohen}, J.~G. and {Norman}, D.},
        title = "{First Results on Shear-selected Clusters from the Deep Lens Survey: Optical Imaging, Spectroscopy, and X-Ray Follow-up}",
      journal = {\apj},
         year = 2006,
        month = may,
       volume = {643},
       number = {1},
        pages = {128-143},
          doi = {10.1086/502621},
archivePrefix = {arXiv},
       eprint = {astro-ph/0507606},
 primaryClass = {astro-ph},
       adsurl = {https://ui.adsabs.harvard.edu/abs/2006ApJ...643..128W}
}

@ARTICLE{2018lee,
       author = {{Lee}, Bomee and {Chary}, Ranga-Ram and {Wright}, Edward L.},
        title = "{Galaxy Ellipticity Measurements in the Near-infrared for Weak Lensing}",
      journal = {\apj},
         year = 2018,
        month = oct,
       volume = {866},
       number = {2},
          eid = {157},
        pages = {157},
          doi = {10.3847/1538-4357/aadfd7},
archivePrefix = {arXiv},
       eprint = {1808.05223},
 primaryClass = {astro-ph.GA},
       adsurl = {https://ui.adsabs.harvard.edu/abs/2018ApJ...866..157L}
}

@ARTICLE{2014umetsu,
       author = {{Umetsu}, Keiichi and {Medezinski}, Elinor and {Nonino}, Mario and {Merten}, Julian and {Postman}, Marc and {Meneghetti}, Massimo and {Donahue}, Megan and {Czakon}, Nicole and {Molino}, Alberto and {Seitz}, Stella and {Gruen}, Daniel and {Lemze}, Doron and {Balestra}, Italo and {Ben{\'\i}tez}, Narciso and {Biviano}, Andrea and {Broadhurst}, Tom and {Ford}, Holland and {Grillo}, Claudio and {Koekemoer}, Anton and {Melchior}, Peter and {Mercurio}, Amata and {Moustakas}, John and {Rosati}, Piero and {Zitrin}, Adi},
        title = "{CLASH: Weak-lensing Shear-and-magnification Analysis of 20 Galaxy Clusters}",
      journal = {\apj},
         year = 2014,
        month = nov,
       volume = {795},
       number = {2},
          eid = {163},
        pages = {163},
          doi = {10.1088/0004-637X/795/2/163},
archivePrefix = {arXiv},
       eprint = {1404.1375},
 primaryClass = {astro-ph.CO},
       adsurl = {https://ui.adsabs.harvard.edu/abs/2014ApJ...795..163U}
}

@ARTICLE{2002bauer,
       author = {{Bauer}, F.~E. and {Alexander}, D.~M. and {Brandt}, W.~N. and {Hornschemeier}, A.~E. and {Miyaji}, T. and {Garmire}, G.~P. and {Schneider}, D.~P. and {Bautz}, M.~W. and {Chartas}, G. and {Griffiths}, R.~E. and {Sargent}, W.~L.~W.},
        title = "{The Chandra Deep Field North Survey. IX. Extended X-Ray Sources}",
      journal = {\aj},
         year = 2002,
        month = mar,
       volume = {123},
       number = {3},
        pages = {1163-1178},
          doi = {10.1086/338903},
archivePrefix = {arXiv},
       eprint = {astro-ph/0112002},
 primaryClass = {astro-ph},
       adsurl = {https://ui.adsabs.harvard.edu/abs/2002AJ....123.1163B}
}

@ARTICLE{2001dawson,
       author = {{Dawson}, Steve and {Stern}, Daniel and {Bunker}, Andrew J. and {Spinrad}, Hyron and {Dey}, Arjun},
        title = "{Serendipitously Detected Galaxies in the Hubble Deep Field}",
      journal = {\aj},
         year = 2001,
        month = aug,
       volume = {122},
       number = {2},
        pages = {598-610},
          doi = {10.1086/321160},
archivePrefix = {arXiv},
       eprint = {astro-ph/0105043},
 primaryClass = {astro-ph},
       adsurl = {https://ui.adsabs.harvard.edu/abs/2001AJ....122..598D}
}

@ARTICLE{1999barger,
       author = {{Barger}, A.~J. and {Cowie}, L.~L. and {Trentham}, N. and {Fulton}, E. and {Hu}, E.~M. and {Songaila}, A. and {Hall}, D.},
        title = "{Constraints on the Early Formation of Field Elliptical Galaxies}",
      journal = {\aj},
         year = 1999,
        month = jan,
       volume = {117},
       number = {1},
        pages = {102-110},
          doi = {10.1086/300675},
archivePrefix = {arXiv},
       eprint = {astro-ph/9809299},
 primaryClass = {astro-ph},
       adsurl = {https://ui.adsabs.harvard.edu/abs/1999AJ....117..102B}
}

@ARTICLE{2023wittman,
       author = {{Wittman}, David and {Stancioli}, Rodrigo and {Finner}, Kyle and {Bouhrik}, Faik and {van Weeren}, Reinout and {Botteon}, Andrea},
        title = "{A New Galaxy Cluster Merger Capable of Probing Dark Matter: A56}",
      journal = {\apj},
         year = 2023,
        month = sep,
       volume = {954},
       number = {1},
          eid = {36},
        pages = {36},
          doi = {10.3847/1538-4357/acdb73},
archivePrefix = {arXiv},
       eprint = {2306.01715},
 primaryClass = {astro-ph.CO},
       adsurl = {https://ui.adsabs.harvard.edu/abs/2023ApJ...954...36W}
}

@ARTICLE{2011george,
       author = {{George}, Matthew R. and {Leauthaud}, Alexie and {Bundy}, Kevin and {Finoguenov}, Alexis and {Tinker}, Jeremy and {Lin}, Yen-Ting and {Mei}, Simona and {Kneib}, Jean-Paul and {Aussel}, Herv{\'e} and {Behroozi}, Peter S. and {Busha}, Michael T. and {Capak}, Peter and {Coccato}, Lodovico and {Covone}, Giovanni and {Faure}, Cecile and {Fiorenza}, Stephanie L. and {Ilbert}, Olivier and {Le Floc'h}, Emeric and {Koekemoer}, Anton M. and {Tanaka}, Masayuki and {Wechsler}, Risa H. and {Wolk}, Melody},
        title = "{Galaxies in X-Ray Groups. I. Robust Membership Assignment and the Impact of Group Environments on Quenching}",
      journal = {\apj},
         year = 2011,
        month = dec,
       volume = {742},
       number = {2},
          eid = {125},
        pages = {125},
          doi = {10.1088/0004-637X/742/2/125},
archivePrefix = {arXiv},
       eprint = {1109.6040},
 primaryClass = {astro-ph.CO},
       adsurl = {https://ui.adsabs.harvard.edu/abs/2011ApJ...742..125G}
}

@ARTICLE{2019gozaliasl,
       author = {{Gozaliasl}, Ghassem and {Finoguenov}, Alexis and {Tanaka}, Masayuki and {Dolag}, Klaus and {Montanari}, Francesco and {Kirkpatrick}, Charles C. and {Vardoulaki}, Eleni and {Khosroshahi}, Habib G. and {Salvato}, Mara and {Laigle}, Clotilde and {McCracken}, Henry J. and {Ilbert}, Olivier and {Cappelluti}, Nico and {Daddi}, Emanuele and {Hasinger}, Guenther and {Capak}, Peter and {Scoville}, Nick Z. and {Toft}, Sune and {Civano}, Francesca and {Griffiths}, Richard E. and {Balogh}, Michael and {Li}, Yanxia and {Ahoranta}, Jussi and {Mei}, Simona and {Iovino}, Angela and {Henriques}, Bruno M.~B. and {Erfanianfar}, Ghazaleh},
        title = "{Chandra centres for COSMOS X-ray galaxy groups: differences in stellar properties between central dominant and offset brightest group galaxies}",
      journal = {\mnras},
         year = 2019,
        month = mar,
       volume = {483},
       number = {3},
        pages = {3545-3565},
          doi = {10.1093/mnras/sty3203},
archivePrefix = {arXiv},
       eprint = {1812.01604},
 primaryClass = {astro-ph.GA},
       adsurl = {https://ui.adsabs.harvard.edu/abs/2019MNRAS.483.3545G}
}

@ARTICLE{2024stancioli,
       author = {{Stancioli}, Rodrigo and {Wittman}, David and {Finner}, Kyle and {Bouhrik}, Faik},
        title = "{A New Dissociative Galaxy Cluster Merger: RM J150822.0+575515.2}",
      journal = {\apj},
         year = 2024,
        month = may,
       volume = {966},
       number = {1},
          eid = {49},
        pages = {49},
          doi = {10.3847/1538-4357/ad3249},
archivePrefix = {arXiv},
       eprint = {2307.10174},
 primaryClass = {astro-ph.CO},
       adsurl = {https://ui.adsabs.harvard.edu/abs/2024ApJ...966...49S}
}

@ARTICLE{2014guennou,
       author = {{Guennou}, L. and {Adami}, C. and {Durret}, F. and {Lima Neto}, G.~B. and {Ulmer}, M.~P. and {Clowe}, D. and {LeBrun}, V. and {Martinet}, N. and {Allam}, S. and {Annis}, J. and {Basa}, S. and {Benoist}, C. and {Biviano}, A. and {Cappi}, A. and {Cypriano}, E.~S. and {Gavazzi}, R. and {Halliday}, C. and {Ilbert}, O. and {Jullo}, E. and {Just}, D. and {Limousin}, M. and {M{\'a}rquez}, I. and {Mazure}, A. and {Murphy}, K.~J. and {Plana}, H. and {Rostagni}, F. and {Russeil}, D. and {Schirmer}, M. and {Slezak}, E. and {Tucker}, D. and {Zaritsky}, D. and {Ziegler}, B.},
        title = "{Structure and substructure analysis of DAFT/FADA galaxy clusters in the [0.4-0.9] redshift range}",
      journal = {\aap},
         year = 2014,
        month = jan,
       volume = {561},
          eid = {A112},
        pages = {A112},
          doi = {10.1051/0004-6361/201321208},
archivePrefix = {arXiv},
       eprint = {1311.6922},
 primaryClass = {astro-ph.CO},
       adsurl = {https://ui.adsabs.harvard.edu/abs/2014A&A...561A.112G}
}

@ARTICLE{2023cfinner,
       author = {{Finner}, Kyle and {Lee}, Bomee and {Chary}, Ranga-Ram and {Jee}, M. James and {Hirata}, Christopher and {Congedo}, Giuseppe and {Taylor}, Peter and {HyeongHan}, Kim},
        title = "{Near-IR Weak-lensing (NIRWL) Measurements in the CANDELS Fields. I. Point-spread Function Modeling and Systematics}",
      journal = {\apj},
         year = 2023,
        month = nov,
       volume = {958},
       number = {1},
          eid = {33},
        pages = {33},
          doi = {10.3847/1538-4357/acfafd},
archivePrefix = {arXiv},
       eprint = {2301.07725},
 primaryClass = {astro-ph.GA},
       adsurl = {https://ui.adsabs.harvard.edu/abs/2023ApJ...958...33F}
}

@ARTICLE{2022sommer,
       author = {{Sommer}, Martin W. and {Schrabback}, Tim and {Applegate}, Douglas E. and {Hilbert}, Stefan and {Ansarinejad}, Behzad and {Floyd}, Benjamin and {Grandis}, Sebastian},
        title = "{Weak lensing mass modeling bias and the impact of miscentring}",
      journal = {\mnras},
         year = 2022,
        month = jan,
       volume = {509},
       number = {1},
        pages = {1127-1146},
          doi = {10.1093/mnras/stab3052},
archivePrefix = {arXiv},
       eprint = {2105.08027},
 primaryClass = {astro-ph.CO},
       adsurl = {https://ui.adsabs.harvard.edu/abs/2022MNRAS.509.1127S}
}

@ARTICLE{2018lee_mass_bias,
       author = {{Lee}, B.~E. and {Le Brun}, A.~M.~C. and {Haq}, M.~E. and {Deering}, N.~J. and {King}, L.~J. and {Applegate}, D. and {McCarthy}, I.~G.},
        title = "{The relative impact of baryons and cluster shape on weak lensing mass estimates of galaxy clusters}",
      journal = {\mnras},
         year = 2018,
        month = sep,
       volume = {479},
       number = {1},
        pages = {890-899},
          doi = {10.1093/mnras/sty1377},
archivePrefix = {arXiv},
       eprint = {1805.12186},
 primaryClass = {astro-ph.CO},
       adsurl = {https://ui.adsabs.harvard.edu/abs/2018MNRAS.479..890L}
}

@ARTICLE{2012george,
       author = {{George}, Matthew R. and {Leauthaud}, Alexie and {Bundy}, Kevin and {Finoguenov}, Alexis and {Ma}, Chung-Pei and {Rykoff}, Eli S. and {Tinker}, Jeremy L. and {Wechsler}, Risa H. and {Massey}, Richard and {Mei}, Simona},
        title = "{Galaxies in X-Ray Groups. II. A Weak Lensing Study of Halo Centering}",
      journal = {\apj},
         year = 2012,
        month = sep,
       volume = {757},
       number = {1},
          eid = {2},
        pages = {2},
          doi = {10.1088/0004-637X/757/1/2},
archivePrefix = {arXiv},
       eprint = {1205.4262},
 primaryClass = {astro-ph.CO},
       adsurl = {https://ui.adsabs.harvard.edu/abs/2012ApJ...757....2G}
}

@ARTICLE{2021oguri,
       author = {{Oguri}, Masamune and {Miyazaki}, Satoshi and {Li}, Xiangchong and {Luo}, Wentao and {Mitsuishi}, Ikuyuki and {Miyatake}, Hironao and {More}, Surhud and {Nishizawa}, Atsushi J. and {Okabe}, Nobuhiro and {Ota}, Naomi and {Plazas Malag{\'o}n}, Andr{\'e}s A. and {Utsumi}, Yousuke},
        title = "{Hundreds of weak lensing shear-selected clusters from the Hyper Suprime-Cam Subaru Strategic Program S19A data}",
      journal = {\pasj},
         year = 2021,
        month = aug,
       volume = {73},
       number = {4},
        pages = {817-829},
          doi = {10.1093/pasj/psab047},
archivePrefix = {arXiv},
       eprint = {2103.15016},
 primaryClass = {astro-ph.CO},
       adsurl = {https://ui.adsabs.harvard.edu/abs/2021PASJ...73..817O}
}

@ARTICLE{2007schirmer,
       author = {{Schirmer}, M. and {Erben}, T. and {Hetterscheidt}, M. and {Schneider}, P.},
        title = "{GaBoDS: the Garching-Bonn Deep Survey. IX. A sample of 158 shear-selected mass concentration candidates}",
      journal = {\aap},
         year = 2007,
        month = feb,
       volume = {462},
       number = {3},
        pages = {875-887},
          doi = {10.1051/0004-6361:20065955},
archivePrefix = {arXiv},
       eprint = {astro-ph/0607022},
 primaryClass = {astro-ph},
       adsurl = {https://ui.adsabs.harvard.edu/abs/2007A&A...462..875S}
}

@ARTICLE{2002miyazaki,
       author = {{Miyazaki}, Satoshi and {Hamana}, Takashi and {Shimasaku}, Kazuhiro and {Furusawa}, Hisanori and {Doi}, Mamoru and {Hamabe}, Masaru and {Imi}, Katsumi and {Kimura}, Masahiko and {Komiyama}, Yutaka and {Nakata}, Fumiaki and {Okada}, Norio and {Okamura}, Sadanori and {Ouchi}, Masami and {Sekiguchi}, Maki and {Yagi}, Masafumi and {Yasuda}, Naoki},
        title = "{Searching for Dark Matter Halos in the Suprime-Cam 2 Square Degree Field}",
      journal = {\apjl},
         year = 2002,
        month = dec,
       volume = {580},
       number = {2},
        pages = {L97-L100},
          doi = {10.1086/345613},
archivePrefix = {arXiv},
       eprint = {astro-ph/0210441},
 primaryClass = {astro-ph},
       adsurl = {https://ui.adsabs.harvard.edu/abs/2002ApJ...580L..97M}
}

@ARTICLE{2007miyazaki,
       author = {{Miyazaki}, Satoshi and {Hamana}, Takashi and {Ellis}, Richard S. and {Kashikawa}, Nobunari and {Massey}, Richard J. and {Taylor}, James and {Refregier}, Alexandre},
        title = "{A Subaru Weak-Lensing Survey. I. Cluster Candidates and Spectroscopic Verification}",
      journal = {\apj},
         year = 2007,
        month = nov,
       volume = {669},
       number = {2},
        pages = {714-728},
          doi = {10.1086/521621},
archivePrefix = {arXiv},
       eprint = {0707.2249},
 primaryClass = {astro-ph},
       adsurl = {https://ui.adsabs.harvard.edu/abs/2007ApJ...669..714M}
}

@ARTICLE{2007gavazzi,
       author = {{Gavazzi}, R. and {Soucail}, G.},
        title = "{Weak lensing survey of galaxy clusters in the CFHTLS Deep}",
      journal = {\aap},
         year = 2007,
        month = feb,
       volume = {462},
       number = {2},
        pages = {459-471},
          doi = {10.1051/0004-6361:20065677},
archivePrefix = {arXiv},
       eprint = {astro-ph/0605591},
 primaryClass = {astro-ph},
       adsurl = {https://ui.adsabs.harvard.edu/abs/2007A&A...462..459G}
}

@ARTICLE{2014utsumi,
       author = {{Utsumi}, Yousuke and {Miyazaki}, Satoshi and {Geller}, Margaret J. and {Dell'Antonio}, Ian P. and {Oguri}, Masamune and {Kurtz}, Michael J. and {Hamana}, Takashi and {Fabricant}, Daniel G.},
        title = "{Reducing Systematic Error in Weak Lensing Cluster Surveys}",
      journal = {\apj},
         year = 2014,
        month = may,
       volume = {786},
       number = {2},
          eid = {93},
        pages = {93},
          doi = {10.1088/0004-637X/786/2/93},
archivePrefix = {arXiv},
       eprint = {1304.4656},
 primaryClass = {astro-ph.IM},
       adsurl = {https://ui.adsabs.harvard.edu/abs/2014ApJ...786...93U}
}

@ARTICLE{2025kim,
       author = {{Kim}, Jinhyub and {Jee}, M. James and {Andreon}, Stefano and {Mroczkowski}, Tony and {Miller}, Lance and {van Marrewijk}, Joshiwa and {Khim}, Hye Gyeong},
        title = "{Exploring the Masses of the Two Most Distant Gravitational Lensing Clusters at Cosmic Noon}",
      journal = {\apj},
         year = 2025,
        month = sep,
       volume = {991},
       number = {1},
          eid = {109},
        pages = {109},
          doi = {10.3847/1538-4357/adf29e},
archivePrefix = {arXiv},
       eprint = {2507.16925},
 primaryClass = {astro-ph.CO},
       adsurl = {https://ui.adsabs.harvard.edu/abs/2025ApJ...991..109K}
}

@ARTICLE{2026scofield,
       author = {{Scofield}, Zachary P. and {Finner}, Kyle and {Joo}, Hyungjin and {Jee}, M. James and {Lee}, Wonki and {Cha}, Sangjun and {Kim}, Jinhyub and {Lin}, Yu-heng and {Chary}, Ranga-Ram and {Faisst}, Andreas and {Lee}, Bomee},
        title = "{An Active Galaxy Cluster Merger at Cosmic Noon Revealed by JWST Weak Lensing and Multiwavelength Probes}",
      journal = {\apjl},
         year = 2026,
        month = mar,
       volume = {999},
       number = {1},
          eid = {L1},
        pages = {L1},
          doi = {10.3847/2041-8213/ae447a},
archivePrefix = {arXiv},
       eprint = {2512.11022},
 primaryClass = {astro-ph.GA},
       adsurl = {https://ui.adsabs.harvard.edu/abs/2026ApJ...999L...1S}
}

@ARTICLE{2018galametz,
       author = {{Galametz}, Audrey and {Pentericci}, Laura and {Castellano}, Marco and {Mendel}, Trevor and {Hartley}, Will G. and {Fossati}, Matteo and {Finoguenov}, Alexis and {Almaini}, Omar and {Beifiori}, Alessandra and {Fontana}, Adriano and {Grazian}, Andrea and {Scodeggio}, Marco and {Kocevski}, Dale D.},
        title = "{Growing up in a megalopolis: environmental effects on galaxy evolution in a supercluster at z {\ensuremath{\sim}} 0.65 in UKIDSS UDS}",
      journal = {\mnras},
         year = 2018,
        month = apr,
       volume = {475},
       number = {3},
        pages = {4148-4169},
          doi = {10.1093/mnras/sty095},
archivePrefix = {arXiv},
       eprint = {1801.03939},
 primaryClass = {astro-ph.CO},
       adsurl = {https://ui.adsabs.harvard.edu/abs/2018MNRAS.475.4148G}
}

@ARTICLE{2022astropy,
       author = {{Astropy Collaboration} and {Price-Whelan}, Adrian M. and {Lim}, Pey Lian and {Earl}, Nicholas and {Starkman}, Nathaniel and {Bradley}, Larry and {Shupe}, David L. and {Patil}, Aarya A. and {Corrales}, Lia and {Brasseur}, C.~E. and {N{\"o}the}, Maximilian and {Donath}, Axel and {Tollerud}, Erik and {Morris}, Brett M. and {Ginsburg}, Adam and {Vaher}, Eero and {Weaver}, Benjamin A. and {Tocknell}, James and {Jamieson}, William and {van Kerkwijk}, Marten H. and {Robitaille}, Thomas P. and {Merry}, Bruce and {Bachetti}, Matteo and {G{\"u}nther}, H. Moritz and {Aldcroft}, Thomas L. and {Alvarado-Montes}, Jaime A. and {Archibald}, Anne M. and {B{\'o}di}, Attila and {Bapat}, Shreyas and {Barentsen}, Geert and {Baz{\'a}n}, Juanjo and {Biswas}, Manish and {Boquien}, M{\'e}d{\'e}ric and {Burke}, D.~J. and {Cara}, Daria and {Cara}, Mihai and {Conroy}, Kyle E. and {Conseil}, Simon and {Craig}, Matthew W. and {Cross}, Robert M. and {Cruz}, Kelle L. and {D'Eugenio}, Francesco and {Dencheva}, Nadia and {Devillepoix}, Hadrien A.~R. and {Dietrich}, J{\"o}rg P. and {Eigenbrot}, Arthur Davis and {Erben}, Thomas and {Ferreira}, Leonardo and {Foreman-Mackey}, Daniel and {Fox}, Ryan and {Freij}, Nabil and {Garg}, Suyog and {Geda}, Robel and {Glattly}, Lauren and {Gondhalekar}, Yash and {Gordon}, Karl D. and {Grant}, David and {Greenfield}, Perry and {Groener}, Austen M. and {Guest}, Steve and {Gurovich}, Sebastian and {Handberg}, Rasmus and {Hart}, Akeem and {Hatfield-Dodds}, Zac and {Homeier}, Derek and {Hosseinzadeh}, Griffin and {Jenness}, Tim and {Jones}, Craig K. and {Joseph}, Prajwel and {Kalmbach}, J. Bryce and {Karamehmetoglu}, Emir and {Ka{\l}uszy{\'n}ski}, Miko{\l}aj and {Kelley}, Michael S.~P. and {Kern}, Nicholas and {Kerzendorf}, Wolfgang E. and {Koch}, Eric W. and {Kulumani}, Shankar and {Lee}, Antony and {Ly}, Chun and {Ma}, Zhiyuan and {MacBride}, Conor and {Maljaars}, Jakob M. and {Muna}, Demitri and {Murphy}, N.~A. and {Norman}, Henrik and {O'Steen}, Richard and {Oman}, Kyle A. and {Pacifici}, Camilla and {Pascual}, Sergio and {Pascual-Granado}, J. and {Patil}, Rohit R. and {Perren}, Gabriel I. and {Pickering}, Timothy E. and {Rastogi}, Tanuj and {Roulston}, Benjamin R. and {Ryan}, Daniel F. and {Rykoff}, Eli S. and {Sabater}, Jose and {Sakurikar}, Parikshit and {Salgado}, Jes{\'u}s and {Sanghi}, Aniket and {Saunders}, Nicholas and {Savchenko}, Volodymyr and {Schwardt}, Ludwig and {Seifert-Eckert}, Michael and {Shih}, Albert Y. and {Jain}, Anany Shrey and {Shukla}, Gyanendra and {Sick}, Jonathan and {Simpson}, Chris and {Singanamalla}, Sudheesh and {Singer}, Leo P. and {Singhal}, Jaladh and {Sinha}, Manodeep and {Sip{\H{o}}cz}, Brigitta M. and {Spitler}, Lee R. and {Stansby}, David and {Streicher}, Ole and {{\v{S}}umak}, Jani and {Swinbank}, John D. and {Taranu}, Dan S. and {Tewary}, Nikita and {Tremblay}, Grant R. and {Val-Borro}, Miguel de and {Van Kooten}, Samuel J. and {Vasovi{\'c}}, Zlatan and {Verma}, Shresth and {de Miranda Cardoso}, Jos{\'e} Vin{\'\i}cius and {Williams}, Peter K.~G. and {Wilson}, Tom J. and {Winkel}, Benjamin and {Wood-Vasey}, W.~M. and {Xue}, Rui and {Yoachim}, Peter and {Zhang}, Chen and {Zonca}, Andrea and {Astropy Project Contributors}},
        title = "{The Astropy Project: Sustaining and Growing a Community-oriented Open-source Project and the Latest Major Release (v5.0) of the Core Package}",
      journal = {\apj},
         year = 2022,
        month = aug,
       volume = {935},
       number = {2},
          eid = {167},
        pages = {167},
          doi = {10.3847/1538-4357/ac7c74},
archivePrefix = {arXiv},
       eprint = {2206.14220},
 primaryClass = {astro-ph.IM},
       adsurl = {https://ui.adsabs.harvard.edu/abs/2022ApJ...935..167A}
}

@ARTICLE{2011koekemoer,
       author = {{Koekemoer}, Anton M. and {Faber}, S.~M. and {Ferguson}, Henry C. and {Grogin}, Norman A. and {Kocevski}, Dale D. and {Koo}, David C. and {Lai}, Kamson and {Lotz}, Jennifer M. and {Lucas}, Ray A. and {McGrath}, Elizabeth J. and {Ogaz}, Sara and {Rajan}, Abhijith and {Riess}, Adam G. and {Rodney}, Steve A. and {Strolger}, Louis and {Casertano}, Stefano and {Castellano}, Marco and {Dahlen}, Tomas and {Dickinson}, Mark and {Dolch}, Timothy and {Fontana}, Adriano and {Giavalisco}, Mauro and {Grazian}, Andrea and {Guo}, Yicheng and {Hathi}, Nimish P. and {Huang}, Kuang-Han and {van der Wel}, Arjen and {Yan}, Hao-Jing and {Acquaviva}, Viviana and {Alexander}, David M. and {Almaini}, Omar and {Ashby}, Matthew L.~N. and {Barden}, Marco and {Bell}, Eric F. and {Bournaud}, Fr{\'e}d{\'e}ric and {Brown}, Thomas M. and {Caputi}, Karina I. and {Cassata}, Paolo and {Challis}, Peter J. and {Chary}, Ranga-Ram and {Cheung}, Edmond and {Cirasuolo}, Michele and {Conselice}, Christopher J. and {Roshan Cooray}, Asantha and {Croton}, Darren J. and {Daddi}, Emanuele and {Dav{\'e}}, Romeel and {de Mello}, Duilia F. and {de Ravel}, Loic and {Dekel}, Avishai and {Donley}, Jennifer L. and {Dunlop}, James S. and {Dutton}, Aaron A. and {Elbaz}, David and {Fazio}, Giovanni G. and {Filippenko}, Alexei V. and {Finkelstein}, Steven L. and {Frazer}, Chris and {Gardner}, Jonathan P. and {Garnavich}, Peter M. and {Gawiser}, Eric and {Gruetzbauch}, Ruth and {Hartley}, Will G. and {H{\"a}ussler}, Boris and {Herrington}, Jessica and {Hopkins}, Philip F. and {Huang}, Jia-Sheng and {Jha}, Saurabh W. and {Johnson}, Andrew and {Kartaltepe}, Jeyhan S. and {Khostovan}, Ali A. and {Kirshner}, Robert P. and {Lani}, Caterina and {Lee}, Kyoung-Soo and {Li}, Weidong and {Madau}, Piero and {McCarthy}, Patrick J. and {McIntosh}, Daniel H. and {McLure}, Ross J. and {McPartland}, Conor and {Mobasher}, Bahram and {Moreira}, Heidi and {Mortlock}, Alice and {Moustakas}, Leonidas A. and {Mozena}, Mark and {Nandra}, Kirpal and {Newman}, Jeffrey A. and {Nielsen}, Jennifer L. and {Niemi}, Sami and {Noeske}, Kai G. and {Papovich}, Casey J. and {Pentericci}, Laura and {Pope}, Alexandra and {Primack}, Joel R. and {Ravindranath}, Swara and {Reddy}, Naveen A. and {Renzini}, Alvio and {Rix}, Hans-Walter and {Robaina}, Aday R. and {Rosario}, David J. and {Rosati}, Piero and {Salimbeni}, Sara and {Scarlata}, Claudia and {Siana}, Brian and {Simard}, Luc and {Smidt}, Joseph and {Snyder}, Diana and {Somerville}, Rachel S. and {Spinrad}, Hyron and {Straughn}, Amber N. and {Telford}, Olivia and {Teplitz}, Harry I. and {Trump}, Jonathan R. and {Vargas}, Carlos and {Villforth}, Carolin and {Wagner}, Cory R. and {Wandro}, Pat and {Wechsler}, Risa H. and {Weiner}, Benjamin J. and {Wiklind}, Tommy and {Wild}, Vivienne and {Wilson}, Grant and {Wuyts}, Stijn and {Yun}, Min S.},
        title = "{CANDELS: The Cosmic Assembly Near-infrared Deep Extragalactic Legacy Survey{\textemdash}The Hubble Space Telescope Observations, Imaging Data Products, and Mosaics}",
      journal = {\apjs},
         year = 2011,
        month = dec,
       volume = {197},
       number = {2},
          eid = {36},
        pages = {36},
          doi = {10.1088/0067-0049/197/2/36},
archivePrefix = {arXiv},
       eprint = {1105.3754},
 primaryClass = {astro-ph.CO},
       adsurl = {https://ui.adsabs.harvard.edu/abs/2011ApJS..197...36K}
}

@ARTICLE{2007leauthaud,
       author = {{Leauthaud}, Alexie and {Massey}, Richard and {Kneib}, Jean-Paul and {Rhodes}, Jason and {Johnston}, David E. and {Capak}, Peter and {Heymans}, Catherine and {Ellis}, Richard S. and {Koekemoer}, Anton M. and {Le F{\`e}vre}, Oliver and {Mellier}, Yannick and {R{\'e}fr{\'e}gier}, Alexandre and {Robin}, Annie C. and {Scoville}, Nick and {Tasca}, Lidia and {Taylor}, James E. and {Van Waerbeke}, Ludovic},
        title = "{Weak Gravitational Lensing with COSMOS: Galaxy Selection and Shape Measurements}",
      journal = {\apjs},
         year = 2007,
        month = sep,
       volume = {172},
       number = {1},
        pages = {219-238},
          doi = {10.1086/516598},
archivePrefix = {arXiv},
       eprint = {astro-ph/0702359},
 primaryClass = {astro-ph},
       adsurl = {https://ui.adsabs.harvard.edu/abs/2007ApJS..172..219L}
}

@ARTICLE{2024bleem,
       author = {{Bleem}, L.~E. and {Klein}, M. and {Abbot}, T.~M.~C. and {Ade}, P.~A.~R. and {Aguena}, M. and {Alves}, O. and {Anderson}, A.~J. and {Andrade-Oliveira}, F. and {Ansarinejad}, B. and {Archipley}, M. and {Ashby}, M.~L.~N. and {Austermann}, J.~E. and {Bacon}, D. and {Beall}, J.~A. and {Bender}, A.~N. and {Benson}, B.~A. and {Bianchini}, F. and {Bocquet}, S. and {Brooks}, D. and {Burke}, D.~L. and {Calzadilla}, M. and {Carlstrom}, J.~E. and {Carnero Rosell}, A. and {Carretero}, J. and {Chang}, C.~L. and {Chaubal}, P. and {Chiang}, H.~C. and {Chou}, T-L. and {Citron}, R. and {Corbett Moran}, C. and {Costanzi}, M. and {Constanzi}, M. and {Crawford}, T.~M. and {Crites}, A.~T. and {da Costa}, L.~N. and {de Haan}, T. and {De Vicente}, J. and {Desai}, S. and {Dobbs}, M.~A. and {Doel}, P. and {Everett}, W. and {Ferrero}, I. and {Flaugher}, B. and {Floyd}, B. and {Friedel}, D. and {Frieman}, J. and {Gallicchio}, J. and {Garc'ia-Bellido}, J. and {Gatti}, M. and {George}, E.~M. and {Giannini}, G. and {Grandis}, S. and {Gruen}, D. and {Gruendl}, R.~A. and {Gupta}, N. and {Gutierrez}, G. and {Halverson}, N.~W. and {Hinton}, S.~R. and {Hinton}, S.~R. and {Holder}, G.~P. and {Hollowood}, D.~L. and {Holzapfel}, W.~L. and {Honscheid}, K. and {Hrubes}, J.~D. and {Huang}, N. and {Hubmayr}, J. and {Irwin}, K.~D. and {Mena-Fern{\'a}ndez}, J. and {James}, D.~J. and {K{\'e}ruzor{\'e}}, F. and {Knox}, L. and {Kuehn}, K. and {Lahav}, O. and {Lee}, A.~T. and {Lee}, S. and {Li}, D. and {Lowitz}, A. and {Marshal}, J.~L. and {McDonald}, M. and {McMahon}, J.~J. and {Menanteau}, F. and {Meyer}, S.~S. and {Miquel}, R. and {Mohr}, J.~J. and {Montgomery}, J. and {Myles}, J. and {Natoli}, T. and {Nibarger}, J.~P. and {Noble}, G.~I. and {Novosad}, V. and {Ogando}, R.~L.~C. and {Padin}, S. and {Patil}, S. and {Pereira}, M.~E.~S. and {Pieres}, A. and {Plazas Malag'on}, A.~A. and {Pryke}, C. and {Reichardt}, C.~L. and {Rodr'iguez-Monroy}, M. and {Romer}, A.~K. and {Ruhl}, J.~E. and {Saliwanchik}, B.~R. and {Salvati}, L. and {Sanchez}, E. and {Saro}, A. and {Schaffer}, K.~K. and {Schrabback}, T. and {Sevilla-Noarbe}, I. and {Sievers}, C. and {Smecher}, G. and {Smith}, M. and {Somboonpanyakul}, T. and {Stalder}, B. and {Stark}, A.~A. and {Suchyta}, E. and {Swanson}, M.~E.~C. and {Tarle}, G. and {To}, C. and {Tucker}, C. and {Veach}, T. and {Vieira}, J.~D. and {Vincenzi}, M. and {Wang}, G. and {Weller}, J. and {Whitehorn}, N. and {Wiseman}, P. and {Wu}, W.~L.~K. and {Yefremenko}, V. and {Zebrowski}, J.~A. and {Zhang}, Y.},
        title = "{Galaxy Clusters Discovered via the Thermal Sunyaev-Zel'dovich Effect in the 500-square-degree SPTpol Survey}",
      journal = {The Open Journal of Astrophysics},
         year = 2024,
        month = feb,
       volume = {7},
          eid = {13},
        pages = {13},
          doi = {10.21105/astro.2311.07512},
archivePrefix = {arXiv},
       eprint = {2311.07512},
 primaryClass = {astro-ph.CO},
       adsurl = {https://ui.adsabs.harvard.edu/abs/2024OJAp....7E..13B}
}

@ARTICLE{2021hilton,
       author = {{Hilton}, M. and {Sif{\'o}n}, C. and {Naess}, S. and {Madhavacheril}, M. and {Oguri}, M. and {Rozo}, E. and {Rykoff}, E. and {Abbott}, T.~M.~C. and {Adhikari}, S. and {Aguena}, M. and {Aiola}, S. and {Allam}, S. and {Amodeo}, S. and {Amon}, A. and {Annis}, J. and {Ansarinejad}, B. and {Aros-Bunster}, C. and {Austermann}, J.~E. and {Avila}, S. and {Bacon}, D. and {Battaglia}, N. and {Beall}, J.~A. and {Becker}, D.~T. and {Bernstein}, G.~M. and {Bertin}, E. and {Bhandarkar}, T. and {Bhargava}, S. and {Bond}, J.~R. and {Brooks}, D. and {Burke}, D.~L. and {Calabrese}, E. and {Carrasco Kind}, M. and {Carretero}, J. and {Choi}, S.~K. and {Choi}, A. and {Conselice}, C. and {da Costa}, L.~N. and {Costanzi}, M. and {Crichton}, D. and {Crowley}, K.~T. and {D{\"u}nner}, R. and {Denison}, E.~V. and {Devlin}, M.~J. and {Dicker}, S.~R. and {Diehl}, H.~T. and {Dietrich}, J.~P. and {Doel}, P. and {Duff}, S.~M. and {Duivenvoorden}, A.~J. and {Dunkley}, J. and {Everett}, S. and {Ferraro}, S. and {Ferrero}, I. and {Fert{\'e}}, A. and {Flaugher}, B. and {Frieman}, J. and {Gallardo}, P.~A. and {Garc{\'\i}a-Bellido}, J. and {Gaztanaga}, E. and {Gerdes}, D.~W. and {Giles}, P. and {Golec}, J.~E. and {Gralla}, M.~B. and {Grandis}, S. and {Gruen}, D. and {Gruendl}, R.~A. and {Gschwend}, J. and {Gutierrez}, G. and {Han}, D. and {Hartley}, W.~G. and {Hasselfield}, M. and {Hill}, J.~C. and {Hilton}, G.~C. and {Hincks}, A.~D. and {Hinton}, S.~R. and {Ho}, S.-P.~P. and {Honscheid}, K. and {Hoyle}, B. and {Hubmayr}, J. and {Huffenberger}, K.~M. and {Hughes}, J.~P. and {Jaelani}, A.~T. and {Jain}, B. and {James}, D.~J. and {Jeltema}, T. and {Kent}, S. and {Knowles}, K. and {Koopman}, B.~J. and {Kuehn}, K. and {Lahav}, O. and {Lima}, M. and {Lin}, Y.-T. and {Lokken}, M. and {Loubser}, S.~I. and {MacCrann}, N. and {Maia}, M.~A.~G. and {Marriage}, T.~A. and {Martin}, J. and {McMahon}, J. and {Melchior}, P. and {Menanteau}, F. and {Miquel}, R. and {Miyatake}, H. and {Moodley}, K. and {Morgan}, R. and {Mroczkowski}, T. and {Nati}, F. and {Newburgh}, L.~B. and {Niemack}, M.~D. and {Nishizawa}, A.~J. and {Ogando}, R.~L.~C. and {Orlowski-Scherer}, J. and {Page}, L.~A. and {Palmese}, A. and {Partridge}, B. and {Paz-Chinch{\'o}n}, F. and {Phakathi}, P. and {Plazas}, A.~A. and {Robertson}, N.~C. and {Romer}, A.~K. and {Carnero Rosell}, A. and {Salatino}, M. and {Sanchez}, E. and {Schaan}, E. and {Schillaci}, A. and {Sehgal}, N. and {Serrano}, S. and {Shin}, T. and {Simon}, S.~M. and {Smith}, M. and {Soares-Santos}, M. and {Spergel}, D.~N. and {Staggs}, S.~T. and {Storer}, E.~R. and {Suchyta}, E. and {Swanson}, M.~E.~C. and {Tarle}, G. and {Thomas}, D. and {To}, C. and {Trac}, H. and {Ullom}, J.~N. and {Vale}, L.~R. and {Van Lanen}, J. and {Vavagiakis}, E.~M. and {De Vicente}, J. and {Wilkinson}, R.~D. and {Wollack}, E.~J. and {Xu}, Z. and {Zhang}, Y.},
        title = "{The Atacama Cosmology Telescope: A Catalog of >4000 Sunyaev-Zel{\textquoteright}dovich Galaxy Clusters}",
      journal = {\apjs},
         year = 2021,
        month = mar,
       volume = {253},
       number = {1},
          eid = {3},
        pages = {3},
          doi = {10.3847/1538-4365/abd023},
archivePrefix = {arXiv},
       eprint = {2009.11043},
 primaryClass = {astro-ph.CO},
       adsurl = {https://ui.adsabs.harvard.edu/abs/2021ApJS..253....3H}
}

@ARTICLE{2011grogin,
       author = {{Grogin}, Norman A. and {Kocevski}, Dale D. and {Faber}, S.~M. and {Ferguson}, Henry C. and {Koekemoer}, Anton M. and {Riess}, Adam G. and {Acquaviva}, Viviana and {Alexander}, David M. and {Almaini}, Omar and {Ashby}, Matthew L.~N. and {Barden}, Marco and {Bell}, Eric F. and {Bournaud}, Fr{\'e}d{\'e}ric and {Brown}, Thomas M. and {Caputi}, Karina I. and {Casertano}, Stefano and {Cassata}, Paolo and {Castellano}, Marco and {Challis}, Peter and {Chary}, Ranga-Ram and {Cheung}, Edmond and {Cirasuolo}, Michele and {Conselice}, Christopher J. and {Roshan Cooray}, Asantha and {Croton}, Darren J. and {Daddi}, Emanuele and {Dahlen}, Tomas and {Dav{\'e}}, Romeel and {de Mello}, Du{\'\i}lia F. and {Dekel}, Avishai and {Dickinson}, Mark and {Dolch}, Timothy and {Donley}, Jennifer L. and {Dunlop}, James S. and {Dutton}, Aaron A. and {Elbaz}, David and {Fazio}, Giovanni G. and {Filippenko}, Alexei V. and {Finkelstein}, Steven L. and {Fontana}, Adriano and {Gardner}, Jonathan P. and {Garnavich}, Peter M. and {Gawiser}, Eric and {Giavalisco}, Mauro and {Grazian}, Andrea and {Guo}, Yicheng and {Hathi}, Nimish P. and {H{\"a}ussler}, Boris and {Hopkins}, Philip F. and {Huang}, Jia-Sheng and {Huang}, Kuang-Han and {Jha}, Saurabh W. and {Kartaltepe}, Jeyhan S. and {Kirshner}, Robert P. and {Koo}, David C. and {Lai}, Kamson and {Lee}, Kyoung-Soo and {Li}, Weidong and {Lotz}, Jennifer M. and {Lucas}, Ray A. and {Madau}, Piero and {McCarthy}, Patrick J. and {McGrath}, Elizabeth J. and {McIntosh}, Daniel H. and {McLure}, Ross J. and {Mobasher}, Bahram and {Moustakas}, Leonidas A. and {Mozena}, Mark and {Nandra}, Kirpal and {Newman}, Jeffrey A. and {Niemi}, Sami-Matias and {Noeske}, Kai G. and {Papovich}, Casey J. and {Pentericci}, Laura and {Pope}, Alexandra and {Primack}, Joel R. and {Rajan}, Abhijith and {Ravindranath}, Swara and {Reddy}, Naveen A. and {Renzini}, Alvio and {Rix}, Hans-Walter and {Robaina}, Aday R. and {Rodney}, Steven A. and {Rosario}, David J. and {Rosati}, Piero and {Salimbeni}, Sara and {Scarlata}, Claudia and {Siana}, Brian and {Simard}, Luc and {Smidt}, Joseph and {Somerville}, Rachel S. and {Spinrad}, Hyron and {Straughn}, Amber N. and {Strolger}, Louis-Gregory and {Telford}, Olivia and {Teplitz}, Harry I. and {Trump}, Jonathan R. and {van der Wel}, Arjen and {Villforth}, Carolin and {Wechsler}, Risa H. and {Weiner}, Benjamin J. and {Wiklind}, Tommy and {Wild}, Vivienne and {Wilson}, Grant and {Wuyts}, Stijn and {Yan}, Hao-Jing and {Yun}, Min S.},
        title = "{CANDELS: The Cosmic Assembly Near-infrared Deep Extragalactic Legacy Survey}",
      journal = {\apjs},
         year = 2011,
        month = dec,
       volume = {197},
       number = {2},
          eid = {35},
        pages = {35},
          doi = {10.1088/0067-0049/197/2/35},
archivePrefix = {arXiv},
       eprint = {1105.3753},
 primaryClass = {astro-ph.CO},
       adsurl = {https://ui.adsabs.harvard.edu/abs/2011ApJS..197...35G}
}
\bibliographystyle{aasjournal}



\end{document}